\documentclass[aps, twocolumn]{emulateapj}

\usepackage{graphicx}
\usepackage{color}
\usepackage{amsmath}
\usepackage{mathrsfs}
\usepackage{braket}
\usepackage[ruled,vlined]{algorithm2e}
\usepackage{algpseudocode}
\usepackage{graphics}
\usepackage{epsfig}

\renewcommand{\footnote}{\fnsymbol{footnote}}

\shorttitle{a Monte Carlo code to solve radiative transfer based on
Neumann Solution} \shortauthors{Yang et al.}

\begin{document}

\title{A New Fast Monte Carlo Code for Solving Radiative Transfer Equations based on Neumann Solution}

\author{Yang Xiao-lin\altaffilmark{1,2,3,4}, Wang
Jian-cheng\altaffilmark{1,2,3,4}, Yang Chu-yuan\altaffilmark{1,2,3}}
\altaffiltext{1} {Yunnan Observatories, Chinese Academy of Sciences,
396 Yangfangwang, Guandu District, Kunming, 650216, P. R. China}
\altaffiltext{2} {Key Laboratory for the Structure and Evolution of
Celestial Objects, Chinese Academy of Sciences, 396 Yangfangwang,
Guandu District, Kunming, 650216, P. R. China} \altaffiltext{3}
{Center for Astronomical Mega-Science, Chinese Academy of Sciences,
20A Datun Road, Chaoyang District, Beijing, 100012, P. R. China}
\altaffiltext{4} {University of Chinese Academy of Sciences,
Beijing, 100049, P. R. China} \altaffiltext{1}{Email:
yangxl@ynao.ac.cn} \altaffiltext{2}{Email: jcwang@ynao.ac.cn}


\begin{abstract}
In this paper, we proposed a new Monte Carlo radiative transport
(MCRT) scheme, which is based completely on the Neumann series
solution of Fredholm integral equation. This scheme indicates that
the essence of MCRT is the calculation of infinite terms of multiple
integrals in Neumann solution simultaneously. Under this perspective
we redescribed MCRT procedure systematically, in which the main work
amounts to choose an associated probability distribution function
(PDF) for a set of random variables and the corresponding unbiased
estimation functions. We can select a relatively optimal estimation
procedure that has a lower variance from an infinite possible
choices, such as the term by term estimation. In this scheme, MCRT
can be regarded as a pure problem of integral evaluation, rather
than as the tracing of random walking photons. Keeping this in mind,
one can avert some subtle intuitive mistakes. In addition the
$\delta$-functions in these integrals can be eliminated in advance
by integrating them out directly. This fact together with the
optimal chosen random variables can remarkably improve the Monte
Carlo (MC) computational efficiency and accuracy, especially in
systems with axial or spherical symmetry. An MCRT code,
Lemon\footnote{ The code is available on GitHub codebase:
\url{https://github.com/yangxiaolinyn/Lemon} and version 1.0 is
archived in Zenodo:
\url{https://doi.org/10.5281/zenodo.4686355}.}({\bf{L}}inear
Integral {\bf{E}}quations' {\bf{M}}onte Carlo S{\bf{o}}lver Based on
the {\bf{N}}eumann solution), has been developed completely based on
this scheme. Finally, we intend to verify the validation of Lemon, a
suite of test problems mainly restricted to flat spacetime have been
reproduced and the corresponding results are illustrated in detail.

\end{abstract}


\section{Introduction}

Radiative transfer (RT) constantly plays an important role in
astrophysical researches. It can not only give us the emergent
spectra and light curves of various astrophysical systems directly
but also participate the co-evolution of these systems
indispensably. Naturally, RT process is completely dictated and
described by the radiative transfer equation (RTE), which is an
integro-differential equation when the scattering contributions are
taken into account \citep{Chand1960, Pom1973}. To solve RTE, various
methods have been proposed over the last decades
\citep{lindquist1966, Connors1980, gorecki1985, hauschildt1991,
Haardt1993, Pout1996a, zane1996, bottcher2001, Dolence2009,
yuan2009, gammie2012, dexter2010, Schnittman2010, younsi2012,
Schnittman2013, dexter2016, taka2017, ryan2020}. Roughly speaking,
these methods can be classified as analytical
\citep{lopez1999,semel1999} and numerical \citep{jan2017a,
jan2017b,jan2019} ones.

The most relevant and widely used numerical method is MC method due
to its remarkable simplicity and powerful efficiency in dealing with
high-dimensional integrations \citep{Pozd1983,whit2011,noeb2019},
which is crucial to solve RTE. Especially, the accuracy of the MC
method is only dependant on the sample size: N, but it is irrelevant
with the dimension of the system. The MC method can deal with
complicated problems provided it can be converted into probability
ones with given PDFs. Usually the sampling algorithms of MC method
are very simple and can be easily implemented by a program language.
Even though with these virtues, the results produced by the MC
method have a famous $N^{-1/2}$ convergent rate, which is quite low
and means that a relatively higher accuracy needs a sufficient large
set of samples.

The MC method is introduced mainly to treat the RT problems with
scattering processes incorporated, which are dictated by the
differentia-integral equations. Up to now, a lot of works based upon
the MC method dedicated to solve such equations (which are even
though not provided explicitly in these works) have been done
\citep{Connors1980, stern1995, Hua1997, Dolence2009, Schnittman2013,
ryan2015, zhangwd2019, mosc2020}. Very early, \citet{Connors1980}
used the MC method to trace the polarized RT around a Kerr black
hole and calculated the spectra emerging from a hot electron cloud
and accretion disk. They adopted the Walker-Penrose
(\citet{walker1970}) complex constant to calculate the parallel
transported polarization vector. This scheme was widely used in
polarized RT later. \citet{Dolence2009} proposed a new scheme and
developed a public available code: grmonty, aiming to calculate the
unpolarized synchrotron spectra of hot plasmas with Compton
scattering considered in full general relativity. In grmonty, the
superphoton, also called as "photon packets", plays a key role.
Later, the scheme of grmonty, especially the scenario of
superphoton, was fully adopted by the codes: bhlight
\citep{ryan2015} and Pandurata \citep{Schnittman2013}. Pandurata is
an MC code aiming for polarized radiation transport around Kerr
black holes, including arbitrary emission and absorption effects, as
well as electron scattering. In Pandurata the superphotons have
broadband energies, while in grmonty they are monoenergetic
\citep{Schnittman2013}. In order to implement a more self-consistent
calculations of Comptonised energy spectra for extended coronae in
Kerr spacetime, an MC polarized RT code, monk, has been developed by
\citet{zhangwd2019}.

The descriptions of these codes and many other analogous ones are in
a very physical intuitive manner, i.e., emitting a superphoton, then
randomly tracing it until either it escapes from the radiative
region or is absorbed by the medium. Of course, during the
propagation, the photon experienced Compton scattering and
absorption in the media. In this physical intuitive description, the
RTE seems unimportant and is rarely mentioned. But one will see
later that this is not true and we believe that the importance of
RTE has been underestimated in the former works in some sense.

In contrast, an RTE without scattering is a differential equation
and can usually be solved based on the ray-tracing approach, rather
than the MC method \citep{broderick2003, lilixin2009, Huang2011,
chen2015, dexter2016, meliani2017, piha2017, piha2018, mosc2018,
bron2018, chan2019, tsun2020, vincent2020}. In curved spacetime, the
geodesics along which the radiation propagates (also including the
polarization vector if the radiation is polarized) should be solved
simultaneously. Hence a fast and accuracy geodesic solver is crucial
\citep{dexter2009, dauser2010, vincent2011, yangxl2013}. Unlike
the scattering incorporated cases, the RTE plays a central role in
these works and is given explicitly.

There are two different paradigms when tracing a radiation, i.e.,
emitter-to-observer \citep{cunning1975, rauch1994, broderick2003,
dov2004, schnittman2006} and observer-to-emitter \citep{laor1990,
kojima1991, Dolence2009, psaltis2012, Schnittman2013}. They are
suitable for RT circumstances with and without scattering
incorporated, respectively. In the emitter-to-observer framework,
the transfer function is crucial and needs to be evaluated
numerically and tabulated in advance to obtain the observational
quantities at infinity \citep{cunning1975,laor1990}.

When one deals with polarized radiation transport, polarization
vector is a relevant concept \citep{Chand1960}. In fact the Stokes
parameters of $I, Q, U, V$ are determined up to a rotation in the
plane perpendicular to the propagation direction of the radiation.
If the reference frame in that plane is rotated anticlockwise by a
angle $\phi$, the components $Q$ and $U$ will change as
$Q'=Q\cos2\phi + U\sin2\phi, U'=-Q\sin2\phi + U\cos2\phi$. It means
that we need a determined frame to fix the values of the Stokes
parameters, which is equivalent to a single vector since the frame
is in a plane. Hence we have to trace the polarization vector along
the ray trajectory and the vector moves in a parallel transport
manner. In flat spacetime, a parallel transported vector remains
unchanged, thus the tracing is trivial. While in curved spacetime,
the parallel transport of a vector $\mathbf{f}=f^\mu\partial_\mu$ is
determined by the equation $D f^\mu / d\lambda=df^\mu/d\lambda +
\Gamma^\mu_{\alpha\beta}f^\alpha k^\beta=0$, where $D f^\mu /
d\lambda$ is the covariant derivative, $\lambda$ and $k^\beta$ are
the affine parameter and tangent vector of the trajectory,
respectively. Solving these equations numerically is time consuming
\citep{chen2015}. In Kerr spacetime, due to the existence of the
complex-valued Wolker-Penrose constant $k_{\text{wp}}$
\citep{walker1970} and the conditions of
$\mathbf{f}\cdot\mathbf{f}=1$ and $\mathbf{f}\cdot\mathbf{k}=0$, the
whole problem amounts to solve a set of linear algebraic equations
\citep{connors1977, Connors1980}.

In curved spacetime, the parallel transport of polarization vector
can be incorporated into the polarized RTEs through several
different but equivalent formalisms. For example, \citet{shche2011}
developed a mechanism, in which a orthogonal tetrad was parallel
transported along the ray from the observer to the black hole, and
the coefficient matrix of absorption, Faraday rotation and
conversion was modified by a rotation. Latter, \citet{gammie2012}
proposed a more generic formalism and demonstrated the equivalence
of approaches adopted by \citet{broderick2003, broderick2004},
\citet{Schnittman2010} and \citet{shche2011}. These formalisms were
recently employed by many authors \citep{dexter2016, jim2018,
piha2018, mosc2018, tsun2020, dexter2020} to obtain more predictable
features for polarized RT in complicated GRMHD simulations, which
could potentially impose more precise constrains on the high quality
observational data. One may notice that in all those works, the
scattering process was not included. Thus for the RT without
scattering, the RTE plays an essential role and is usually solved by
a non-MC method. But for the RT with scattering, the MC method is
employed and described in a quite physically intuitive manner, and
the RTEs are mentioned rarely. Motivated by this unsymmetrical and
unsatisfactory situation for the RTEs and MC method, we intend to
develop an RT scheme which employs the MC method and RTEs equally
and simultaneously, or a MCRT scheme based upon the RTEs. As we will
see that the RT without scattering can also be solved by this new
scheme appropriately, even though not so efficiently.

This motivation is also inspired by several other reasons. First,
the MC method initially introduced in the RT aims to evaluate the
infinite terms of multiple integrals in the Neumann solution of RTE
\citep{Davison1957}. Regarding MC method from this mathematical
perspective can remarkably improve the calculation efficiency and
accuracy compared to the superphoton scheme. Second, since the whole
work is integral evaluation, we are flexible to choose various PDFs
for sampling and weight functions for unbiased estimations. These
choices are equivalent but with different computational efficiency
and variance. Therefore we can select a relatively optimal PDF to
improve the accuracies of integral evaluation. Especially our scheme
can avoid subtle mistakes that may be caused by physical intuitions.
Third, the RTE and its Neumann solution are oriented in our scheme.
They are also the starting point of the whole framework. If the RTE
can be greatly simplified at the very beginning under a given
condition, such as the axial symmetry of the system, the
corresponding MC sampling procedure will be simplified as well. And
the PDFs for scattering sampling could be totally nonphysical but
mathematically correct.

This paper is organized as follows. In section \ref{methods_2}, we
give a detailed description of our scheme built upon the RTE and its
Neumann solution, including the sampling procedures for position
transport, scattering and observational quantity estimations. Next
we extend this scheme to deal with polarized radiative transfer
processes in Section \ref{sec3_PRT}. We verify our scheme through
its applications in various radiative transfer problems in Section
\ref{sec4_verify}. Finally a brief discussion on our scheme and its
limitations is presented in Section \ref{sec5_discussion}.

\section{Methods}
\label{methods_2}

In this section we will illustrate the scheme through a simple
example, which contains all relevant ingredients. We will
demonstrate that the observational quantities can be expressed as an
infinite series of multiple integrals via the Neumann solution and
the MC method is introduced for evaluating these integrals.

\subsection{Neumann Series Solution and Recording Function}
The RT is essentially a particle transport process described exactly by
the Boltzmann's equation (BE). Therefore an RTE is actually
identical with a BE. The difference is that the particles (here is
photon) do not interact with each other except acting with the
external medium. We also do not consider the inducing process
causing the RTE to be non-linear \citep{Pom1973}. Taking the
function $N(\nu, \mathbf{\Omega}, \mathbf{r})$ to describe the
photon number of unpolarized radiation distributed over position
$\mathbf{r}$, frequency $\nu$ and direction $\mathbf{\Omega}$.
Without lost of generality, we assume that the radiative field is
time independent, the RTE simply reads \citep{Pom1973}
\begin{eqnarray}
\label{eqa1}
\begin{split}
& \mathbf{\Omega}\cdot\nabla N(\nu, \mathbf{\Omega}, \mathbf{r})
+\sigma(\nu, \mathbf{\Omega}, \mathbf{r})N(\nu, \mathbf{\Omega}, \mathbf{r})=S(\nu,\mathbf{\Omega}, \mathbf{r})\displaystyle \\
&+\int_0^\infty\int_{4\pi}\sigma_s(\nu'\rightarrow\nu,\mathbf{\Omega}'\rightarrow\mathbf{\Omega},\mathbf{r})
N(\nu',\mathbf{\Omega}', \mathbf{r})d\nu'd\mathbf{\Omega}',
\end{split}
\end{eqnarray}
where $\sigma = \sigma_a + \sigma_s$, $\sigma_a$ and $\sigma_s$ are
the absorption and scattering coefficients, respectively and
\begin{eqnarray}
\begin{split}
&\sigma_s(\nu', \mathbf{\Omega}') =
\int\sigma_s(\nu'\rightarrow\nu,\mathbf{\Omega}'\rightarrow\mathbf{\Omega},
\mathbf{r})d\nu d\mathbf{\Omega}.
\end{split}
\end{eqnarray}
$S(\nu,\mathbf{\Omega}, \mathbf{r})$ is the emissivity of the
medium. Eq. \eqref{eqa1} is an integro-differential equation. In
order to solve it numerically by the MC method, we need to recast it
into an integral equation. To accomplish this, we define the right
hand side of Eq. \eqref{eqa1} as an auxiliary quantity
$\mathbf{\psi}$, i.e.,
\begin{eqnarray}\label{eqa2}
\begin{split}
&\displaystyle \mathbf{\psi}(\nu,\mathbf{\Omega},
\mathbf{r})=S(\nu,\mathbf{\Omega}, \mathbf{r}) \displaystyle
\\&+\int_0^\infty\int_{4\pi}\sigma_s(\nu'\rightarrow\nu,\mathbf{\Omega}'\rightarrow\mathbf{\Omega},
\mathbf{r}) N(\nu',\mathbf{\Omega}',
\mathbf{r})d\nu'd\mathbf{\Omega}'.
\end{split}
\end{eqnarray}
Obviously $\mathbf{\psi}$ represents the total photon number flux of
the radiation field at a given point, direction and frequency.
Taking $\mathbf{\psi}$ as a quantity already known and solving Eq.
\eqref{eqa1} formally, we have \citep{Pom1973}
\begin{eqnarray}\label{eqa3}
\begin{split}
&\displaystyle N(\nu, \mathbf{\Omega}, \mathbf{r})=
\int_{0}^{|\mathbf{r}-\mathbf{r}_0|} \psi\left(\nu, \mathbf{\Omega},
\mathbf{r}-s\mathbf{\Omega}\right)\\
&\times\exp\left(-\int^{s}_0\sigma\left(\nu, \mathbf{\Omega},
\mathbf{r}-s'\mathbf{\Omega} \right)ds'\right)ds.
\end{split}
\end{eqnarray}
This expression is very useful and its physical implication is
explicit, i.e., the photon number density at a given place and
direction equals the contributions of $\psi$ integrated along
the ray backward with a exponential attenuation factor multiplied at
each point, in which the factor accounts for the absorption and
scattering effects. Substituting Eq. \eqref{eqa3} into Eq.
\eqref{eqa2} to eliminate $N(\nu, \mathbf{\Omega}, \mathbf{r})$, we
obtain an integral equation for $\psi$
\begin{eqnarray}
\label{eqa4}
\begin{split}
 \mathbf{\psi}&\displaystyle(\nu,\mathbf{\Omega},
\mathbf{r})=S(\nu,\mathbf{\Omega}, \mathbf{r}) \displaystyle +
\int_0^\infty d\nu'\int_{4\pi}d\mathbf{\Omega}'\int_{0}^{\infty}ds\\
&\times \psi\left(\nu', \mathbf{\Omega}',
\mathbf{r}-s\mathbf{\Omega}'\right)
\sigma_s(\nu'\rightarrow\nu,\mathbf{\Omega}'\rightarrow\mathbf{\Omega}, \mathbf{r})\\
& \times\exp\left(-\int^{s}_0\sigma\left(\nu', \mathbf{\Omega}',
\mathbf{r}-s'\mathbf{\Omega}' \right)ds'\right).
\end{split}
\end{eqnarray}
To simplify the above equation, we introduce the normalized
scattering and transport kernels denoted by $C$ and $T_s$, and
\begin{eqnarray}\label{eqa5}
\begin{split}
&\displaystyle C(\nu'\rightarrow\nu,\mathbf{\Omega}'\rightarrow\mathbf{\Omega}| \mathbf{r})=
\frac{\sigma_s(\nu'\rightarrow\nu,\mathbf{\Omega}'\rightarrow\mathbf{\Omega}, \mathbf{r})}
{\sigma_s(\nu', \mathbf{\Omega}', \mathbf{r})},
\end{split}
\end{eqnarray}

\begin{eqnarray}\label{eqa6}
\begin{split}
 T_s(\mathbf{r}_s\rightarrow\mathbf{r}&|\nu',
\mathbf{\Omega}')=\displaystyle\frac{1}{A}\sigma_s(\nu',
\mathbf{\Omega}', \mathbf{r})\\&
\times\exp\left(-\int^{s}_0\sigma_s\left(\nu', \mathbf{\Omega}',
\mathbf{r}_{s'} \right)ds'\right),
\end{split}
\end{eqnarray}
where $\mathbf{r}_s=\mathbf{r}-s\mathbf{\Omega}'$ and $A$ is the
normalization factor for $T_s$. Since both $C$ and $T_s$ are
normalized, they will be taken as PDFs for photon transport in
momentum and position spaces respectively. For simplicity, we define
the total transport kernel $K=T_s\cdot C\cdot w_s$, where
$\displaystyle w_s=A\cdot\exp\left(-\int^{s}_0\sigma_a\left(\nu',
\mathbf{\Omega}', \mathbf{r}_{s'} \right)ds'\right)$ is weight which
will be used as estimation for observational quantity. Using the
total absorption coefficient $\sigma$, we can choose another
position transport function $T$ given by
\begin{eqnarray}
\begin{split}
&\displaystyle
T(\mathbf{r}_s\rightarrow\mathbf{r}|\nu', \mathbf{\Omega}')=\\&\frac{1}{A}\sigma(\nu', \mathbf{\Omega}', \mathbf{r})
\exp\left(-\int^{s}_0\sigma\left(\nu', \mathbf{\Omega}', \mathbf{r}_{s'} \right)ds'\right),
\end{split}
\end{eqnarray}
and the corresponding weight function $w=A\sigma_s / \sigma$. Also
we have $K=T\cdot C\cdot w$. The two choices of $T$ correspond two
different sampling algorithms for position transport. We prefer to
choose $T_s$ and $w_s$ because they are more convenient for
sampling. Meanwhile the effect of absorption is considered in the
factor $w_s$. If we choose $T$ as the transport sampling PDF, the
weight becomes $A\sigma_s / \sigma$ and $\sigma_s / \sigma$ is the
probability that scattering happens.

If we denote $P=(\nu, \mathbf{\Omega}, \mathbf{r})$, then Eq.
\eqref{eqa4} can be written as
\begin{eqnarray}\label{eqa8}
\begin{split}
&\displaystyle  \psi(P)=S(P) + \int \psi\left(P'\right)K(P'\rightarrow P)dP',
\end{split}
\end{eqnarray}
which is mathematically called as the Fredholm integral equation of
second kind. It is well known that Eq. \eqref{eqa8} has the Neumann
series solution, which can be given directly by \citep{Davison1957}
\begin{eqnarray}\label{eqa9}
\begin{split}
&\displaystyle  \psi(P)=\sum_{m=0}^{\infty}\psi_m(P),
\end{split}
\end{eqnarray}
where
\begin{equation}\label{eqa10}
\begin{split}
&\displaystyle \psi_0(P)=S(P),\\
&\displaystyle \psi_1(P)=\int_{P_1}\psi_0(P_1)K(P_1\rightarrow P)dP_1\\
&=\int_{P_1}S(P_1)K(P_1\rightarrow P)dP_1,\\
&\displaystyle \psi_2(P)=\int_{P_2}\psi_1(P_2)K(P_2\rightarrow P)dP_2\\
&=\int_{P_1}\int_{P_1}S(P_1)K(P_1\rightarrow P_2)K(P_2\rightarrow P)dP_1dP_2,\\
&\cdots\\
&\displaystyle \psi_m(P)=\int_{P_m}\psi_{m-1}(P_m)K(P_m\rightarrow P)dP_m\\
\end{split}
\end{equation}
\begin{equation*}
\begin{split}
&=\int_{P_1}\cdots\int_{P_m}S(P_1)K(P_1\rightarrow P_2)K(P_2\rightarrow P_3)\cdots \\
&K(P_m\rightarrow P)dP_1dP_2\cdots dP_m.\\
\end{split}
\end{equation*}
The Neumann solution has a very simple physical
interpretation that the photon number flux of the radiation
field at $P$ is the sum of all photons travel to there after zero,
one, two and many times transportations.

To obtain the observed quantities at infinity in a special direction
$\mathbf{\Omega}_{\text{obs}}$, such as the total photon number flux
$F_{\text{obs}}$, we introduce the recording function $f(P)$ defined
as
\begin{eqnarray}
\label{f_rc}
\begin{split}
\displaystyle  f(\nu, \mathbf{\Omega},
\mathbf{r})=&\exp\left(-\int^{s}_0\sigma\left(\nu, \mathbf{\Omega},
\mathbf{r}-s'\mathbf{\Omega} \right)ds'\right)\\&
\times\delta(\mathbf{\Omega}-\mathbf{\Omega}_{\text{obs}}).
\end{split}
\end{eqnarray}
Noting Eq. \eqref{eqa3}, we have
\begin{eqnarray}
\begin{split}
I=\int_P \psi(P)f(P)dP.
\end{split}
\end{eqnarray}
Substituting the Neumann solution into above equation, we obtain
\begin{eqnarray}
\begin{split}
\label{I_mseq1}
\displaystyle I=\sum_{m=0}^{\infty} I_m,
\end{split}
\end{eqnarray}
where
\begin{equation}
\label{I_mseq2}
\begin{split}
&\displaystyle I_m=\int_{P_1}\cdots\int_{P_m}S(P_1)K(P_1\rightarrow P_2)K(P_2\rightarrow P_3) \\
&\cdots \times K(P_m\rightarrow P)f(P)dP_1dP_2\cdots dP_mdP.\\
\end{split}
\end{equation}
There are infinite number of multiple integrals needed to be
evaluated simultaneously, this work can be done by MC method in an
efficient way. In the next subsections, we will discuss the MC
method and the procedure to evaluate those integrals in detail.

\subsection{The Monte Carlo Method}
The strategy of the MC to calculate an integral $I=\int_a^b f(x)dx$
is very simple, i.e., the integral $I$ is regarded as the
expectation value of a random variable $y$ with a PDF: $p(y)$. On
the other hand, the expectation value can be approximated by the
algebraic average of a set of samples of $y$. Suppose that $y$ is a
function of another random varible $x$, i.e., $y=g(x)$, then the PDF
of $x$, $p(x)$ satisfies $p(y)dy=p(x)dx$. The expectation value of
$y$ is given approximately by
\begin{eqnarray}\label{eqa14}
\begin{split}
&\displaystyle \int yp(y)dy =\int g(x)p(x)dx \approx
\frac{1}{N}\sum_{i=1}^{N}g(x_i),
\end{split}
\end{eqnarray}
where $x_i$ is a set of samples of $x$ obtained by sampling $p(x)$.
Since $p(y)dy=p(x)dx$, $y_i=g(x_i)$ are samples of $y$. Then one can
split $f(x)$ as the product of two functions: $f(x)=g(x)p(x)$, where
$p(x)$ is normalized and will be used as a PDF to generate a set of
samples of $x$, we have
\begin{eqnarray}\label{eqa15}
\begin{split}
\displaystyle I=\int_a^b f(x)dx =\int_a^b g(x)p(x)dx \approx
\frac{1}{n}\sum_{i=1}^{n}g(x_i).
\end{split}
\end{eqnarray}
Notice that only $g(x)$ appears in the final expression, which is
usually called as the weight or the unbiased estimation of $I$. The
number of ways to split $f(x)$ is infinite. The various splitting
has different computational efficiency and variance. But
there exists an optimal choice for $f(x)\ge 0$, where $p(x)=f(x)/A$
and $A=\int f(x)dx$ is the normalization factor (if $f(x)$ can take
negative values, the optimal choice is different), for this
choice the variance is exactly vanished. But the problem is that
we do not know $A$ in advance and its value is exactly what we want
to calculate. It means that the optimal choice is impossible for
practical implementation. However this result tells us that we
should choose a $p(x)$ as similar with $f(x)$ as possible, the error
will also be relatively smaller. This is the essence of the
so-called important sampling.

\subsection{Transport Game: an Example of Illustration}
In the above section, we discuss how to calculate a single integral
by the MC method. But the quantity we want to evaluate given in Eq.
\eqref{I_mseq1} is an infinite series of multiple integrals. To
calculate these integrals simultaneously by the MC method, we need a
similar but extended scheme. Such scheme has been extensively
studied in nuclear physics sector, where neutron transport is
heavily concerned \citep{Davison1957}. \citet{Goer1958} proposed a
novel and systematic strategy, called as transport game, to evaluate
such a quantity. \citet{Spanier1959} put this scheme on a stringent
mathematical foundation and its validation can be proved by
probability and measure theories.

\begin{algorithm}[t]
   \label{algo0}
    \caption{Steps of transport game}
    \textcircled{1} generate $P_1$ by sampling $S(P_1)$,\\
    \textcircled{2} For any $l\ge 2$, compute $\tilde{p}(P_{l-1})=\displaystyle \int K(P_{l-1}\rightarrow P_l)dP_l$ for a given $P_{l-1}$
    and take $\tilde{p}(P_{l-1})$ as a survival probability,\\
    \textcircled{3} If $\xi \le \tilde{p}(P_{l-1})$ generate $P_l$ by sampling $\displaystyle\tilde{K}(P_l|P_{l-1})=
       \frac{K(P_{l-1}\rightarrow P_l)}{\displaystyle\int K(P_{l-1}\rightarrow P_l)dP_l}$,\\
       Repeat steps \textcircled{2} and \textcircled{3}\\
    \textcircled{4} If $\xi > \tilde{p}(P_M)$, then terminate the game.
\end{algorithm}

To illustrate how does the transport game work, we shall first
demonstrate a toy model containing the key ingredients. The steps of
the algorithm for the toy model are presented in Algorithm.
\ref{algo0}. After implementing these steps, one can obtain a state
sequence: $P_1, P_2, \cdots, P_M$. One of the weights associated
with this algorithm is given by
\begin{equation}
\begin{split}
&\displaystyle F_1 = \frac{f(P_M)}{1-\tilde{p}(P_M)}.
\end{split}
\end{equation}
Now we will show that $F_1$ is an unbiased estimation of the
photon number $I$. First, the probability corresponding to such
a state sequence is obviously given by
\begin{equation}\label{eqa19}
\begin{array}{lll}
& H(P_1, P_2, \cdots, P_M)dV&\\
=&\displaystyle S(P_1)\left(\prod_{l=2}^M\tilde{p}(P_{l-1})\tilde{K}(P_{l-1}\rightarrow P_l)\right)
\times(1-\tilde{p}(P_M))dV&\\
=&\displaystyle S(P_1)\left(\prod_{l=2}^MK(P_{l-1}\rightarrow P_l)\right)
\times(1-\tilde{p}(P_M))dV,&
\end{array}
\end{equation}
where $dV=dP_1dP_2\cdots dP_M$. Thus the expectation value of $F_1$
for an explicit $M$ is given by
\begin{equation}\label{eqa20}
\begin{array}{lll}
&\langle F_1\rangle_M=\displaystyle\int\cdots\int H(P_1, P_2, \cdots, P_M)F_1(P_M)dV&\\
=&\displaystyle\int\cdots\int S(P_1) \displaystyle\left(\prod_{l=2}^M
K(P_{l-1}\rightarrow P_l)\right)f(P_M)dV=I_M.&
\end{array}
\end{equation}
Then the average of $I$ is given by the sum of $\langle
F_1\rangle_M$ for all $M$, i.e., $\displaystyle
E[F_1]=\sum_{M=0}^\infty\langle F_1\rangle_M=\sum_{M=0}^\infty
I_M=I$.

Similarly one can show that
\begin{equation}\label{eqa21}
\begin{split}
&\displaystyle F_2 = \displaystyle\sum_{m=1}^M f(p_m),
\end{split}
\end{equation}
is another unbiased estimation of $I$, i.e., $\displaystyle
E[F_2]=\sum_{M=0}^\infty\langle F_2\rangle_M=\sum_{M=0}^\infty
I_M=I$. $F_1$ and $F_2$ are called as the final event and item by
item estimations of $I$, respectively. In addition, there are many
other unbiased estimations of $I$. For example, we can modify
Algorithm. \ref{algo0} by introducing a weight factor $w$ and obtain
Algorithm. \ref{algo01},
\begin{algorithm}[t]
   \label{algo01}
    \caption{Steps of transport game}
    \textcircled{1} generate $P_1$ by sampling $S(P_1)$ and set $w=1$;\\
    \textcircled{2} For any $l\ge 2$, compute $C(P_{l-1})=\displaystyle \int K(P_{l-1}\rightarrow P_l)dP_l$ for a given $P_{l-1}$,
   and take $C(P_{l-1})$ as the normalization factor;\\
  \textcircled{3} generate $P_l$ by sampling $\displaystyle\tilde{K}(P_l|P_{l-1})=\frac{1}{C(P_{l-1})}K(P_{l-1}\rightarrow P_l$);\\
  \textcircled{4} Update $w$ as $w=w\cdot C(P_{l-1})$;\\
    Repeat steps \textcircled{2},\,\,\textcircled{3} and \textcircled{4}.
\end{algorithm}
where, the unbiased estimations for $I$ is $F_3^m=w\cdot f(P_m)$.
In Algorithm. \ref{algo01}, the state sequence will not be
truncated but with a decreased weight $w$.

We can also understand the transport game from the MC splitting
scheme directly. Notice that the transport kernel is $K=T_s\cdot
C\cdot w_s$, for the $m$-th term $I_m$ given by Eq. \eqref{I_mseq2},
we can separate a associated PDF for all random variables: $P_1,
\cdots, P_m, P$ from the integrand as
\begin{equation}
\begin{split}
&\displaystyle \widetilde{f}(P1, \cdots, P_m, P) = S(P_1)\cdot
T_s(\mathbf{r}_1\rightarrow\mathbf{r}_2)\cdot C(\nu_1,\mathbf{\Omega}_1\rightarrow\\
&\nu_2, \mathbf{\Omega}_2|\mathbf{r}_2)\cdots
T_s(\mathbf{r}_m\rightarrow\mathbf{r})\cdot
C(\nu_m,\mathbf{\Omega}_m\rightarrow\nu,
\mathbf{\Omega}|\mathbf{r}).
\end{split}
\end{equation}
The remaining part of the integrand will be taken as the weight or
estimation functions for $I_m$, i.e.,
\begin{equation}
\label{F_estim}
\begin{split}
&\displaystyle F  = w_{s,1}\cdot w_{s,2} \cdots w_{s,m}\cdot f(P),
\end{split}
\end{equation}
where $w_s=A\cdot \exp(-\tau_a)$, $A$ is the normalization factor of
$T_s$, and $\tau_a$ is the absorption optical depth between any two
consecutive scattering points. Utilizing the associated PDF
$\widetilde{f}$, we can generate a set of samples for all of the
random variables and these samples can be regarded as a random
sequence (or Markov Chain) in the phase space. The detailed sampling
procedure will be discussed in the next section.

One can see that once a random sequence with $N$ components is
obtained, the estimations for all $I_m$ (where $m\le N$) can be
immediately obtained as well. Since the weight $w_s<1$, $I_m$ will
decrease as $m$ increases. Therefore we can set a tolerance value
$\varepsilon$ and once the condition $I_m\le\varepsilon$ is
satisfied, we can terminate the generation of the random sequence
immediately.

One may notice that there is a $\delta$-function in the recording
function $f(P)$, i.e.,
$\delta(\mathbf{\Omega}-\mathbf{\Omega}_{\text{obs}})$. If
$\mathbf{\Omega}\ne\mathbf{\Omega}_{\text{obs}}$, $F$ always
vanishes. In the traditional treatment for this difficult, one
chooses a bin with finite and small lengths $\Delta\theta\Delta\phi$
around $\mathbf{\Omega}_{\text{obs}}$. If $\mathbf{\Omega}$ falls
into it, then the contribution of $F$ is recorded, otherwise
rejected. Since all $\mathbf{\Omega}$ in the sequence are randomly
generated, few of them will make contributions to the evaluation of
$I$. Hence the significant portion of the random sequence is
absolutely discarded and wasted. We believe this is exactly the
reason why the estimation procedure of the traditional MC method has
a quite low computational efficiency and accuracy for the RT
solving. While in our scheme, these $\delta$-functions can be
eliminated directly by integrating them out before any concrete
calculations. Then the expressions of $I_m$ will be modified and we
can reselect the associated PDF and estimation function. They will
not contain any $\delta$-functions at all. The detailed discussions
will be presented later.

\subsection{Random Sequence Generation}

Now we discuss how to generate a random sequence by sampling the
emissivity function $S(P)$, position transport and scattering
kernels $T_s$ and $C$, respectively. The procedure is stated as
follows.

By sampling the source function $S(P)$, we can obtain the first
component of the sequence $P_1$. If $S(P)$ is too complicated to be
as a PDF, many sampling algorithms will become unfeasible except the
rejection of sampling method, but with a quite low efficiency
\citep{Dolence2009}. Therefore we will adopt the method proposed by
\citet{Pozd1983}, i.e., we sample $P_1$ from an uniform PDF instead
of $S(P)$ and add $S(P_1)$ to the weight function $F$, which is
revised as $F\cdot S(P_1)$. This method can simplify the way for
sampling $S(P)$ and the compensation is that the variance will be
increased for a little amount.

\subsubsection{the Position Transport}
\label{distancsample} 
\begin{figure}[t]
  \includegraphics[scale=0.75]{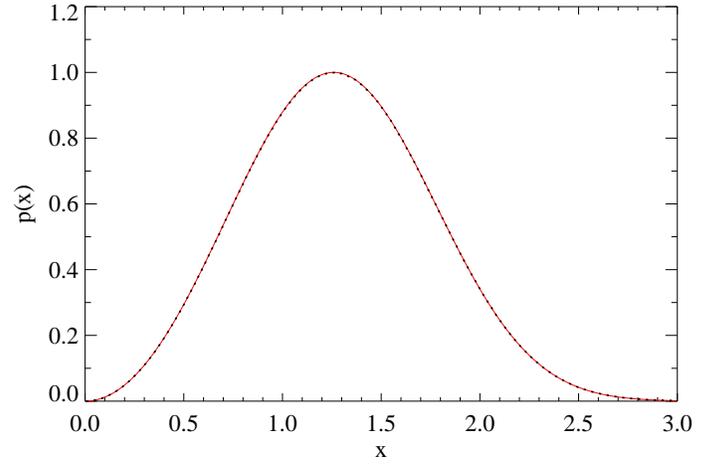}
\caption{A test of Algorithm. \ref{algo1} for a PDF with
$\sigma=x^2$. The red solid and black dotted lines are plotted
according to the function $p(x)$ and sampling results obtained by
Algorithm. \ref{algo1}, respectively.}
  \label{myfig1eps}
\end{figure}

Now we discuss how to produce the next
component $P_l$ of the sequence as the former $P_{l-1}$ is provided,
which is determined by the transfer kernel $K(P_{l-1}\rightarrow
P_l)$. For simplicity, we will use $P$ and $P'$ to replace $P_l$ and
$P_{l-1}$ respectively. First we consider the position transport
process. When $P'=P'(\mathbf{r}', \mathbf{\Omega}', \nu')$ is
specified, the PDF for position transport from $\mathbf{r}'$ to
$\mathbf{r}=\mathbf{r}'+s\mathbf{\Omega}'$ simply reads
\begin{eqnarray}
\label{eqa23}
\begin{split}
p(s)ds = \frac{1}{A}\exp[-\tau_s(s)]d\tau_s(s),
\end{split}
\end{eqnarray}
where $\tau_s(s)=\int^{s}_0\sigma_s\left(\mathbf{\Omega}', \nu',
\mathbf{r}'+t\mathbf{\Omega}' \right)dt$ is the scattering optical
depth and $A=1-\exp[-\tau_s(s_b)]$ is the normalization factor, and
$s_b$ is the distance from $\mathbf{r}'$ to the boundary surface of
the radiative region along $\mathbf{\Omega}'$. Since $p(s)$ is
integrable, we can sample it by the so-called inverse cumulative
distribution function (CDF) method, and the transfer distance is
determined by a random number (all random numbers through out this
manuscript are generated by a code adopted from the public available
code CosmoMC \citep{lewis2002}) $\xi$ through the following equation
\begin{eqnarray}\label{eqa27}
 \displaystyle \int^{s}_0 \sigma_s\left(\mathbf{\Omega}',\nu', \mathbf{r}'+t\mathbf{\Omega}' \right)dt=-\ln(1-\xi A).
\end{eqnarray}
For an uniform and isotropic media, $\sigma_s$ is a constant, we have
\begin{eqnarray}\label{eqa28}
 \displaystyle s=-\frac{1}{\sigma_s}\ln(1-\xi A).
\end{eqnarray}
If the integral on the LHS of Eq. \eqref{eqa27} is too complicated
to be integrated or the inverse procedure to solve the equation is
unfeasible, one can try an alternative procedure given in Algorithm.
\ref{algo1}, where $N= 1-\exp(-\sigma_{\text{max}}s_b)$,
$\sigma_\text{max}$ is the maximum of $\sigma_s(s)$, and $\xi_1$,
$\xi_2$, $\xi_3$ are random numbers. To illustrate the correctness
of Algorithm. \ref{algo1}, we employ it to sample a PDF given as:
$p(x)=\sigma(x)\exp(-\int^x_0\sigma(t)dt)=x^2\exp(-x^3/3)$ and
$x\in[0, 3]$, thus $\sigma_{\text{max}}=9$. The result is shown in
Figure \ref{myfig1eps}. One can find that the sampling result agrees
with $p(x)$ very well. A detailed proof of Algorithm. \ref{algo1} is
presented in Appendix \ref{app1}.
\begin{algorithm}[t]
   \label{algo1}
    \caption{Scattering Distance Sampling for Finite Region}
    \textcircled{1} let $s=0, i=0$\\
    \textcircled{2} let $s_1 = -\ln(1-\xi_1\cdot N)/\sigma_{\text{max}}, i=i+1$\\
    \textcircled{3} let $s = s + s_1$, \\
    \If{$s>s_\text{max}$}{
        goto \textcircled{1}
    }
    \ElseIf{$ \xi_2 \le \sigma(s)/\sigma_{\text{max}}$}{
        \If{i=1}{accept s}
        \ElseIf{$\xi_3 \le N$}{accept s}
        \Else{ goto \textcircled{1}}
    }
    \Else{
        \If{i=1}{goto \textcircled{2}}
        \ElseIf{$\xi_3 \le N$}{goto \textcircled{2}}
        \Else{goto \textcircled{1}}
    }
\end{algorithm}

\subsubsection{the Scattering Transport}
\label{compton_section}

Now we discuss the momentum transport caused by the scattering
kernel
$C(\nu'\rightarrow\nu,\mathbf{\Omega}'\rightarrow\mathbf{\Omega}|
\mathbf{r})$ (Eq. \eqref{eqa5}). We first point out that the
scattering kernel $C$ equals to the normalized scattering coefficient
$\sigma_s$, which is related with the medium properties and also the
different scattering mechanisms (or scattering cross sections). For
example, in an uniform and isotropic medium, the scattering
mechanism is Rayleigh scattering, then the scattering coefficient
$\sigma_s$ is given by
$n_e\sigma(\mathbf{\Omega}'\rightarrow\mathbf{\Omega})$, where $n_e$
is the electron number density and $\sigma$ is the Rayleigh
scattering cross section.

In the former discussions we took an implicit assumption that the
scattering medium was static. This is an appropriate approximation
for low energy or non-relativistic scattering medium, where the
speed of the plasma can be neglected. While in more practical
circumstances the effects induced by the motion or the velocity
distribution of the scattering particles can not be ignored. In the
simplest case, the velocities of the particles are isotropic and
distribute according to the relativistic Maxwell distribution which
is given by \citep{Synge1957}
\begin{eqnarray}
\label{maxwelldis}
\begin{split}
&\displaystyle N_e(\gamma, T_e)=n_e\frac{v\gamma^2\exp(-\gamma/\Theta)}{\Theta K_2(1/\Theta)},
\end{split}
\end{eqnarray}
where $\Theta=kT_e/m_ec^2$ is the dimensionless temperature of the
gas, $\gamma$ is the Lorentz factor, and $K_2$ is the modified
Bessel function of second kind of order two.

\begin{figure}[t]
\begin{center}
  \includegraphics[scale=0.65]{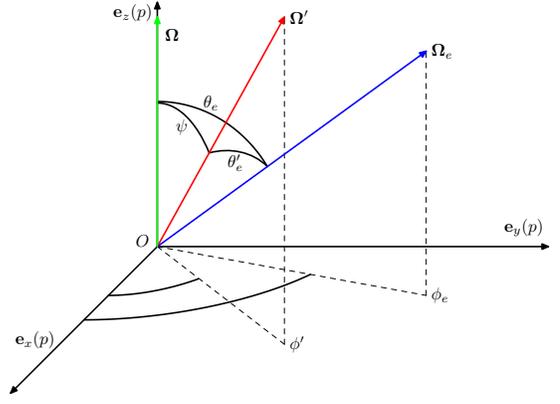}
\caption{The schematic demonstration of the tetrad
($\mathbf{e}_i(p)$, $i=x,y,z$) attached to the momentum direction
$\mathbf{\Omega}$ of the incident photon. $\mathbf{\Omega}'$ and
$\mathbf{\Omega}_e$ represent the momentum directions of scattered
photon and the scattering electron, respectively.}
\label{photo_tetrad}
\end{center}
\end{figure}

\begin{figure}[t]
\begin{center}
  \includegraphics[scale=0.75]{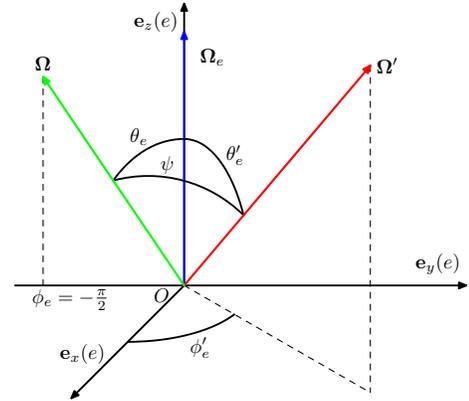}
\caption{The schematic demonstration of the tetrad
($\mathbf{e}_i(e)$, $i=x,y,z$) attached to the momentum direction
$\mathbf{\Omega}_e$ of the scattering electron. $\mathbf{\Omega}$
and $\mathbf{\Omega}'$ represent the momentum directions of incident
and scattered photons, respectively.} \label{ele_tetrad}
\end{center}
\end{figure}

A photon with four momentum $\mathbf{p}=(E, \mathbf{\Omega})$
transporting in a hot electron gas will experience an averaged
Compton scattering process, which can be described by a modified
differential cross section. The quantities in this cross section are
defined with respect to three different frames, i.e., the static
frame whose basis vectors are $\mathbf{e}_x, \,\mathbf{e}_y,
\,\mathbf{e}_z$, in which the incident and scattered directions are
denoted by $\mathbf{\Omega}$ and $\mathbf{\Omega}'$. The second
frame is attached to $\mathbf{\Omega}$, whose basis vectors are
$\mathbf{e}_z(p)=\mathbf{\Omega}$,
$\mathbf{e}_y(p)=\mathbf{\Omega}\times\mathbf{e}_z/|\mathbf{\Omega}\times\mathbf{e}_z|$,
$\mathbf{e}_x(p)=\mathbf{e}_y(p)\times\mathbf{e}_z(p)$ (see Fig.
\ref{photo_tetrad}), in which the electron's momentum vector
$\mathbf{p}_e$ is defined and characterized by $(\gamma, \mu_e,
\phi_e)$. The final frame is attached to $\mathbf{p}_e$, i.e.,
$\mathbf{e}_z(e)=\mathbf{p}_e$,
$\mathbf{e}_x(e)=\mathbf{p}_e\times\mathbf{e}_z(p)/|\mathbf{p}_e\times\mathbf{e}_z(p)|$
and $\mathbf{e}_y(e)=\mathbf{e}_z(e)\times\mathbf{e}_x(e)$ (see Fig.
\ref{ele_tetrad}). In this frame, the expression of Compton
scattering differential cross section can be given appropriately.
The incident and scattered directions are denoted by $(\mu_e,
\phi_e)$ and $(\mu_e', \phi_e')$ respectively, and obviously we have
$\phi_e\equiv-\pi/2$ (see Fig. \ref{ele_tetrad}). Any two of these
frames are simply connected by an orthonormal transformation.

Then the scattering coefficient is given by the averaged
Klein-Nishima (KN) differential cross section \citep{Canfield1987},
i.e.,
\begin{eqnarray}
\label{eqa5811}
\begin{split}
&\displaystyle \sigma_s(\nu\rightarrow\nu', \mathbf{\Omega}\rightarrow \mathbf{\Omega}')=
\frac{1}{4\pi}\int_1^\infty d\gamma N_e(\gamma)\int_{-1}^1d\mu_e\\
&\times(1-\mu_e v)\int_0^{2\pi}d\phi_e
\frac{d\sigma_{\text{KN}}}{d\mu_e'd\phi_e'} \delta\left[\nu'-g(\nu,
\mu_e, \mu_e', \psi)\right],
\end{split}
\end{eqnarray}
where $d\sigma_{\text{KN}}/d\mu_e'd\phi_e'$ is the KN cross section \citep{Akhiezer1969}
\begin{eqnarray}
\label{eqa59}
\begin{split}
&\displaystyle \frac{d\sigma_{\text{KN}}}{d\mu_e'\phi_e'} = \frac{3\sigma_T}{16\pi}\frac{1}{\gamma^2}
\frac{\chi}{(1-\mu_e v)^2}\left(\frac{\nu'}{\nu}\right)^2,
\end{split}
\end{eqnarray}
where
\begin{eqnarray}
\label{scKN1}
\begin{split}
\left\{
\begin{array}{ll}
&\displaystyle \chi = \frac{\epsilon'}{\epsilon} + \frac{\epsilon}{\epsilon'} +
\frac{4}{\epsilon}\left(1-\frac{\epsilon}{\epsilon'}\right) +
\frac{4}{\epsilon^2}\left(1-\frac{\epsilon}{\epsilon'}\right)^2;\\
&\displaystyle\epsilon=\frac{2h\nu}{m_ec^2}\gamma(1-\mu_e v),
\quad\epsilon'=\frac{2h\nu'}{m_ec^2}\gamma(1-\mu_e' v);\\
&\displaystyle g(\nu, \mu_e, \mu_e', \psi)=\frac{(1-\mu_e
v)\nu}{1-\mu_e' v+(h\nu/\gamma m_ec^2)(1-\mu_\psi)},
\end{array}
\right.
\end{split}
\end{eqnarray}
where $\nu$ and $\nu'$ are the incident and scattered frequencies
respectively, $\psi$ is angle between the incident and scattered
directions and $\mu_\psi=\cos\psi$. From Eq. \eqref{eqa5811} we can
obtain the total scattering coefficient by integrating out all of
the scattered directions $\mathbf{\Omega}'$ and frequencies $\nu'$ as
\begin{eqnarray}
\label{sigma_int}
\begin{split}
&\displaystyle \sigma_s(\nu, T_e)=\int_0^\infty\int_{4\pi}
\sigma_s(\nu\rightarrow\nu', \mathbf{\Omega}
\rightarrow\mathbf{\Omega}')d\nu'd\mathbf{\Omega}'.
\end{split}
\end{eqnarray}
Then the normalized scattering kernel is given by
\begin{eqnarray}
\label{C_normal}
\begin{split}
&\displaystyle
C(\nu\rightarrow\nu',\mathbf{\Omega}\rightarrow\mathbf{\Omega}') =
\frac{\sigma_s(\nu\rightarrow\nu',
\mathbf{\Omega}\rightarrow\mathbf{\Omega}')}{\sigma_s(\nu, T_e)}.
\end{split}
\end{eqnarray}

The integral given by Eq. \eqref{sigma_int} can be computed either
in the frame of $\mathbf{e}_i(p)$ or $\mathbf{e}_i(e)$. The
components of $\mathbf{\Omega}'$ in $\mathbf{e}_i(p)$ and
$\mathbf{e}_i(e)$ are given by $(\psi, \phi')$ and $(\theta'_e,
\phi'_e)$ respectively. Then from the geometrical relationships
depicted in Figure. \ref{photo_tetrad} and \ref{ele_tetrad}, we have
\begin{eqnarray}
\label{eqa60}
\begin{split}
\left\{
\begin{array}{ll}
&\displaystyle \mu_e'=\mu_\psi\mu_e+\sqrt{1-\mu_\psi^2}\sqrt{1-\mu_e^2}\cos(\phi_e-\phi'), \\
&\displaystyle
\mu_\psi=\mu_e\mu_e'+\sqrt{1-\mu_e^2}\sqrt{1-\mu_e'^2}\cos\phi_e'.
\end{array}
\right.
\end{split}
\end{eqnarray}
From those we can obtain the Jacobian of the transformation given by
\begin{eqnarray}
\begin{split}
&\displaystyle J=\left|
\begin{array}{ll}
\displaystyle\frac{\partial (\mu_\psi, \phi)}{\partial (\mu_e',
\phi_e')}
\end{array}
\right|=1,
\end{split}
\end{eqnarray}
hence the integral of $\mathbf{\Omega}'$ amounts to
$\displaystyle \int\frac{d\sigma_{\text{KN}}}{d\mu_e'd\phi_e'} d\mu_e'd\phi_e'$, which gives
\begin{eqnarray}
\begin{split}
&\displaystyle
\sigma_{\text{KN}}(\epsilon)=\frac{3\sigma_T}{4}\frac{1}{\epsilon}
\left[\left(1-\frac{4}{\epsilon}-\frac{8}{\epsilon^2}\right)\ln(1+\epsilon)+\frac{1}{2}\right.\\
&\left.+\frac{8}{\epsilon}-\frac{1}{2(1+\epsilon)^2}\right],
\end{split}
\end{eqnarray}
then we have \citep{Canfield1987}
\begin{eqnarray}
\begin{split}
&\displaystyle \sigma_s(\nu, T_e)=\frac{1}{2}\int_1^\infty d\gamma
\int_{-1}^1d\mu_e(1-\mu_e v)N_e(\gamma)\sigma_{\text{KN}}(\epsilon),
\end{split}
\end{eqnarray}
where the integral of $d\phi_e$ has been finished trivially
since $\sigma_{\text{KN}}(\epsilon)$ is independent on $\phi_e$. For
its frequent request we can tabulate the values of $\sigma_a(\nu,
T_e)$ in terms of a properly divided grid of $\nu$ and $T_e$, and
evaluate the value of $\sigma_a(\nu, T_e)$ through the linear
interpolation \citep{Hua1997}. There are also some other schemes to
evaluate $\sigma_a(\nu, T_e)$ numerically \citep{Wienke1985}.

Now we need to sample a scattered photon from $C$ (eq.
\eqref{C_normal}) with given incident $\nu$ and $\mathbf{\Omega}$.
While $C$ involves a multiple integral in terms of the distribution
function of hot electron gas. To illustrate the structure of $C$ in
a more succinct form, we recast it as
\begin{eqnarray}
\label{eqa61}
\begin{split}
&\displaystyle C(P')=\int f(P'|P_e)f_e(P_e)dP_e,
\end{split}
\end{eqnarray}
where $P'=(\nu', \mathbf{\Omega}')$, $\displaystyle
f(P'|P_e)=d\sigma_{\text{KN}}/d\Omega'/\sigma_{\text{KN}}(\epsilon)$,
$f_e(P_e)=N_e(\gamma) (1-\mu_e
v)\sigma_{\text{KN}}(\epsilon)/\sigma_s(\nu, T_e)$ and $P_e=(\gamma,
\mu_e, \phi_e)$. Both $f$ and $f_e$ have been normalized with
factors $\sigma_{\text{KN}}(\epsilon)$ and $\sigma_s(\nu, T_e)$,
thus they can be taken as PDFs for sampling. An algorithm called
as composition sampling has been proposed by \citet{Kahn1954} to
deal with PDFs given with an integral. The algorithm simply
reads
\begin{equation*}
    \begin{array}{ll}
        \text{\textcircled{1}}&\text{Get a sample }Y_e\text{ for $P_e$ by sampling } f_e(P_e);\\
        \text{\textcircled{2}}&\text{Substitute $Y_e$ back into $f(P'|P_e)$ and
          sample } \\&f(P'|Y_e)\text{ to get a sample $X$ for $P'$}.
   \end{array}
\end{equation*}
One can readily show that $X$ is a sample of $C(P')$. The sampling
of $f_e(P_e)$ is usually called as selecting an electron to scatter
off the photon. The algorithms to sample $f_e(P_e)$ have already
been proposed and here we adopt a combined one consisting of that
proposed by \citet{Canfield1987} and by \citet{Hua1997}
respectively. Since the former one involves an acceptation
probability that equals to the ratio of $\sigma_{\text{KN}}$ to the
Thomson cross sections $\sigma_T$, which becomes quite low for high
energy photon scattering due to KN effect. While this drawback can
be overcome by the algorithm proposed by \citet{Hua1997} properly.


The sampling of $f(P'|Y_e)$ is more subtle than that of $f_e(P_e)$.
Before the scattering, all relevant quantities should be transformed
into the electron rest frame, where the formula of scattering cross
section is greatly simplified. After the scattering, they need to be
transformed back into the static frame again.

Here we try to understand this sampling procedure mathematically
from the so-called transformation sampling method. The relevant
formulae are useful in our estimation scheme. We demonstrate it
through an example where the PDF $f(u, v)$ has two random variables,
$u$ and $v$. Through a bijection transformation: $u=\varphi_1(x, y),
v=\varphi_2(x, y)$, we can obtain the PDF for two new random
variable $x$ and $y$ as $g(x, y)=f[\varphi_1(x, y), \varphi_2(x,
y)]J$, where $J=|\partial (u, v)/\partial(x, y)|$ is the Jacobian of
the transformation. Hopefully $g(x, y)$ can be simplified and
readily sampled. Once the samples of $x, y$: $x_i, y_j$ are
obtained, the corresponding samples of $u, v$ are immediately given
by $u=\varphi_1(x_i, y_j), v=\varphi_2(x_i, y_j)$. Notice that the
inverse CFD method is actually a special case of transformation
sampling method, where the transformed PDF is uniformly
distributed.

Now the Lorentz transformation sampling procedure for $f(P'|Y_e)$ is
easily understandable. The expression of $f(P'|Y_e)$ given by Eq.
\eqref{eqa59} is a complicated function of $\mu_e'$ and $\phi_e'$.
Fortunately there is a transformation in terms of two new variables,
$\Psi$ and $\Phi$ given by
\begin{eqnarray}
\label{eqa62}
\begin{split}
   \left\{\begin{array}{ll}
&\displaystyle  \mu_\Psi = \widetilde{\mu}_e\widetilde{\mu}_e'-
\sqrt{1-\widetilde{\mu}_e'^2}\sqrt{1-\widetilde{\mu}_e^2}\cos(\phi_e-\phi_e'),\\
&\displaystyle \cos\Phi=\frac{\widetilde{\mu}_e'-\mu_\Psi\widetilde{\mu}_e}
{\sqrt{1-\widetilde{\mu}_\Psi^2}\sqrt{1-\widetilde{\mu}_e^2}},
   \end{array}
   \right.
\end{split}
\end{eqnarray}
where $\mu_\Psi=\cos\Psi$, and
\begin{eqnarray}
\label{eqa63}
\begin{split}
&\displaystyle  \widetilde{\mu}_e=\frac{\mu_e - v}{1-\mu_e v},
\quad\widetilde{\mu}_e'=\frac{\mu_e' - v}{1-\mu_e' v}.
\end{split}
\end{eqnarray}
Obviously, $\Psi$ and $\Phi$ are the azimuth angles of the scattered
direction defined with respect to the incident direction of the
photon in the rest frame of the electron. After some tedious
calculations, one can obtain the Jacobian of this transformation as
\begin{eqnarray}
\label{my_jacobi}
\begin{split}
&\displaystyle  J=\left|\frac{\partial(\mu_\Psi, \Phi)}
{\partial(\mu_e', \phi_e')}\right|=\frac{1}{\gamma^2(1-\mu_e'v)^2}.
\end{split}
\end{eqnarray}
And the quantity $\epsilon'/\epsilon$ becomes
\begin{eqnarray}
\begin{split}
&\displaystyle  \frac{\epsilon'}{\epsilon}= \frac{1}{1+
\displaystyle\epsilon/2(1-\mu_\Psi)}.
\end{split}
\end{eqnarray}
Then the PDF in terms of $\mu_\Psi$ and $\Phi$ reads
\begin{eqnarray}
\begin{split}
&\displaystyle  f(\mu_\Psi, \Phi|Y_e)= \frac{r_e^2}{2}
\left(\frac{\epsilon'}{\epsilon}\right)^2\left(\frac{\epsilon'}{\epsilon}+
\frac{\epsilon}{\epsilon'}-\sin^2\Psi\right),
\end{split}
\end{eqnarray}
which is exactly the differential cross section of Compton
scattering. The sampling procedure for this simplified PDF will be
easier and we will adopt the one provided by \citet{Hua1997}. Once
we obtain $\Psi$ and $\Phi$, $\mu_e', \phi_e'$ can also be obtained
immediately through the inverse transformation given by Eq.
\eqref{eqa62} and \eqref{eqa63}, which is the Lorentz transformation
obviously.

In our estimation scheme, we need to evaluate the scattering
kernel $C$ (given by Eqs. \eqref{C_normal} or \eqref{eqa61}) with
$\mathbf{\Omega}'(=\mathbf{\Omega}_{\text{obs}})$, $\nu$ and
$\mathbf{\Omega}$ are specified and take it as a weight. One can
see that with these quantities are given, the remaining thing is to
calculate the integral in terms of $P_e$, which can be accomplished
by the MC method, i.e., we obtain a set of values of $Y_e=(\gamma,
\mu_e, \phi_e)$ by sampling $f(P_e)$ and then take
$f(P'|Y_e)=d\sigma_{\text{KN}}/d\Omega'/\sigma_{\text{KN}}(\epsilon)$
as weight. And $f(P'|Y_e)$ can be computed either by Eq.
\eqref{eqa59} directly or by $f(\mu_\Psi, \Phi|Y_e)J$.

In this section we mainly discussed the sampling procedure for the
unpolarized Compton scattering with hot electron gas. The procedure
for the polarized Compton and Rayleigh scattering with Stokes
parameters involved will be presented later.

\subsection{the estimation of observable quantities}
Now we will discuss how to calculate the observational quantities at
infinity by using the weight functions, especially how to deal with
the case where a $\delta$ function appears in the recording
function. These quantities include energy spectrum, light curve,
angular-dependent photon number flux, etc.. Basically we divide the
energy (or any other variables) section, on which the spectra are
distributed, into a set of bins. Then we count and accumulate the
contributions made by each component of the random sequence
according to which bin they belong to.

As we mentioned that the final observed quantities can not be
constructed from radiative flux $\psi(\mathbf{r}, \mathbf{\Omega},
\nu)$ directly, since they are confined in the radiative region
where the emission and scattering processes are fulfilled. However,
$\psi$ absolutely plays a relevant role in the construction. The
quantity received by the observer can be expressed as an integral of
$\psi$ and recording function $f(P)$ (see Eq. \eqref{f_rc}) given by
Eqs. \eqref{I_mseq1} and \eqref{I_mseq2}. Thus the problem is
reduced to evaluate all of $I_m$. Since a random sequence of
states $P_1, \cdots, P_N$ is generated, we can immediately get the
estimations $F_m$, given by Eq. \eqref{F_estim}, for any $I_m$ with
$m\le N$. We then accumulate the $F_m$ to the bin where the
frequency $\nu_{m}$ belongs to and the magnitude of the spectrum in
the $i$-th bin is given by
\begin{eqnarray}
\label{eqa33}
\begin{split}
&F_i=\displaystyle \sum_{j}I_{m}^j(\nu_i),
\end{split}
\end{eqnarray}
where superscript $j$ is used to indicate the contributions made by
different components that may come from a same random sequence or
different ones.

The strategy discussed here can be called as term by term
estimation, since each term of the random sequence can make the
contribution to the observational quantities (even though it may be
rejected). Comparing to the photon tracing scheme, this strategy has
a lower variance. While for each $I_m$ we need to compute an
additional quantity $\exp[-\tau(\mathbf{r}_{m})]$, where
$\tau(\mathbf{r}_{m})$ is the total optical depth and
\begin{eqnarray}
\label{eqa34}
\begin{split}
&\displaystyle \tau(\mathbf{r}_{m})=\int_0^{s_{\text{max}}}
\sigma(\mathbf{\Omega}_m, \mathbf{r}_m+t\mathbf{\Omega}_m, \nu_m)dt,
\end{split}
\end{eqnarray}
where $s_{\text{max}}$ is the distance along $\mathbf{\Omega}_m$
from $\mathbf{r}_m$ to the boundary surface of the radiative region.
In the photon tracing scheme, a different estimation function is
chosen and the calculation of $\exp[-\tau(\mathbf{r}_{m})]$ is
actually replaced by the position transport samplings. It can be
called as the final event estimation scheme and will be discussed in
the next subsection.

Obviously if $\mathbf{\Omega}_{m}=\mathbf{\Omega}_{\text{obs}}$, we
record the contribution of $I_m$, otherwise it will be rejected.
This fact is appropriately described by the $\delta$ function in the
recording function. Since $\mathbf{\Omega}_{m}$ in a random sequence
are totally stochastic and unexpected, hence almost all of them will
be rejected. This drawback however can be overcome naively in our
scheme, i.e., we eliminate the $\delta$ function by integrating it
out directly. After that integration the integral of $I_m$ becomes
\begin{equation}
\begin{split}
\displaystyle I_m=&\int_{P_1}\cdots\int_{P_m}S(P_1)K(P_1\rightarrow P_2)
\cdots K(P_m\rightarrow \mathbf{r}, \nu, \mathbf{\Omega}_{\text{obs}}) \\
& \times\exp[-\tau(\mathbf{r}, \nu,
\mathbf{\Omega}_{\text{obs}})]dP_1\cdots dP_md\mathbf{r}d\nu.
\end{split}
\end{equation}
Then we can choose a new recording function given by
\begin{equation}
\begin{split}
&\displaystyle f_\text{new} = C(\nu_{m},
\mathbf{\Omega}_m\rightarrow \nu, \mathbf{\Omega}_{\text{obs}})\cdot
\exp[-\tau(\mathbf{r}_{m+1}, \nu, \mathbf{\Omega}_{\text{obs}})],
\end{split}
\end{equation}
where the scattering kernel $C$ is completely determined since $\nu$
is fixed as the incident $\nu_m, \mathbf{\Omega}_m$ and scattered
direction $\mathbf{\Omega}_{\text{obs}}$ are specified. Also notice
that for Compton scattering with averaged cross section, there is an
extra weight factor, $\sigma_s(\nu_m, T_e)$, needs to be included.
It arises from the procedure of sampling the electron distribution
function $f(P_e)$. Then the estimation function for $I_m$ given by
Eq. \eqref{F_estim} becomes
\begin{equation}
\label{F_estim2}
\begin{split}
&\displaystyle F  = w_{s,1}\cdot w_{s,2} \cdots w_{s,m}\cdot
f_\text{new}.
\end{split}
\end{equation}
With this strategy, one can see that any term of a random sequence
can always make contributions to the observational quantities. This
can significantly improve the calculation efficiency and accuracy.
And this strategy is very natural from the perspective of integral
evaluation, since the value of a $\delta$ function involved integral
can be obtained directly, i.e., $\int_a^b
f(x)\delta(x-x_0)dx=f(x_0)$. This is absolutely one of the most
important advantages to build the MCRT based on of Neumann
solution.

Particularly, in the calculation of the angular dependent spectrum,
we can obtain the values corresponding to all poloidal angles
$\theta_i$ simultaneously, i.e., we evaluate $f_\text{new}$ for $n$
times with different $\mu_i=\cos\theta_i$. This procedure can
genuinely increase the computational efficiency and accuracy (see
the discussions in the section of scheme verification).

\subsection{final event estimation}
Now we discuss how to understand the photon tracing scheme from the
perspective of Neumann solution and demonstrate that it actually
corresponds to a special choice of estimation function, i.e., the
final event estimation. Where a photon (or superphoton) is generated
and traced until to its ending, either escaping from the radiative
region, or absorbed by the medium. These steps can be derived from
the estimation function.

Physically speaking both the absorption and scattering coefficients
$\sigma_a$ and $\sigma_s$ vanish outside the radiative
domain, which implies that the photons actually can never escape
from the region, since
$p(s)=\exp[-\int_0^s\sigma_s(t)dt]\sigma_s(s)/A\equiv0$, if
$s>s_{b}$. But mathematically we have the flexibility to choose an
arbitrary function as PDF, provided its corresponding weight can
give a correct unbiased estimation for the final result. Thus we
extend the definition zone of $\sigma_s$ to the whole space by
filling the outer vacuum with an auxiliary medium, and in which
$\sigma_s$ can be an arbitrary function, provided $\tau_s(s)$ blows
up as $s$ approaches infinity, i.e.,
$\lim\limits_{s\rightarrow\infty}\tau_s(s)\rightarrow\infty$. With
$\sigma_s$, both the scattering optical depth $\tau_s(s)$ and $p(s)$
are non-vanishing in the whole position space. The form of $p(s)$
now becomes $p(s)ds=\exp(-\tau_s)d\tau_s$ and $A=1$. We use $p(s)$
to sample the scattering distance $s$ and the photon has the
probability to escape from the radiative region. Then one can
readily show the following identity as
\begin{eqnarray}
\label{eqa37}
\begin{split}
\exp[-\tau_s(s_b)]=\int^\infty_0 H(s'-s_{b})p(s')ds',
\end{split}
\end{eqnarray}
where $H(s'-s_{b})$ is the Heaviside function. With that the
recording function $f(P)$ can be rewritten as
\begin{eqnarray}
\begin{split}
f(P) =&\delta(\mathbf{\Omega}-\mathbf{\Omega}_{\text{obs}})\exp[-\tau_a(s_b)]\\
 &\times \int^\infty_0 H(s'-s_{b})p(s')ds',
\end{split}
\end{eqnarray}
where $\tau_a(s_b)$ is the absorption optical depth. Substituting
$f(P)$ into Eq. \eqref{I_mseq2}, we can reselect the associated PDF
and estimation function respectively as
\begin{equation}
\begin{split}
&\displaystyle \widetilde{f}(P1, \cdots, P_m, P) = S(P_1)\cdot
T_s(\mathbf{r}_1\rightarrow\mathbf{r}_2)\cdot C(\nu_1,\mathbf{\Omega}_1\rightarrow\\
&\nu_2, \mathbf{\Omega}_2)\cdots
T_s(\mathbf{r}_m\rightarrow\mathbf{r})\cdot
C(\nu_m,\mathbf{\Omega}_m\rightarrow\nu, \mathbf{\Omega})\cdot
T_s(\mathbf{r}\rightarrow\mathbf{r}'),
\end{split}
\end{equation}
\begin{equation}
\begin{split}
\displaystyle F  = &w_{s,1}\cdot w_{s,2} \cdots w_{s,m}\cdot
f_{\text{new}},
\end{split}
\end{equation}
where $T_s=\exp(-\tau_s)d\tau_s$, $w_{s,i}=\exp(-\tau_a)$,
$\mathbf{r}'=\mathbf{r}+s'\mathbf{\Omega}$ and
\begin{eqnarray}
\begin{split}
f_{\text{new}}
=&\delta(\mathbf{\Omega}-\mathbf{\Omega}_{\text{obs}})\exp[-\tau_a(s_b)]H(s'-s_{b}).
\end{split}
\end{eqnarray}
Due to the factor $H(s'-s_{b})$, $F$ always vanishes and the
corresponding contribution is rejected, unless the scattering
distance $s > s_b$. And $s > s_b$ means that the photon has escaped
the radiative region. This is what the final event estimation means.
While from Eq. \eqref{eqa37} one can can see that the final event
estimation is nothing but evaluating the factor $\exp[-\tau_s(s_b)]$
by sampling $p(s)$ and taking $H(s_i-s_{b})$ as weight, where $s_i$
are the samples of $s$, and
\begin{eqnarray*}
\begin{split}
\displaystyle\exp[-\tau_s(s_b)]\approx \sum_{i=1}^NH(s_i-s_{b})=
\frac{M}{N},
\end{split}
\end{eqnarray*}
where $N$ is the number of samples and $M$ is the number of $s_i$
greater than $s_b$.

Compared to the term by term estimation, the final event estimation
has been applied more widely in the former RT researches (e.g.,
\citet{Dolence2009}, \citet{Schnittman2013}, etc.). However there
are also some shortages for the final event estimation. For example
its variance is higher than term by term estimation, since for any
sequence there is only one chance to record the contribution.
Especially in an optically thin system, where the photons are prone
to escape rather than scattering. To overcome this difficulty,
\citet{Dolence2009} adopted a biased PDF for position transport
sampling, i.e., $p(s)=\exp(-b\tau_s)d(b\tau_s)$, where $b$ is called
as bias parameter, by tuning the its value one can implement the
sampling process in a better way and improve the poor signal to
noise performance. Obviously an additional weight
$w_b=\exp[-\tau_s(1-b)]/b$ is needed to balance the biased sampling.

While in the term by term estimation scheme, such problem will not
plagues us anymore, since the radiation sampling is mandatorily
confined in the radiative region. This is a variation of the
weighted sampling technique discussed by \citet{Pozd1983}. The low
optical depth exerts its effect on the term by term estimation
scheme through the normalization factor $A=1-\exp(-\tau_s)\approx
\tau_s$. Thus the contributions made by the sufficiently scattered
radiations will diminish rapidly.

\section{the polarized radiative transfer}
\label{sec3_PRT}

In the above section we have demonstrated how to solve an RTE
systematically by evaluating the multiple integrals stem from its
Neumann solution, where we ignored the polarization effects. The
purpose of this section is to include these effects by extending the
scheme. The treatment of polarized RT surely becomes more
complicated. However taking the former discussion as a foundation,
we can also treat the polarized RT in a consistent way, where the
Neumann solution also plays a significant role.

The polarizations of radiation are appropriately described by the
Stokes parameters of $I, Q, U, V$ \citep{Chand1960} and the
polarized RTE becomes a set of integro-differential equations on
these parameters. Similarly, by introducing four auxiliary
quantities (called as radiation flux), these equations can be
transformed into a set of integral equations. Then their Neumann
solutions are naturally obtained and observational quantities can be
evaluated directly by these solutions.

\subsection{the Stokes Parameters}
\label{polarized_Compton_sect}

To describe polarization, we must introduce the Stokes parameters
(SPs) of $I, Q, U, V$, where $I$ is the radiation intensity, $Q, U$
and $V$ depict the linear and circular polarizations, respectively.
It is convenient to group them as a column vector, i.e.,
$\mathbf{I}=(I, Q, U, V)^T$. The components $Q$ and $U$ are defined
with respect to a reference in the plane that is perpendicular to
the direction of propagation. If the frame is rotated by an angle
$\Phi$ anticlockwise in that plane, the SPs will change as
\begin{eqnarray}\label{eqa_142}
\begin{split}
&\displaystyle \mathbf{I}'=\mathbf{M}(\Phi)\mathbf{I},
\end{split}
\end{eqnarray}
where
\begin{equation}\label{eq3}
\mathbf{M}(\Phi) = \left(
   \begin{array}{cccc}
      1    &  0             & 0           & 0 \\
      0    &  \quad\cos 2\Phi   & \sin 2\Phi  & 0 \\
      0    & -\sin 2\Phi   & \cos 2\Phi  & 0 \\
      0    & 0              & 0          & 1
    \end{array}
  \right).
\end{equation}
A vector used to fix a frame in the transverse plane is called as
polarization vector. In order to determine the SPs at any position,
we need to parallel transport the polarization vector along the ray
trajectory and trace their changes after any scattering as well. In
flat spacetime, the parallel transport of a vector along a straight
line is trivial and the vector remains unchanged. In scattering
process, the scattering matrix is implicitly defined with respect to
the scattering plane, which is determined by the incident and
scattered directions of the photon. In the description of RT of
\citet{Chand1960}, the polarization vector was always defined in a
meridian plane that was determined by the wave vector $\mathbf{k}$
and the base vector $\mathbf{e}_z$ of the static reference. Thus the
SPs experienced two rotation transformations just before and after
any scattering process respectively (see Eq. \eqref{eqa39}).

For Rayleigh scattering, the scattering matrix $\mathbf{R}$ defined
with respect to the scattering plane is given by \citep{Chand1960}
\begin{eqnarray}\label{eqa40}
\begin{split}
&\mathbf{R} = \frac{3}{16\pi}\delta(\nu-\nu')\times \\
      &\times \left(
   \begin{array}{cccc}
      1+\cos^2\Theta & - \sin^2\Theta     & 0  & 0 \\
      - \sin^2\Theta    &  1+\cos^2\Theta  & 0  & 0 \\
             0        & 0          & 2\cos\Theta & 0 \\
             0        & 0          & 0          & 2\cos\Theta
    \end{array}
  \right),
\end{split}
\end{eqnarray}
where $\Theta$ is the scattering angle.

The description of polarized photon scattered by unpolarized
electron is more complicated \citep{nagirner1993, poutanen1993}. The
polarization vector of a photon is actually the spin wave function
$\Ket{\mathbf{e}}$, which can be expressed as the sum of two base
states:
$\Ket{\mathbf{e}}=c_x\Ket{\mathbf{e}_x}+c_y\Ket{\mathbf{e}_y}$,
where $\Ket{\mathbf{e}_x}$ and $\Ket{\mathbf{e}_y}$ can either be
linear or (left and right) circular polarization states
respectively. $c_x$ and $c_y$ are two complex numbers and their
squared absolute values are the probabilities that the photon is
found in the corresponding states. We can further introduce the
density matrix defined as:
$\rho_{xy}=c_xc_y^*=\Braket{\mathbf{e}_x|\mathbf{e}}\Braket{\mathbf{e}|\mathbf{e}_y}$.
Using Pauli matrices $\sigma_i$, $\rho_{xy}$ can be expressed as
$\rho=\frac{1}{2}(I+\xi_1\sigma_1+\xi_2\sigma_2+\xi_3\sigma_3)$,
where $I$ is the unit matrix. $\xi_i$ are the SPs of a photon
given by \citep{koso1996}
\begin{eqnarray}
\label{xi2xy}
\begin{split}
  & \xi_1=c_xc_y^*+c_x^*c_y,\\
  & \xi_2=i(c_xc_y^*-c_x^*c_y),\\
  & \xi_3=|c_x|^2-|c_y|^2,
\end{split}
\end{eqnarray}
which are connected with the normalized SPs, i.e., $\xi_1=U/I,
\xi_2=V/I, \xi_3=Q/I$. If we do a measurement, then the probability
that we find the photon stays in the $\Ket{\mathbf{e}'}$ state (the
corresponding SPs are $\xi_i'$) is given by
\begin{eqnarray}
\label{eqs_p}
\begin{split}
 p=|\Braket{\mathbf{e}|\mathbf{e}'}|^2=\frac{1}{2}(1+\xi_1\xi_1'+\xi_2\xi_2'+\xi_3\xi_3').
\end{split}
\end{eqnarray}
Suppose a polarized photon with $\xi_i$ is scattered by an
unpolarized and static electron into a polarization state with
$\xi_i^{(f)}$ in direction $(\mu, \varphi)$, then the probability
(or the differential KN cross section) that a detector finds the
scattered photon in the polarized state with $\xi_i'$ is given by
\citep{fano1949}:
\begin{eqnarray}
\label{scraw}
\begin{split}
  &\displaystyle d\sigma=\frac{r_e^2}{4}\left(\frac{\epsilon'}{\epsilon}\right)^2[F_0+F_3
\widetilde{\xi_3}+F_{11}\widetilde{\xi}_1\xi_1'+F_{22} \xi_2\xi_2'\\
  &  +(F_{33}\widetilde{\xi_3}+F_3)\xi_3']d\mu d\varphi,
\end{split}
\end{eqnarray}
where $\epsilon$ and $\epsilon'$ are the energies of incident and
scattered photons, $r_e$ is the classic electron radius, and the
other parameters are given by \citep{fano1949,fano1957}
\begin{eqnarray}
  \begin{array}{ll}
    &\displaystyle F_0=\frac{\epsilon'}{\epsilon}+\frac{\epsilon}
 {\epsilon'}-\sin^2\vartheta,\quad F_3=\sin^2\vartheta,\quad F_{11}=2\cos\vartheta\\
    &\displaystyle  F_{22}=\left(\frac{\epsilon'}{\epsilon}+
  \frac{\epsilon}{\epsilon'}\right)\cos\vartheta, \quad F_{33}=1+\cos^2\vartheta,
  \end{array}
\end{eqnarray}
and
\begin{eqnarray}
\begin{split}
  &\displaystyle\widetilde{\xi_3}=-(\xi_3\cos2\varphi+\xi_1\sin2\varphi),\\
  &\widetilde{\xi_1}=-(-\xi_3\sin2\varphi+\xi_1\cos2\varphi).
\end{split}
\end{eqnarray}
Performing a summation over all possible polarized states with
$\xi_i'$ in the direction $(\mu, \varphi)$ , we obtain the total KN
cross section as
\begin{eqnarray}
\label{eq:sc_formulas_1}
  \begin{array}{ll}
   d\sigma&\displaystyle=\frac{r_e^2}{2}\left(\frac{\epsilon'}{\epsilon}\right)^2(F_0+F_3\widetilde{\xi_3})d\mu d\varphi.
  \end{array}
\end{eqnarray}
Comparing the coefficients before $\xi_i'$ in Eqs. \eqref{eqs_p} and
\eqref{scraw}, we have
\begin{eqnarray}
\label{eq:sp_new_old}
  \left\{\begin{array}{ll}
    &\displaystyle \xi^{(f)}_1=\frac{F_{11}\widetilde{\xi_1}}{F_0+F_3\widetilde{\xi_3}},\\
    &\displaystyle \xi^{(f)}_2=\frac{F_{22}\xi_2}{F_0+F_3\widetilde{\xi_3}},\\
    &\displaystyle \xi^{(f)}_3=\frac{F_3+F_{33}\widetilde{\xi_3}}{F_0+F_3\widetilde{\xi_3}}.
  \end{array}\right.
\end{eqnarray}
Combining the above equations and the relations between $\xi_i$ and
$(I, Q, U, V)$, we can obtain the transformation formula for
incident and scattered SPs as \citep{mosc2020}
\begin{eqnarray}
\label{eq:sp_new_old2}
  \begin{pmatrix}
    I'\\ Q' \\ U' \\ V'
  \end{pmatrix}=\left(\frac{\epsilon'}{\epsilon}\right)^2
  \begin{pmatrix}
    F_0 & F_3 & 0 & 0 \\ F_3 & F_{33} & 0 & 0\\0 & 0 & F_{11} & 0 \\ 0 & 0 & 0 & F_{22}
  \end{pmatrix}
  \begin{pmatrix}
    \widetilde{I} \\ \widetilde{Q} \\ \widetilde{U} \\ \widetilde{V}
  \end{pmatrix},
\end{eqnarray}
where $\widetilde{\mathbf{I}}=\mathbf{M}(\varphi)\mathbf{I}$. One
may notice that the above formulae are valid just in the rest frame
of the electron. However the corresponding formulae in an arbitrary
frame can easily be obtained through a Lorentz transformation.
Utilizing the Jacobian of a Lorentz transformation given by Eq.
\eqref{my_jacobi}, we have
\begin{eqnarray}
\label{eq:sp_new_old3}
    \mathbf{I}'=\frac{1}{\gamma^2(1-\mu_e v)^2}\left(\frac{\nu'}{\nu}\right)^2 \mathbf{F}\mathbf{M}(\varphi) \mathbf{I},
\end{eqnarray}
which is very useful in our scheme for polarized Compton scattering,
i.e., as the scattered direction is replaced by
$\mathbf{\Omega}_{\text{obs}}$, the results on the RHS will be used
as weights for observed SPs. \citet{kraw2012} gave a systematical
but complicated discussions on the transformation procedures for
such scattering, especially the transformations and rotations for
SPs. In our scheme, the corresponding treatment can be somehow
simplified greatly.

\subsection{the Neumann Solution for Polarized RTEs}
Now we begin to demonstrate how to solve the polarized RTEs by the
method based upon the Neumann solution as discussed before. We will
take the polarized RTEs of \citet{Chand1960} in flat spacetime as an
example. These equations were first appropriately formulated by
\citet{Chand1960} (also see \citet{Pom1973} for a more pedagogical
introduction), in which the Faraday rotation and conversion effects
were not taken into consideration. So the total absorption
coefficient matrix will take the form of $\sigma(\nu)E$, where $E$
is the unit matrix, $\sigma(\nu)$ is the total interaction
coefficient. The polarized RTEs with scattering process incorporated
read \citep{Pom1973}
\begin{eqnarray}\label{eqa39}
\begin{split}
&\displaystyle\mathbf{\Omega}\cdot\nabla \mathbf{I}(\nu,
\mathbf{\Omega}, \mathbf{r})
+\sigma(\nu, \mathbf{\Omega}, \mathbf{r})\mathbf{I}(\nu, \mathbf{\Omega}, \mathbf{r})\\
&=\mathbf{S}(\nu,\mathbf{\Omega}, \mathbf{r})\displaystyle
+\int_0^\infty d\nu'
\int_{4\pi}d\mathbf{\Omega}'\sigma_s(\nu', \mathbf{\Omega}', \mathbf{r})\frac{\nu}{\nu'} \\
&\times \mathbf{M}(\pi-\psi)\mathbf{R}(\nu'\rightarrow\nu,
\cos\Theta) \mathbf{M}(\Phi)\mathbf{I}(\nu',\mathbf{\Omega}',
\mathbf{r}),
\end{split}
\end{eqnarray}
where $\cos\Theta=\mathbf{\Omega}\cdot\mathbf{\Omega}'$ is the
cosine of scattering angle, $\Phi$ is the angle between the incident
meridian plane and the scattering plane,
$\sigma(\nu)=\sigma_a(\nu)+\sigma_s(\nu)$, $\sigma_a(\nu)$ and
$\sigma_s(\nu)$ are the absorption and scattering coefficients at
frequency $\nu$, respectively. As aforementioned $\mathbf{I}=(I, Q,
U, V)^T$ represents the SPs vector and $\mathbf{S}=(S_I, S_Q, S_U,
S_V)^T$ represents the emissivity vector, whose components
characterize the polarized radiations emitted by the medium
spontaneously. Since we prefer to use $I, Q$ rather than $I_l, I_r$
in MCRT, the scattering matrix $\mathbf{R}$ is different from that
given by \citet{Chand1960} (see the discussion in Appendix
\ref{app2}). Since the polarization vector always lies in the
meridian plane and the scattering matrix $\mathbf{R}$ is defined
with respect to the scattering plane, we need use the rotation
matrix $\mathbf{M}(\Phi)$ to transform the SPs from the incident
meridian plane to the scattering plane and the matrix
$\mathbf{M}(\pi-\psi)$ to implement the inverse transformation. From
the geometrical relations, one can easily show that $\psi$ is a
function of $\Theta$ and $\Phi$. Thus we can simplify the expression
of the scattering term by introducing a matrix $\mathbf{K}_c$
defined as $\mathbf{K}_c(\Theta,
\Phi)=\mathbf{M}(\pi-\psi)\mathbf{R}(\cos\Theta)\mathbf{M}(\Phi)$.

On the other hand, we can recast the transfer equations with respect
to the static frame, in which the incident and scattered directions
of the radiation are denoted by $\theta', \phi'$ and $\theta, \phi$
respectively. One can show that $\Theta, \Phi$ are the triangular
functions of $\theta, \phi$ with given $\theta', \phi'$
\citep{Chand1960}, implying that $\mathbf{K}_c$ is also a function
of $\theta, \phi$. Hence we can either use $\Theta, \Phi$ or
$\theta, \phi$ to describe the scattering process. In our scheme we
prefer to use $\theta, \phi$, even though the elements of matrix
$\mathbf{K}_c$ are very complicated functions of them (see Appendix
\ref{app2}). Because we need not consider any complicated rotations
and transformations in terms of $\mathbf{M}(\Phi)$ and
$\mathbf{M}(\pi-\psi)$ and the related tetrad constructions.

Now we can implement the similar procedures to transform the RTEs
into a set of integral equations on an auxiliary quantity
$\mathbf{\Psi}(\nu,\mathbf{\Omega}, \mathbf{r})$, which exactly
equals to the RHS of Eq. \eqref{eqa39}, i.e.,
\begin{eqnarray}\label{eqa41}
\begin{split}
&\displaystyle \mathbf{\Psi}(\nu,\mathbf{\Omega}, \mathbf{r})=
\mathbf{S}(\nu,\mathbf{\Omega}, \mathbf{r})\displaystyle +\\
&\int_0^\infty d\nu'\int_{4\pi}d\mathbf{\Omega}'\sigma_s(\nu', \mathbf{\Omega}', \mathbf{r})
\frac{\nu}{\nu'}\mathbf{K}_c(\Theta, \Phi) \mathbf{I}(\nu',\mathbf{\Omega}', \mathbf{r}).
\end{split}
\end{eqnarray}
Notice that the gradient derivative on the LHS of Eq. \eqref{eqa39}
can be recast into a total derivative with respect to the distance
$s$ along the radiation, i.e.,
\begin{eqnarray}\label{eqa42}
\begin{split}
&\displaystyle -\frac{d}{ds}\mathbf{I}\left(\nu, \mathbf{\Omega},
\mathbf{r}-s\mathbf{\Omega} \right)= \mathbf{\Omega}\cdot\nabla
\mathbf{I}\left(\nu, \mathbf{\Omega}, \mathbf{r}-s\mathbf{\Omega} \right).
\end{split}
\end{eqnarray}
Then Eq. \eqref{eqa39} becomes (see the page 27 of \citet{Pom1973})
\begin{eqnarray}
\label{eqa43}
\begin{split}
&\displaystyle -\frac{d}{ds}\mathbf{I}\left(\nu, \mathbf{\Omega},
\mathbf{r}-s\mathbf{\Omega}\right)
+\sigma\left(\nu, \mathbf{\Omega}, \mathbf{r}-s\mathbf{\Omega}\right) \\
&\times\mathbf{I}\left(\nu, \mathbf{\Omega}, \mathbf{r}-s\mathbf{\Omega}\right)=
\mathbf{\Psi}\left(\nu, \mathbf{\Omega}, \mathbf{r}-s\mathbf{\Omega},\right).
\end{split}
\end{eqnarray}
Integrating the above equation formally, we have
\begin{eqnarray}\label{eqa44}
\begin{split}
&\displaystyle\mathbf{I}(\nu, \mathbf{\Omega}, \mathbf{r})=
\int_{0}^{|\mathbf{r}-\mathbf{r}_0|} ds'\mathbf{\Psi}\left(\nu, \mathbf{\Omega},
\mathbf{r}-s'\mathbf{\Omega}, \right) \\
&\times\exp\left[-\int^{s'}_0\sigma\left(\nu, \mathbf{\Omega},
\mathbf{r}-s''\mathbf{\Omega} \right)ds''\right],
\end{split}
\end{eqnarray}
where $\mathbf{r}_0$ is the position vector of the starting
point of the ray trajectory on the boundary. If the radiative region
is infinite, we have $|\mathbf{r}-\mathbf{r}_0|=\infty$. We should
keep in mind that $\mathbf{I}$ is a column vector with four
components, thus the above equation means that the four components
share a common exponential attenuation factor as the radiation
propagates. For the RT in a magnetized plasma where the Faraday
rotation and conversion effects should be considered
\citep{Shcherbakov2011, Huang2011}, the absorption coefficient will
be replaced by a matrix, which gives rise to the failure of the
integration procedure of Eq. \eqref{eqa43}. However we can still
employ it by introducing a $\sigma(\nu)I$ term and alter the
coefficient matrix (see the discussion below).

Substituting Eq. \eqref{eqa44} into the RHS of Eq. \eqref{eqa41}, we
obtain the integral equation of $\mathbf{\Psi}$
\begin{eqnarray}\label{eqa45}
\begin{split}
&\displaystyle \mathbf{\Psi} (\nu,\mathbf{\Omega}, \mathbf{r})=
\mathbf{S}(\nu,\mathbf{\Omega}, \mathbf{r})\displaystyle +
\int_0^\infty d\nu'\int_{4\pi}d\mathbf{\Omega}'\int_0^{|\mathbf{r}-\mathbf{r}_0|} ds'\\
&\times\sigma_s(\nu',\mathbf{\Omega}', \mathbf{r})\frac{\nu}{\nu'}\mathbf{K}_c(\Theta, \Phi)
\mathbf{\Psi}(\nu',\mathbf{\Omega}', \mathbf{r}_0+s'\mathbf{\Omega}) \\
&\times\exp\left[-\int^{|\mathbf{r}-\mathbf{r}_0|}_{s'}\sigma\left(\nu', \mathbf{\Omega}',
\mathbf{r}_0+s''\mathbf{\Omega} \right)ds''\right].
\end{split}
\end{eqnarray}
Comparing to Eq. \eqref{eqa44}, we have reversed the direction of
integration, i.e., from $\mathbf{r}_0$ to $\mathbf{r}$ for the sake
of convenience. Denote $P=(\nu, \mathbf{\Omega}, \mathbf{r})$, the
above equation can be expressed in a compact form as
\begin{eqnarray}\label{eqa46}
\begin{split}
&\displaystyle \mathbf{\Psi}(P)=\mathbf{S}(P)\displaystyle +
\int_{P'}\mathbf{K}(P'\rightarrow P)\mathbf{\Psi}(P')dP',
\end{split}
\end{eqnarray}
where $\mathbf{K}(P'\rightarrow P)=T_s\mathbf{C}w_s$ is the total
transport kernel, $T_s$, $\mathbf{C}$ are the position transport and
Compton scattering kernels respectively, $w_s$ is the weight factor
for counting the effect of absorption, and
\begin{eqnarray}\label{eqa47}
\begin{split}
&\displaystyle T_s(\mathbf{r}_{s'}\rightarrow\mathbf{r}|\nu',
\mathbf{\Omega}')=\frac{1}{A}\sigma_s(\nu',\mathbf{\Omega}', \mathbf{r})\\
&\times\exp\left[-\int^{|\mathbf{r}-\mathbf{r}_0|}_{s'}\sigma_s\left(\nu', \mathbf{\Omega}',
\mathbf{r}_0+s''\mathbf{\Omega} \right)ds''\right],\\
&\mathbf{C}(\mathbf{\Omega}'\rightarrow\mathbf{\Omega}, \nu'\rightarrow\nu|\mathbf{r})=
\frac{\nu}{\nu'}\mathbf{K}_c(\Theta, \Phi),\\
&w_s(s')=A\exp\left[-\int^{|\mathbf{r}-\mathbf{r}_0|}_{s'}\sigma_\alpha\left(\nu', \mathbf{\Omega}',
\mathbf{r}_0+s''\mathbf{\Omega} \right)ds''\right],
\end{split}
\end{eqnarray}
where $A$ is the normalization factor of $T_s$. Eq. \eqref{eqa46} is
a set of Fredholm integral equations of second kind and their
Neumann series expansion can be directly obtained as
\begin{eqnarray}
\label{eqa48}
\begin{split}
&\displaystyle \mathbf{\Psi}(P)=\sum_{m=0}^\infty \mathbf{\Psi}_m(P),
\end{split}
\end{eqnarray}
where
\begin{eqnarray}
\label{eqa49}
\begin{split}
\left\{
   \begin{array}{ll}
&\displaystyle  \mathbf{\Psi}_0(P)=\mathbf{S}(P),\\
&\displaystyle\mathbf{\Psi}_1(P)=\int\mathbf{K}(P_0\rightarrow P)\mathbf{\Psi}_0(P_0)dP_0\\
&\displaystyle=\int\mathbf{K}(P_0\rightarrow P)\mathbf{S}(P_0)dP_0,\\
&\displaystyle\cdots,\\
&\displaystyle\mathbf{\Psi}_m(P)=\int\mathbf{K}(P_{m-1}\rightarrow P)\mathbf{\Psi}_{m-1}(P_{m-1})dP_{m-1}\\
&\displaystyle=\idotsint\displaystyle\mathbf{K}(P_{m-1}\rightarrow P)\cdots\mathbf{K}(P_{1}\rightarrow P_2)\\
&\times\mathbf{K}(P_0\rightarrow P_1) \mathbf{S}(P_0)dP_0dP_1 \cdots dP_{m-1}.
    \end{array}
  \right.
\end{split}
\end{eqnarray}
Using Eq. \eqref{eqa44}, the SPs observed in the direction
$\mathbf{\Omega}_{\text{obs}}$ can be expressed as an integral over
the whole emissive region
\begin{eqnarray}
\label{eqa50}
\begin{split}
\mathbf{I}&\displaystyle=\int_{P}\mathbf{\Psi}(P)f(P)dP=\sum_{m=0}^\infty\mathbf{I}_m,
\end{split}
\end{eqnarray}
where
\begin{eqnarray}
\label{eqa50s4}
\begin{split}
\mathbf{I}_m&\displaystyle=\idotsint\displaystyle f(P)\mathbf{K}(P_{m-1}\rightarrow P)\cdots\mathbf{K}(P_{1}\rightarrow P_2)\\
&\times\mathbf{K}(P_0\rightarrow P_1) \mathbf{S}(P_0)dP_0dP_1 \cdots
dP_{m-1}dP,
\end{split}
\end{eqnarray}
and
\begin{eqnarray}
\begin{split}
&\displaystyle f(P)=\exp\left[-\int^{s_b}_0\sigma\left(\nu, \mathbf{\Omega},
\mathbf{r}+s''\mathbf{\Omega} \right)ds''\right]
\delta(\mathbf{\Omega}-\mathbf{\Omega}_{obs}).
\end{split}
\end{eqnarray}
which is the recording function. Hence the problem of polarized RT
can also be converted to the calculations of infinite multiple
integrals arising from Neumann solution. Next we will demonstrate
how to evaluate these integrals by the MC method.

\subsection{the Generation of Random Sequence}
\label{section4.3}Similarly to the unpolarized situations, in order
to calculate the integrals given by Eq. \eqref{eqa50s4}
simultaneously, we need to generate a scattering sequence: $P_0,
P_1, \cdots,$ by sampling an associated PDF separated from the
integrand of $\mathbf{I}_m$. Meanwhile the remaining part is taken
as weight functions for the estimation of $\mathbf{I}_m$. The kernel
$\mathbf{K}(P'\rightarrow P)$ is a matrix, which will mix the
components of the column vector $\mathbf{\Psi}$ inevitably after its
action on it, the evolution of them are coupled. Therefore we must
appropriately find a PDF shared by the four components of
$\mathbf{\Psi}$. For position transport, the four components share a
common kernel: $T_s(\mathbf{r}_{s}\rightarrow\mathbf{r}|\nu,
\mathbf{\Omega})$. While for scattering we have to choose a common
PDF for all components. In addition, we define a weight variable $w$
and set its initial value to be $1$.

Our start point is also to generate $P_0$ by sampling the source
function $\mathbf{S}(P)$. If $\mathbf{S}(P)$ is too complicated, we
will sample an uniform PDF to get $P_0$ instead and take
$\mathbf{S}(P_0)$ as the weight for estimations, i.e.,
$\mathbf{\Psi}_0=\mathbf{S}(P_0)$.

Next we need to generate $P_m$ when $P_{m-1}$ is provided by
sampling the total transfer kernel $\mathbf{K}(P_{m-1}\rightarrow
P_m)$. Since $\mathbf{K}=T_s\mathbf{C}w_s$, we can use $T_s$ as the
position transport PDF $p(s)$ for all components of $\mathbf{\Psi}$
and it can be sampled by using either inverse CDF method or the
Algorithm. \ref{algo1}. With the distance sample $s$, we have
$\mathbf{r}_{m}=\mathbf{r}_{m-1}+s\mathbf{\Omega}_{m-1}$.
Accompanying each position transport, an extra weight $w_s$ (see Eq.
\eqref{eqa47}) must be multiplied to the weight: $w=w\cdot w_s$.

To obtain the PDF for scattering, we must carry out the
multiplication of matrix $\mathbf{K}_c$ and $\mathbf{\Psi}_{m-1}$.
After that, we get four updating relationships between the
components of $\mathbf{\Psi}_{m-1}$ and $\mathbf{\Psi}_{m}$, i.e.
($\mathbf{\Omega}_m$ is determined by $\Theta_{m}, \Phi_m$, for
simplicity we will drop the subscript $m$ from now on),
\begin{eqnarray}
\label{eqa51}
\begin{split}
\left\{
   \begin{array}{ll}
&\displaystyle \Psi^{(I)}_m= f_I(\Theta, \Phi)\Psi^{(I)}_{m-1},\,\,
\displaystyle \Psi^{(Q)}_m= f_Q(\Theta, \Phi)\Psi^{(I)}_{m-1},\\
&\displaystyle \Psi^{(U)}_m= f_U(\Theta, \Phi)\Psi^{(I)}_{m-1},\,\,
\displaystyle \Psi^{(V)}_m= f_V(\Theta, \Phi)\Psi^{(V)}_{m-1},
    \end{array}
  \right.
\end{split}
\end{eqnarray}
where
\begin{eqnarray}
\label{eqa52}
\begin{split}
\left\{
   \begin{array}{ll}
&\displaystyle f_I(\Theta, \Phi) = [1+\cos^2\Theta-\sin^2\Theta\times \\
&\quad\quad\quad\times(q_{m-1}\cos2\Phi+u_{m-1}\sin2\Phi)]c_1,\\
&\displaystyle f_Q(\Theta, \Phi) = [-\sin^2\Theta+(1+\cos^2\Theta)\times\\
&\quad\quad\quad\times(q_{m-1}\cos2\Phi+u_{m-1}\sin2\Phi)]c_1,\\
&\displaystyle f_U(\Theta, \Phi) = 2\cos\Theta (-q_{m-1}\sin2\phi + u_{m-1}\cos2\phi) c_1,\\
&\displaystyle f_V(\Theta, \Phi) = 2\cos\Theta c_1,\\
    \end{array}
  \right.
\end{split}
\end{eqnarray}
where $c_1=3/16\pi$, and
\begin{eqnarray}
\label{eqa53}
\begin{split}
\left.
   \begin{array}{ll}
\displaystyle q_{m-1} =
\frac{\Psi^{(Q)}_{m-1}}{\Psi^{(I)}_{m-1}},\quad \displaystyle
u_{m-1} = \frac{\Psi^{(U)}_{m-1}}{\Psi^{(I)}_{m-1}}.
    \end{array}
  \right.
\end{split}
\end{eqnarray}
Then we can choose $f_I(\Theta, \Phi)$ as a PDF to sample $\Theta$
and $\Phi$, since $f_I$ is always positive. While the other three
functions can not be taken as a PDF, because they take negative
values. Since $f_I$ is automatically normalized, then the PDF of
$\Theta$ and $\Phi$ is given by
\begin{eqnarray}
\label{eqa54}
\begin{split}
&\displaystyle p(\Theta, \Phi) = \frac{3}{16\pi}[1+\cos^2\Theta-\sin^2\Theta(q_{m-1}\cos2\Phi\\
&\quad\quad\quad\quad\quad+u_{m-1}\sin2\Phi)].
\end{split}
\end{eqnarray}
After separating $p(\Theta, \Phi)$ from each component of
$\mathbf{\Psi}$, we take the remaining parts as
weights and obtain new updating relations for
$\mathbf{\Psi}_m$ as
\begin{eqnarray}
\label{eqa55}
\begin{split}
\left\{
   \begin{array}{ll}
&\displaystyle \Psi^{(I)}_m= \Psi^{(I)}_{m-1},\,\,
\displaystyle \Psi^{(Q)}_m= \frac{f_Q(\Theta, \Phi)}{f_I(\Theta, \Phi)}\Psi^{(I)}_{m-1},\\
&\displaystyle \Psi^{(U)}_m= \frac{f_U(\Theta, \Phi)}{f_I(\Theta,
\Phi)}\Psi^{(I)}_{m-1},\,\, \displaystyle \Psi^{(V)}_m=
\frac{f_V(\Theta, \Phi)}{f_I(\Theta, \Phi)}\Psi^{(V)}_{m-1}.
    \end{array}
  \right.
\end{split}
\end{eqnarray}
To keep the equations unchanged, a factor $f_I$ is divided for all
components of $\mathbf{\Psi}$.

The algorithm to randomly select a direction from $p(\Theta, \Phi)$
can be stated readily as follows. We first obtain the marginalized
PDF of $\Theta$, $p(\Theta)$, by integrating $p(\Theta, \Phi)$ over
the variable $\Phi$, and we have $p(\Theta)=3/8(1+\cos^2\Theta)$.
Then we can sample $p(\Theta)$ by the inverse CDF method
\citep{Schnittman2013}. Substituting the selected $\Theta_m$ into
Eq. \eqref{eqa54} and using the Bayes formula: $p(\Phi|\Theta_m) =
p(\Theta_m, \Phi)/p(\Theta_m)$, we get the PDF of $\Phi$ as
\begin{equation}
p(\Phi) =
\frac{1}{2\pi}\left[1-\frac{1-\mu^2}{1+\mu^2}(q_{m-1}\cos2\Phi+
u_{m-1}\sin2\Phi)\right],
\end{equation}
where $\mu=\cos\Theta_m$. $p(\Phi)$ can be expressed in a more
compact form: $p(\Phi) = A-B\cos2\Phi-C\sin2\Phi$, where $A=1/2\pi$,
$B=(1-\mu^2)q_{m-1}/[2\pi(1+\mu^2)]$, and
$C=(1-\mu^2)u_{m-1}/[2\pi(1+\mu^2)]$. After some algebraic
calculations, $p(\Phi)$ can be transformed as $p(\Phi) =
A-\sqrt{B^2+C^2}\cos2(\Phi-\Phi_0)$, which can be sampled by the
inverse CDF method. It equals to solve a Kepler's equation
\citep{Schnittman2013}. While if we notice that the triangular
functions in $p(\Phi)$, such as $\cos\Phi, \sin\Phi$, are piecewise
central symmetrical functions, we can sample it by a completely new
method, which is given in Algorithm. \ref{algo2}. Most importantly,
the difficulty of solving the Kepler equation can be avoided and the
sampling efficiency remains 100$\%$.

As $\Theta$ and $\Phi$ are obtained, we can construct the scattered
direction $\mathbf{\Omega}$ (or $\mathbf{\Omega}_m$) immediately.
With them we can proceed to carry out the next position transport
and scattering, and repeat this procedure until the accumulated
weight $w$ is smaller than the tolerance value. After each
scattering, we can calculate the contribution from that point by
using the incident and scattered photon momentums, the weight $w$
and the recording function $f(P)$. The $\delta$-function in $f(P)$
can also be eliminated to increase the estimation efficiency by
integrating it out directly and reselecting the estimation functions
as before. The estimations of the four components are given
by
\begin{eqnarray}
\label{eqa51}
\begin{split}
\left\{
   \begin{array}{ll}
&\displaystyle F^{(X)}_m=
w\cdot f_X(\mathbf{\Omega}_{\text{obs}})\cdot\Psi^{(I)}_{m}\cdot\exp(-\tau),\\
& F^{(V)}_m= w\cdot
f_V(\mathbf{\Omega}_{\text{obs}})\cdot\Psi^{(V)}_{m}\cdot\exp(-\tau),
    \end{array}
  \right.
\end{split}
\end{eqnarray}
where $X$ can be $I, Q, U$ and
\begin{eqnarray}
\begin{split}
&\displaystyle \tau= \int^{s_b}_0\sigma\left(\nu_{\text{obs}},
\mathbf{\Omega}_{\text{obs}},
\mathbf{r}+s'\mathbf{\Omega}_{\text{obs}} \right)ds',
\end{split}
\end{eqnarray}
which is the total optical depth from $\mathbf{r}$ to the boundary
along $\mathbf{\Omega}_{\text{obs}}$.

We have demonstrated the procedure to solve the polarized RTEs and
evaluate the observational quantities. The key point is that we
should choose an associated PDF shared by the four components of
$\mathbf{\Psi}$ due to their coupled evolution. Thus they will
transport in the same way both in position and momentum space. In
addition, we chose the scattering PDF as a function of $\Theta$ and
$\Phi$. Under this choice, we must construct a tetrad associated
with the incident direction and make some transformations for each
scattering. As it is aforementioned that the scattering kernel can
also be expressed as functions of $\theta$ and $\varphi$ (or
$\mu=\cos\theta$), which are the polar and azimuth angles of the
photon momentum in the static frame. Even though these functions are
very complicated, they can also be treated by the MC method
appropriately (see the discussion given in the Appendix \ref{app2}),
especially any tetrad related constructions and transformations are
no longer needed. For the configurations with geometrical
symmetries, the RTEs can usually be simplified in some sense. For
example, a plane-parallel atmosphere has an axial symmetry, the RTE
can be simplified by integrating $\varphi$ out (see
\citet{Ports2004} for a detailed derivation) and the only relevant
quantities are $I$ and $Q$ (or $I_l, I_r$). The simplified RTE can
be solved by our new scheme readily and efficiently.

\subsection{Polarized RT with Faraday Rotation and Conversion}
\label{RTwithFRC} In the above subsections, we have discussed the
procedure to solve the polarized RTEs, where only the absorption and
scattering effects are taken into account and the four components of
the Stokes parameters share a common total absorption coefficient
$\sigma(\nu)$. While as we compute the polarized RT of synchrotron
radiation in a plasma where the magnetic field plays a important
role \citep{shche2011, dexter2016}, the Faraday Rotation and
conversion effects must be considered. Now we extend our scheme to
include these effects, with which the RTEs can be written as
\citep{mosc2020}
\begin{eqnarray}
\label{eq:symchrotroneq}
       \frac{d}{d\lambda}
    \begin{pmatrix}
        I\\ Q\\ U\\ V
    \end{pmatrix}=
    \begin{pmatrix}
       j_I\\j_Q\\j_U\\j_V
    \end{pmatrix}-
    \begin{pmatrix}
       \alpha_I & \alpha_Q & \alpha_U & \alpha_V\\
       \alpha_Q & \alpha_I & \rho_V   &  -\rho_U\\
       \alpha_U & -\rho_V &  \alpha_I & \rho_Q \\
       \alpha_V & \rho_U &   -\rho_Q  & \alpha_I
    \end{pmatrix}
    \begin{pmatrix}
       I\\ Q\\ U\\ V
    \end{pmatrix},
\end{eqnarray}
where $j_X$ (where $X$ can be $I, Q, U, V$) are Stokes emissivity,
$\alpha_X$ and $\rho_X$ are the absorption and Faraday
rotation/conversion coefficients respectively. To solve Eq.
\eqref{eq:symchrotroneq} by the MC method, we need also convert them
into a set of integral equations. For simplicity, we will denote the
column vectors and matrix by bold face characters. First we
introduce an auxiliary vector $\mathbf{\Psi}$ and a positive scaler
parameter $\alpha$ and recast Eq. \eqref{eq:symchrotroneq} into
\begin{eqnarray}
\label{my_Faradayeq}
   \frac{d\mathbf{I}}{d\lambda}+ \alpha\mathbf{I} = \mathbf{j} - \mathbf{R} \mathbf{I}=\mathbf{\Psi},
\end{eqnarray}
where
\begin{eqnarray}
    \mathbf{R}=
    \begin{pmatrix}
       \alpha_I-\alpha & \alpha_Q & \alpha_U & \alpha_V\\
       \alpha_Q & \alpha_I-\alpha & \rho_V   &  -\rho_U\\
       \alpha_U & -\rho_V &  \alpha_I-\alpha & \rho_Q \\
       \alpha_V & \rho_U &   -\rho_Q  & \alpha_I-\alpha
    \end{pmatrix}.
\end{eqnarray}
In order to make the final results of MC calculations to be
convergent, $\alpha$ must be greater than any absolute values of the
elements of matrix $\mathbf{R}$, i.e., $\alpha\ge|\alpha_X|$ and
$\alpha\ge|\rho_X|$. After the transformation of Eq.
\eqref{my_Faradayeq}, the four components of SPs will have a same
absorption optical depth. It is important to notice that we can
change $\alpha$ to adjust the compromise between converging speed
and calculation accuracy. From Eq. \eqref{my_Faradayeq}, we directly
obtain
\begin{eqnarray}
   \mathbf{I} = \int_0^\lambda\mathbf{\Psi}(\lambda')\exp[-\alpha(\lambda-\lambda')]d\lambda' +
   \mathbf{I}(0)\exp(-\alpha\lambda),
\end{eqnarray}
where $\mathbf{I}(0)$ is the initial condition. Substituting the
above equation into $\mathbf{\Psi}=\mathbf{j} - \mathbf{R}
\mathbf{I}$ and eliminating $\mathbf{I}$, we can obtain the integral
equation for $\mathbf{\Psi}$ as
\begin{eqnarray}
   \label{voltera1}
   \mathbf{\Psi}(\lambda) = \mathbf{J}(\lambda)  - \int_0^\lambda \mathbf{K}(\lambda, \lambda') \mathbf{\Psi}(\lambda')d\lambda',
\end{eqnarray}
where $\mathbf{K}(\lambda,
\lambda')=\mathbf{R}(\lambda)\exp(-\lambda'\alpha)$,
$\mathbf{J}(\lambda)=\mathbf{j}(\lambda)-\mathbf{j_0}(\lambda)$ and
$\mathbf{j}_0(\lambda)=\mathbf{R}(\lambda)\mathbf{I}(0)\exp(-\lambda\alpha)$.
Eq. \eqref{voltera1} is called as the Volterra integral equation of
second kind, we can obtain the corresponding Neumann series solution
as
\begin{eqnarray}
\label{eq:psi_series1}
\begin{array}{l}
   \displaystyle \mathbf{\Psi}_{0}(\lambda)=\pmb{J}(\lambda),\quad\mathbf{\Psi}_{1}(\lambda)=-\int_0^\lambda\mathbf{K}(\lambda, \lambda_0)
   \mathbf{\Psi}_{0}(\lambda_0) d\lambda_0 ,\\ \cdots,\\
   \displaystyle \mathbf{\Psi}_{m+1}(\lambda)=(-1)^{m+1}\int_0^\lambda \mathbf{K}(\lambda, \lambda_m)\mathbf{\Psi}_{m}(\lambda_m)d\lambda_m\\
   \displaystyle=(-1)^{m+1}\int_0^\lambda \mathbf{K}(\lambda, \lambda_{m})\int_0^{\lambda_m}\mathbf{K}(\lambda_m, \lambda_{m-1})\int_0^{\lambda_{m-1}}
   \\\displaystyle \cdots \mathbf{K}(\lambda_{2}, \lambda_{1})
   \int_0^{\lambda_1}\mathbf{K}(\lambda_1, \lambda_0)\mathbf{J}(\lambda_0)  d\lambda_{0}\cdots d\lambda_{m-1}d\lambda_m.
\end{array}
\end{eqnarray}
Utilizing the similar strategy, we split the integrand of each
$\mathbf{\Psi}_m(\lambda)$ into two parts, one is used as an
associated PDF to construct a random sequence, and the other is used
as the estimation weight for final results. From the expression of
transfer kernel $\mathbf{K}(\lambda, \lambda')$, we can see that the
position transport is simply determined by
\begin{eqnarray}
\label{my_Tsss1}
   T_s(\lambda)=\frac{1}{A}\exp(-\lambda\alpha)\alpha,
\end{eqnarray}
where $A$ is the normalization factor. If the initial position is
$\lambda'$, the transported position $\lambda$ is randomly
determined by $\lambda=\lambda'-\ln(1-\xi A)/\alpha$, where $\xi$ is
a random number. At the new position, the components of
$\mathbf{\Psi}$ will experience an action exerted by the matrix
$\mathbf{R}/\alpha$ (the factor $\alpha$ is to balance the same
factor appeared in Eq. \eqref{my_Tsss1}). This action can be
regarded as a special scattering which does not change the frequency
and momentum of the photon. Finally the normalization factor $A$
will be added to the estimation factor. Later, we will demonstrate
the validation of this procedure by a simple RT problem which has
analytical solution (see discussions given in Section
\ref{testofRT_FRC}).

When the scattering is included, the RTE becomes
\begin{eqnarray}
\label{eq:symchrotroneqfs}
      &\displaystyle \frac{d\mathbf{I}}{d\lambda} + \alpha\mathbf{K} = \mathbf{j} - \mathbf{R} \mathbf{I}+
       \int \mathbf{S}(\mathbf{p}'\rightarrow\mathbf{p}) \mathbf{I}(\mathbf{p}')d\mathbf{p}',
\end{eqnarray}
where $\mathbf{p}=(\nu, \mathbf{\Omega})$, $\mathbf{S}$ is the
scattering matrix, the diagonal elements of $\mathbf{R}$ are
replaced by $\alpha_I + \alpha_s-\alpha$, and $\alpha_s$ is the
scattering coefficient. Implementing the same strategy, Eq.
\eqref{eq:symchrotroneqfs} can be converted into
\begin{eqnarray}
\label{eq:symchrotroneqfs2}
  & \displaystyle \mathbf{\Psi} = \mathbf{J} + \int \widetilde{\mathbf{S}}(\mathbf{p}'\rightarrow \mathbf{p})
  & \displaystyle \mathbf{\Psi}(\mathbf{p}', \tau')\exp(-\tau')d\tau'd\mathbf{p}',
\end{eqnarray}
where $\tau=\lambda\alpha$ is the optical depth, and
\begin{eqnarray}
\label{eq:symchrotroneqfs3}
 \widetilde{\mathbf{S}}(\mathbf{p}'\rightarrow\mathbf{p})=p_1\frac{\mathbf{R}(\lambda)\delta(\mathbf{p}'-\mathbf{p})}{\alpha-\alpha_s}
 +p_2\frac{\mathbf{S}(\mathbf{p}'\rightarrow\mathbf{p})}{\alpha_s},
\end{eqnarray}
where $p_1=(\alpha-\alpha_s)/\alpha$, $p_2= \alpha_s/\alpha$. Also,
we choose $\alpha$ such that all absolute values of elements of
matrix $\mathbf{R}(\lambda)/(\alpha-\alpha_s)$ are smaller than 1.
We have $p_1>0, p_2>0$ and $p_1+p_2=1$. Thus $p_1$ and $p_2$ can be
regarded as two probabilities that correspond the scattering
processes described by $\delta(\mathbf{p}'-\mathbf{p})$ and
$\mathbf{S}(\mathbf{p}'\rightarrow\mathbf{p})$ to take place,
respectively. We generate a random number $\xi$, if $\xi\le p_1$,
the momentum $\mathbf{p}$ keeps unchanged and $\mathbf{\Psi}$ is
updated by multiplying the matrix $\mathbf{R}(\lambda)$. If
$\xi>p_1$, we will sample the scattering kernel
$\mathbf{S}(\mathbf{p}'\rightarrow\mathbf{p})$ to get the scattered
direction and frequency. The algorithm for position transport is
exactly the same with that given by Eq. \eqref{my_Tsss1}.

\section{Verifications of the scheme}
\label{sec4_verify}

In the last two sections, we have introduced our new MC scheme based
on Neumann solution to solve the RTEs with or without polarizations
in detail. We emphasize to understand the MCRT scheme mathematically
and the whole thing is nothing but to evaluate the infinite terms of
multiple integrals appeared in the Neumann solution simultaneously.
One of the most important advantages is that one can eliminate any
$\delta$ functions in these integrals in advance, especially the
$\delta$ function in the recording function that connects the
Neumann solution and the observational quantities. After this
elimination, each sample of a scattering sequence can make
contribution to the observational quantity. This can greatly improve
the signal to noise performance of MCRT calculation. Even further
for each $P_i$, one can use it to estimate the observational
quantities at any directions. For example, if one wants to calculate
the angular dependent emergent spectrum $I(\mu)$ for $n$ directions
$\mu_j, \,\,j=1,\cdots,n$. One can evaluate $C(P_i, \mu_j)$ for $n$
times with different $\mu_j$ and take them as estimations for all
$I(\mu_j)$ simultaneously.

To verify the validation of this new MCRT scheme, we will apply it
to various RT problems in this section. The comparison between our
and former results shows its excellence performances in dealing RT
problems.

\subsection{Optically Thin and Thick Synchrotron Radiations from a Spherical Cloud}
\begin{figure}[b]
\begin{center}
\includegraphics[scale=0.45]{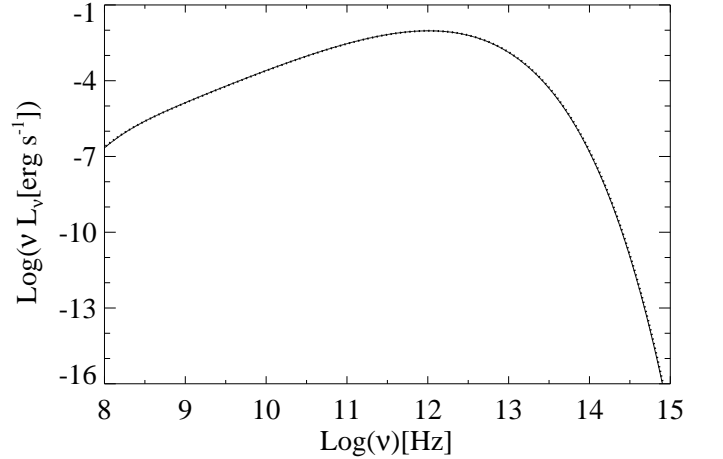}
\caption{The synchrotron emission spectrum emerging from an optical thin sphere threaded with
vertical magnetic fields and the line of sight of the observer
is perpendicular to the field lines. The solid and doted lines
represent the results produced by the analytic formula and our MC
code, respectively.}
  \label{sphere1}
\end{center}
\end{figure}

\begin{figure}[t]
\begin{center}
  \includegraphics[scale=0.45]{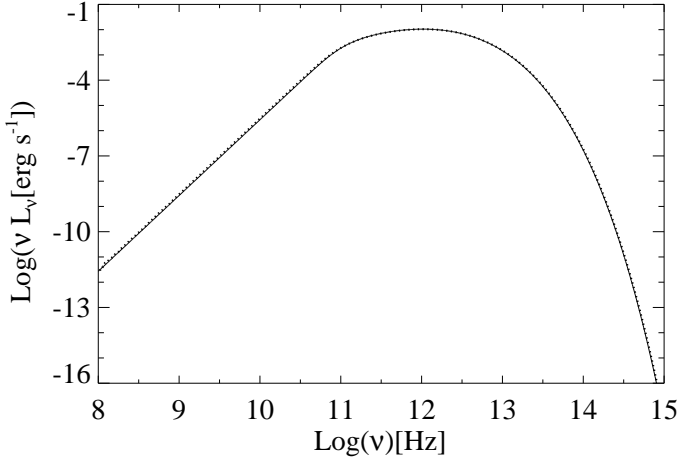}
\caption{The same spectrum with Fig. \ref{sphere1} except for
that the electron number density is increased by a $10^5$ magnitude
and thus the cloud is optically thick.} \label{sphere2}
\end{center}
\end{figure}
In the first example, we will reproduce the energy spectra emerged
from an optically thin and thick spherical cloud with an unit volume
and uniformly distributed magnetic fields along the vertical
direction. This simple example has been discussed by
\cite{Dolence2009}. Both the emissivity and absorption coefficients
are uniform and no scattering processes are considered. With these
conditions, the RTE can be solved analytically and the observed
intensity is given by
\begin{eqnarray}
\begin{split}
&\displaystyle  I_\nu=\frac{j_\nu}{\alpha_\nu}(1-e^{-\alpha_\nu L}),
\end{split}
\end{eqnarray}
where $L$ is the trajectory length of the ray measured in the
sphere, $j_\nu$ is the emissivity, which is given by the Eq. (4) of
\citet{Dolence2009}, $\alpha_\nu$ is absorption coefficient and
$\alpha_\nu=j_\nu/B_\nu$, where $B_\nu$ is the blackbody emissivity.
For optically thin case the above expression reduces to $
I_\nu\approx j_\nu L $.

With $I_\nu$, the value of the observed energy spectrum at frequency
$\nu$ can be obtained by integrating out the whole solid angle
spanned by the sphere with respect to the observer, and in order to
obtain the energy spectrum, this integration should be repeated for
each $\nu$ of a grid. In the calculation, all parameters are taken
as the same values with that given in \citet{Dolence2009}.

While in our scheme, we prefer to calculate this integral by the MC
method. Obviously, the integral is given by
\begin{eqnarray}
\begin{split}
&\displaystyle  I_\nu= \int j_\nu(\nu, \mathbf{\Omega})f(\mathbf{r},
\nu, \mathbf{\Omega}) d\mathbf{\Omega}d\mathbf{r}d\nu,
\end{split}
\end{eqnarray}
where $f=\exp[-\alpha_\nu
L(\mathbf{r},\mathbf{\Omega})]\delta(\mathbf{\Omega}-\mathbf{\Omega}_{\text{obs}})$
is the recording function,
$L=-\mathbf{r}\cdot\mathbf{\Omega}+\sqrt{(\mathbf{r}\cdot\mathbf{\Omega})^2+1-r^2}$.
Eliminating the $\delta$-function, we obtain
\begin{eqnarray}
\begin{split}
&\displaystyle  I_\nu= \int j_\nu(\nu,
\mathbf{\Omega}_{\text{obs}})\exp[-\alpha_\nu
L(\mathbf{r},\mathbf{\Omega}_{\text{obs}})] d\mathbf{r}d\nu.
\end{split}
\end{eqnarray}
To sample the frequency $\nu$, we introduce a new variable $y$
defined as $10^y=\nu$ and $y\in[8,15]$, the above integral becomes
\begin{eqnarray}
\begin{split}
&\displaystyle  I_\nu= \int j_\nu(10^y,
\mathbf{\Omega}_{\text{obs}})\exp(-\alpha_\nu L)\ln(10) 10^y
d\mathbf{r}dy,
\end{split}
\end{eqnarray}
where $\mathbf{r}=(r,\mu, \phi)$ and $y$ are sampled according to
the uniform PDFs and we directly have: $r=\xi_1^{1/3}, \mu=1-2\xi_2,
\phi=2\pi\xi_3$, and $y=8+7\xi_4$, where $\xi_i$ are random numbers.
The weight is $w= C j_\nu(10^y,
\mathbf{\Omega}_{\text{obs}})e^{-\alpha_\nu
L(\mathbf{r},\mathbf{\Omega}_{\text{obs}})}10^y$, where $C$ is the
product of normalization factors of these uniform PDFs. The
weight is recorded to a bin to which the $\nu$ belongs. The spectra
for optical thin and thick cases are shown in Figs. \ref{sphere1}
and \ref{sphere2}. One can see that they agree with each other very
well.

\subsection{the Comptonlization of Low Energy Radiations by hot Plasma}

\begin{figure}[b]
\begin{center}
  \includegraphics[scale=0.65]{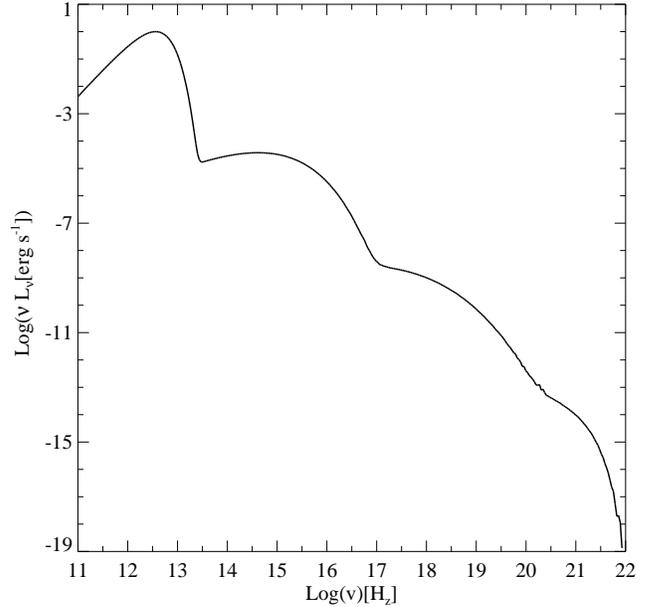}
\caption{Comptonized spectrum of radiations emerge from a
spherical cloud of hot plasma. The thermal point source is located
at the center of the cloud. The values of dimensionless parameters
are $\Theta_e=4$ and $\Theta_s=10^{-8}$ and $\tau=10^{-4}$,
respectively. Compare to Figure 7 of \citet{Dolence2009}.}
\label{sphere3}
\end{center}
\end{figure}

\begin{figure}[b]
\begin{center}
  \includegraphics[scale=0.65]{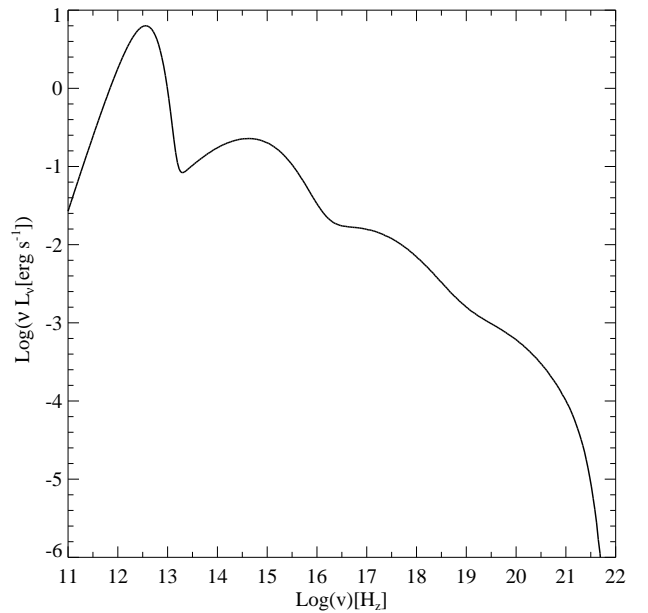}
\caption{Same as in Figure \ref{sphere3}, but for $\tau$ = 0.1.
Compare to Figure 8 of \citet{Dolence2009}.} \label{sphere4}
\end{center}
\end{figure}

\begin{figure}[t]
\begin{center}
  \includegraphics[scale=0.65]{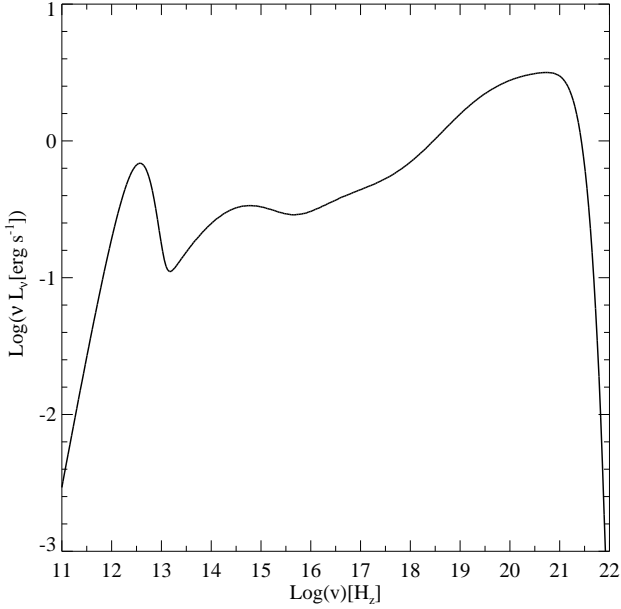}
\caption{Same as in Figure \ref{sphere3}, but for $\tau$ = 3.0.
Compare to Figure 9 of \citet{Dolence2009}.} \label{sphere5}
\end{center}
\end{figure}

Next we hope to reproduce the Comptonlized spectrum of soft photons
emitted by a point  thermal source at the center of a spherical
cloud of plasma with various optical depth and compare it with that
computed by \citet{Dolence2009}. There is no any absorption and
emission in the cloud and the RT process is unpolarized. The
central source is isotropic and then the emissivity of the cloud can
be written as $j_\nu(\nu)=B_\nu(\Theta_s)\delta(r)$, where
$\Theta_s=kT_s/(m_ec^2)$ is the dimensionless temperature of the
source. The soft photons are scattered out by the thermal electrons
of the cloud with dimensionless temperature
$\Theta_e=kT_e/(m_ec^2)=4$, radius $R$, number density $n_e$ and
Thomson optical depth $\tau=R\sigma_Tn_e$.

From $j_\nu(\nu)$ we can directly obtain the starting point of a
scattering sequence as: $r=0$ and momentum direction $(\mu=1-2\xi_1,
\phi=2\pi\xi_2)$. The initial frequency $\nu$ is sampled by
$\nu_\xi=10^{n_1+(n_2-n_1)\xi}$ and $w=Cj_\nu(\nu_\xi)\nu_\xi$ is
taken as the initial weight. The scattering distance is randomly
selected according to a normalized PDF $p(l)=n_e\sigma_s(\nu,
T_e)\exp[-n_e\sigma_s(\nu, T_e)l]/A$, where
$A=[1-\exp(-n_e\sigma_s(\nu, T_e)L)]$ is the normalization factor
and $L$ is the path length extended to the cloud boundary. With
$p(l)$, the distance is simply determined by
$l=-\ln(1-A\xi)/[n_e\sigma_s(\nu, T_e)]$. After each position
transport, the weight is updated as $w=w\cdot A$.

In our scheme, when the direction of the observer
$\mathbf{\Omega}_{\text{obs}}$ is specified, the contribution made
by the $i$-th sample of the sequence can be written as:
$F_i=wC(\nu_i,\mathbf{\Omega}_i\rightarrow\nu_{\text{obs}},
\mathbf{\Omega}_\text{obs})\exp[-\tau_s(s_b,\mathbf{\Omega}_\text{obs}
)]$. Then each scattering will make contribution to the spectrum.
While, due to the completely spherical symmetry possessed both by
the cloud and the source, any direction is actually valid and
equivalent in the estimation of the spectrum. The scattered
direction $\mathbf{\Omega}_i$ is actually a better choice which has
lower variance comparing to our scheme. And the contribution can be
written as: $F_i=w \exp[-\tau_s(s_b, \mathbf{\Omega}_i)]$. In this
example we prefer to employ the later estimation procedure. Since
$A<1$, $F_i$ will decrease gradually as the scattering number
increases. As $F_i$ is lower than a threshold, its contribution can
be ignored completely and scattering sequence is truncated as well.

The results are plotted in Figure. \ref{sphere3}-\ref{sphere4} for
$\tau=10^{-4}$, $0.1$, $3$ respectively. Compare with the
results of \citet{Dolence2009} one can find that they agree with
each other very well.

\subsection{Chandrasekhar's Limit}
\label{section5.3}

\begin{figure*}[t]
\begin{center}
  \includegraphics[scale=0.46]{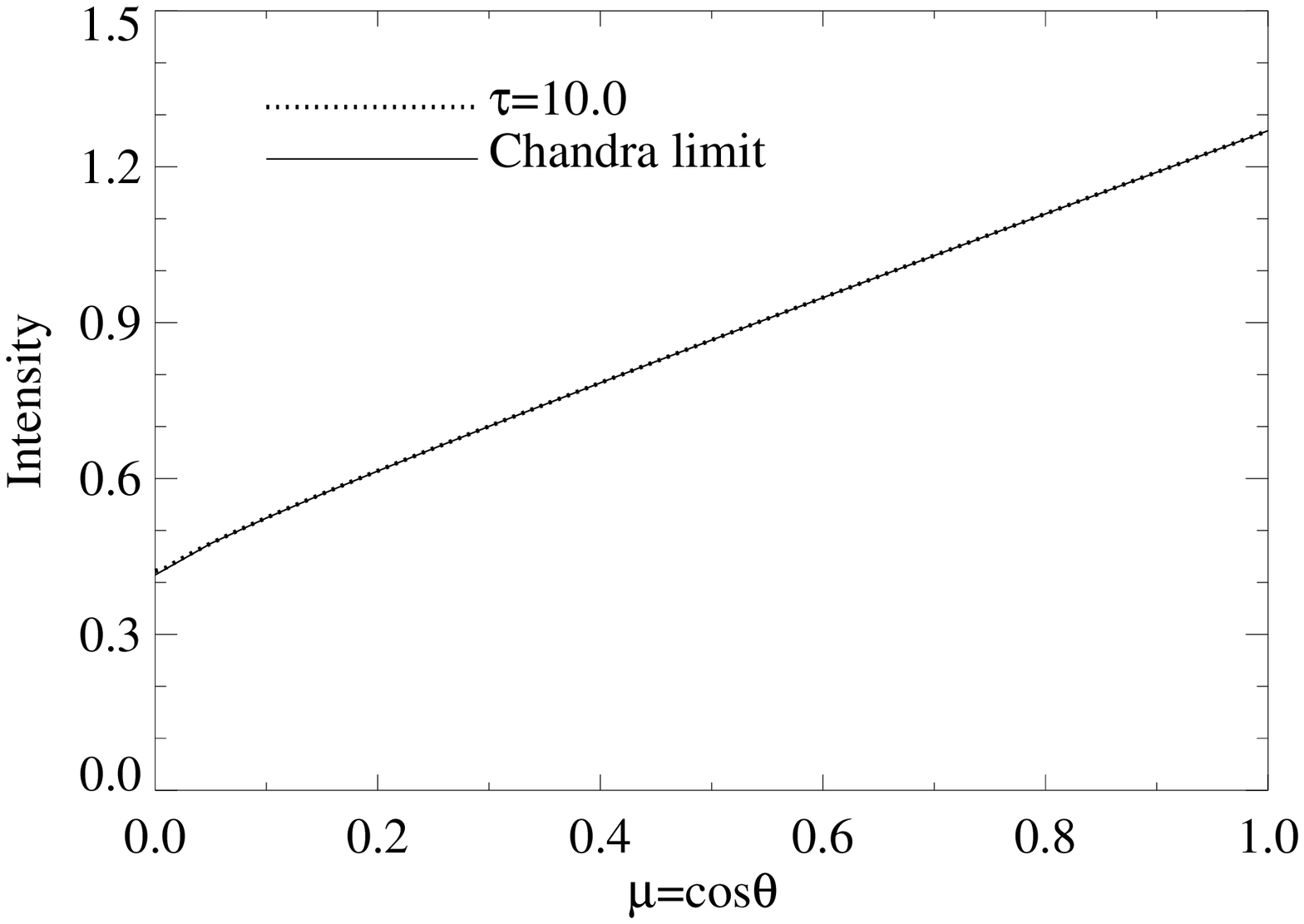}
  \includegraphics[scale=0.46]{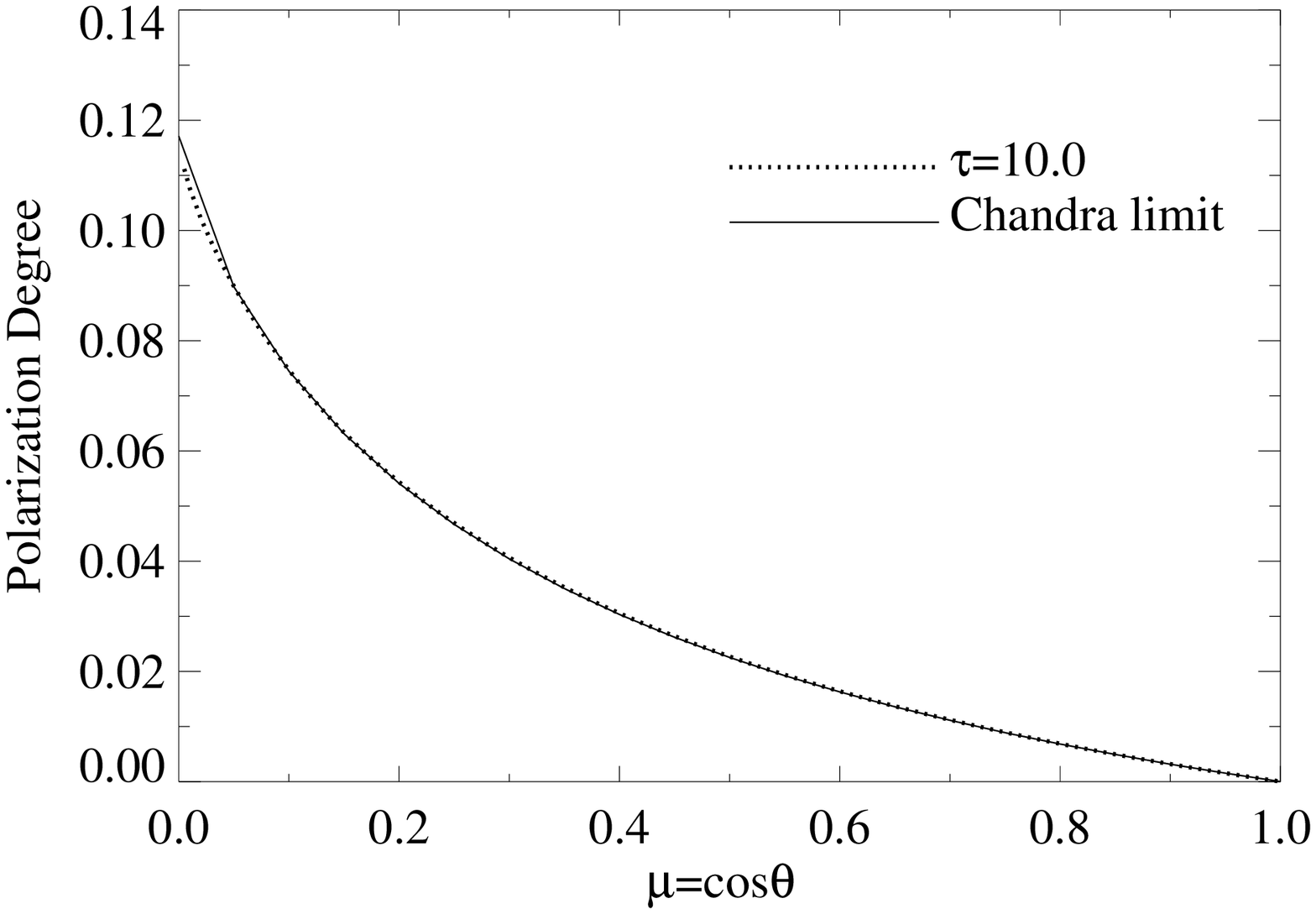}
\caption{The angular-dependent radiation intensity (left) and
polarization degree (right) emerge from a semi-infinite
plane-parallel atmosphere with Rayleigh scattering dominating. The
dotted and solid lines represent the results calculated by our MC
scheme and the semi-analytical Table XXIV of \cite{Chand1960},
respectively.} \label{chand1}
\end{center}
\end{figure*}

Now we begin to test our scheme for the polarized RT with
scattering. In the first example, we will calculate the angular
dependent polarization degrees of radiation emerging from the
surface of a semi-infinite plane-parallel atmosphere with an
infinite large optical depth $\tau$, and this result is
exactly what the Table XXIV of \citet{Chand1960} presented. The
atmosphere is dominated by Rayleigh scattering and its optical depth
$\tau$ is measured from the up boundary (where $\tau=0$)
downward. Due to the axial symmetry of the plane-parallel
geometry of the atmosphere, the polarized RTE can be simplified and
only $I$ and $Q$ are sufficient to describe the RT process.
The RTE is given by \citep{Chand1960,Ports2004}
\begin{eqnarray}
\label{eqas1}
\begin{split}
&\displaystyle  \mu\frac{d\mathbf{I}(\tau, \mu)}{d\tau}= \mathbf{I}(\tau, \mu)-\frac{3}{16}
\int_{-1}^1 \mathbf{P}(\mu;\mu') \mathbf{I}(\tau, \mu')d\mu',
\end{split}
\end{eqnarray}
where $\mathbf{I}=(I, Q)^T$, and
\begin{eqnarray}
\begin{split}
&\displaystyle  \mathbf{P} =\left(
   \begin{array}{lr}
       3+3\mu^2\mu'^2-\mu^2-\mu'^2 & \,\, (1-\mu'^2)(1-3\mu^2)\\
       (1-\mu^2)(1-3\mu'^2)        & \,\, 3(1-\mu'^2)(1-\mu^2)
   \end{array}\right),
\end{split}
\end{eqnarray}
is the simplified scattering matrix. Since we adopt $I$ and $Q$ to
describe the RT, rather than $I_r$ and $I_l$, the scattering matrix
$\mathbf{P}$ is different from that of \citet{Chand1960}. One can
see that the transfer equations given by Eq. \eqref{eqas1} has no
emissivity term. This is certainly a severe trouble which fails our
scheme, since if the emissivity vanishes, the Neumann solution
vanishes as well (see Eq. \eqref{eqa49}) and we can not even
initiate the construction of the stochastic sequence without an
emission source. In order to apply our scheme, we have to introduce
an emissivity term, which is given by
\begin{eqnarray}
  \left\{\begin{array}{ll}
  &\displaystyle J_I(\tau, \mu) = F_0\exp\left(-\frac{\tau_1-\tau}{\mu}\right), \\
    &J_Q(\tau, \mu)=0,
  \end{array}\right.
\end{eqnarray}
where $F_0$ and $\tau_1$ are two constants. In our practical
calculations, the total optical depth of the atmosphere can not be
taken as infinite, instead we always choose a finite but sufficient
large $\tau_1$. This emissivity $\mathbf{J}$ is actually equivalent
to a boundary condition, i.e., an isotropic incident radiation is
designated at the bottom boundary at $\tau=\tau_1$ with constant
intensity $F_0$. Using $\mathbf{J}$, we can solve the RTE and
calculate the emerging spectrum according to the procedures
discussed in Section \ref{section4.3}. First, we sample
$\mathbf{J(\mu)}$ to generate the first point of a sequence, i.e.,
$P_0=(\tau_0, \mu_0)$, where $\tau_0=\tau_1\xi_2$, $\mu_0=1-2\xi_1$.
Since the scattering distance is equivalent to the optical depth
$\tau$ and the corresponding PDF $p(\tau'\rightarrow\tau)$ is given
by
\begin{eqnarray}
\label{eaq_tau1}
&\displaystyle p(\tau'\rightarrow\tau) = \left\{
  \begin{array}{lr}
    \displaystyle \frac{1}{A_1}\exp\left(-\frac{\tau'-\tau}{\mu}\right), \quad \mu >0,\\
    \displaystyle \frac{1}{A_2}\exp\left(-\frac{\tau-\tau'}{|\mu|}\right), \quad \mu <0,
  \end{array} \right.
\end{eqnarray}
where $A_1=1-\exp(-\tau'/\mu)$ and
$A_2=1-\exp[-(\tau_1-\tau')/|\mu|]$ are the normalization factors of
$p(\tau'\rightarrow\tau)$ for $\mu>0$ and $\mu<0$ respectively. By
sampling $p(\tau'\rightarrow\tau)$, we can get the next
scattering position $\tau_m$ from the initial position $\tau_{m-1}$
as
\begin{eqnarray}
\label{eaq_tau2}
&\displaystyle  \tau_m=\left\{
             \begin{array}{lr}
               \tau_{m-1} + \mu\ln(1-A_1\xi), \quad \mu>0,\\
               \tau_{m-1} - |\mu|\ln(1-A_2\xi), \quad \mu <0.
             \end{array} \right.
\end{eqnarray}
After each position transport, the weight factor $w$ is updated as
$w=w\cdot A_1$ for $\mu>0$ and $w=w\cdot A_2$ for $\mu<0$.

The scattering PDF of $\mu$ can be abstracted from the
scattering matrix after its action on the vector $(\Psi^{(I)}_{m-1},
\Psi^{(Q)}_{m-1})^T$ with given incident angle $\mu'=\mu_{m-1}$,
i.e.,
\begin{eqnarray}
\begin{split}
 \displaystyle  \Psi^{(I)}_m = f_I(\mu;\mu')\Psi^{(I)}_{m-1}, \,\, \Psi^{(Q)}_m =
f_Q(\mu;\mu') \Psi^{(I)}_{m-1}.
\end{split}
\end{eqnarray}
where
\begin{eqnarray}
\begin{split}
 \displaystyle  & f_I(\mu;\mu') = P_{11} + P_{12}\xi_Q, \,\,
 f_Q(\mu;\mu')=  P_{21} + P_{22}\xi_Q,
\end{split}
\end{eqnarray}
and $\xi_Q=\Psi^{(Q)}_{m-1}/\Psi^{(I)}_{m-1}$. Then we can choose the normalized function of
$f_I(\mu;\mu_{m-1})$ as the PDF for $\mu$:
\begin{eqnarray}
\begin{split}
 \displaystyle  p(\mu_{m-1}\rightarrow\mu) &= f_I(\mu;\mu_{m-1}),
\end{split}
\end{eqnarray}
Since $p(\mu_{m-1}\rightarrow\mu)$ is a quadratic function of $\mu$,
we can sample it by the inverse CDF method, which amounts to solve a
cubic equation. With the sampled $\mu_m$, similar to Eq.
\eqref{eqa55}, the updating relations for $\mathbf{\Psi}$ is
given by
\begin{eqnarray}
\begin{split}
  \left\{\begin{array}{ll}
     \displaystyle  \Psi^{(I)}_m = \Psi^{(I)}_{m-1}, \\
    \displaystyle\Psi^{(Q)}_m = \frac{f_Q(\mu_m;\mu_{m-1})}
         {f_I(\mu_m;\mu_{m-1})}\Psi^{(I)}_{m-1}.
    \end{array}\right.
\end{split}
\end{eqnarray}

Repeating this transport and scattering process, we can obtain
a sequence consisting of $P_m$ and $\mathbf{\Psi}_m$, and
$\Psi_0^{(I)}=J_I(\tau_0, \mu_0)$ and $\Psi_0^{(Q)}=0$. The
recording function is given by
\begin{eqnarray}
 \displaystyle f_{r,m}(\tau_m,\mu_m)=\exp\left(-\frac{\tau_m}{\mu_m}\right) \frac{1}{\mu_m}\delta(\mu_m-\mu_{\text{obs}}).
\end{eqnarray}
After eliminating the $\delta$ function, the above function takes a
new form as
\begin{eqnarray}
\begin{split}
\left\{\begin{array}{ll}
 &f_{r,m}^{(I)}(\tau_m,\mu_{\text{obs}})= \displaystyle\frac{f_I(\mu_{\text{obs}};\mu_{m-1})}{\mu_{\text{obs}}}\exp\left(-\frac{\tau_m}{\mu_{\text{obs}}}\right),\\
 &f_{r,m}^{(Q)}(\tau_m,\mu_{\text{obs}})= \displaystyle\frac{f_Q(\mu_{\text{obs}};\mu_{m-1})}{\mu_{\text{obs}}}\exp\left(-\frac{\tau_m}{\mu_{\text{obs}}}\right).
    \end{array}\right.
\end{split}
\end{eqnarray}
The estimations for $I_m$ and $Q_m$ can be written as
$F^{(I)}_m=w\Psi^{(I)}_{m-1} f_{r,m}^{(I)}$ and
$F^{(Q)}_m=w\Psi^{(Q)}_{m-1} f_{r,m}^{(Q)}$. Especially, for $m=0$,
we have
\begin{eqnarray}
\begin{split}
\left\{\begin{array}{ll}
 & F^{(I)}_0=\displaystyle J_I(\tau_0, \mu_{\text{obs}})\exp\left(-\frac{\tau_0}{\mu_{\text{obs}}}\right) \frac{1}{\mu_{\text{obs}}},\\
 & F^{(Q)}_0=0.
 \end{array}\right.
\end{split}
\end{eqnarray}
Obviously the above procedure can calculate the estimations for any
direction with $\mu$, provided $\mu_{\text{obs}}$ is replaced by
$\mu$. Therefore, if we input a set of observational cosines:
$\mu_{\text{obs}}^i=i/n, i=0, 1, \cdots, n$, we can obtain the
estimations for all of them simultaneously. It can greatly improve
the calculation efficiency. The comparison of our results with that
of Table XXIV of \citet{Chand1960} is shown in Fig. \ref{chand1},
where we take $\tau_1=10$. One can see that they agree well.

\subsection{Polarizations of a Scattering Dominated Plane-Parallel Disk Atmosphere}
\label{section5.4}

\begin{figure*}[t]
\begin{center}
  \includegraphics[scale=0.47]{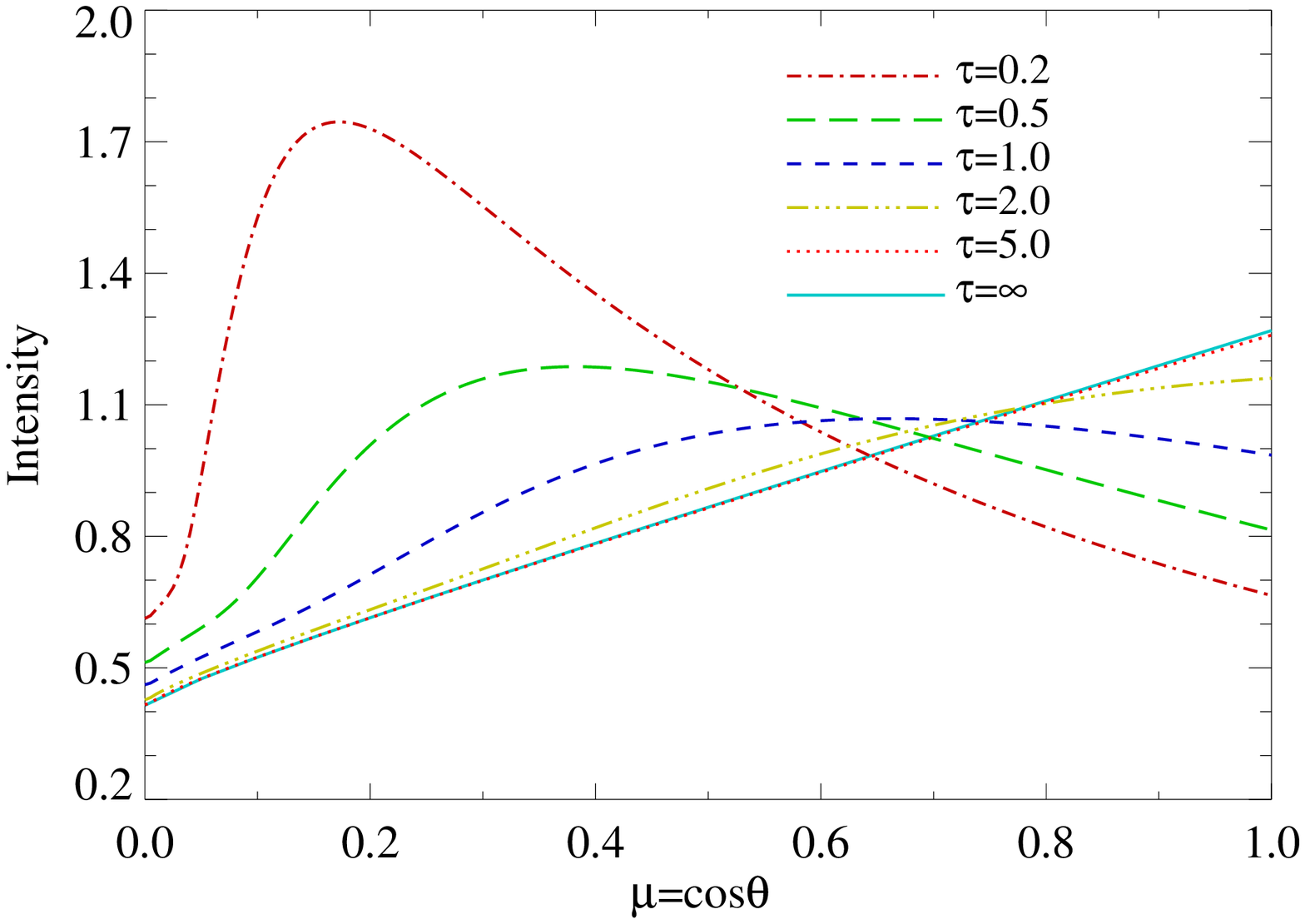}
  \includegraphics[scale=0.47]{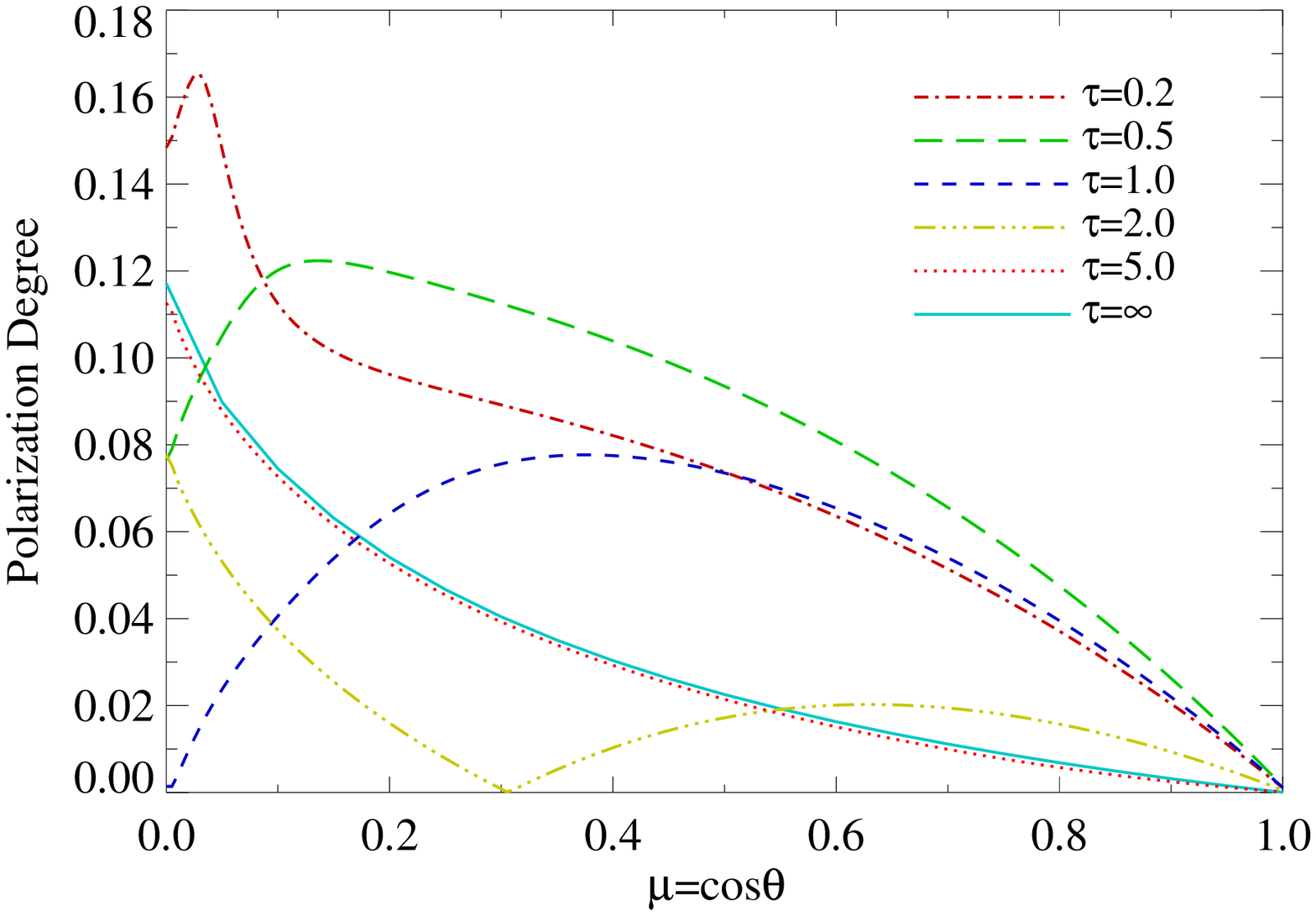}
\caption{The polarized spectrum (left) and polarization degree
(right) of radiations emerge from the plane-parallel atmosphere of a
disk without thickness. The primary photons are isotropically
emitted from the central plane of the whole system where the disk
located. Radiations can penetrate the disk freely as transporting.
Curves with different line styles represent various optical depth of
the atmosphere, i.e., $\tau=0.2,0.5,1.0,2.0,5.0,\infty$. Compare to
Figure.1 of \cite{Dov2008}.} \label{plane1}
\end{center}
\end{figure*}

With the procedure presented in the above subsection, we can proceed
to calculate the angular distribution of polarization degrees of
radiation emerging from the atmosphere of a disk with a finite
optical depth $\tau_1$ (see the configuration shown in panel (a) of
Fig. 1 of \citet{sun1985}). The RTE of this system is exactly the
same with Eq. \eqref{eqas1}, except that the primary source is
provided explicitly. The source is isotropically and uniformly
located on the disk surface (where $\tau=\tau_1/2$). The emissivity
can be written as $\mathbf{J}(\tau, \mu)=F_0\delta(\tau-\tau_1/2)(1,
0)^T$, where $F_0$ is a constant. Repeating the same procedure, we
can reproduce the results of Fig. 1 of \citet{Dov2008}, which were
calculated by the public available code STOKES \citep{Goosmann2007}.
Our results are demonstrated in Fig. \ref{plane1}, where
$\tau=\tau_1/2$. One can compare it with that of \citet{Dov2008}
and find that they are in a good agreement. From the figure one can
see that as the optical depth increases, the results approach the
Chandrasekhar's Limit eventually.

As we have mentioned in \ref{section4.3}, one can calculate the
results equivalently through another scheme, in which the direction
is described by $\Theta, \Phi$. We need trace the polarization
vector and construct tetrad, in which the scattering takes place.
Thus, this scheme is more complicated and time consuming Compared
with the above one.

\subsection{Radiations Diffusely Reflected from a Semi-Infinite Plane-Parallel Atmosphere}
\label{diffuseReflec}

In this section, we will calculate the angular-dependent
polarizations of radiations diffusely reflected from a semi-infinite
plane-parallel atmosphere. The incident radiation is a parallel beam
with net flux $\mathbf{F}$ per unit area perpendicular to the
propagating direction $(-\mu_0, \varphi_0)$. Notice that the
transport of the fourth component $V$ of the SPs is decoupled from
the other three components, we denote $\mathbf{I}=(I, Q, U)^T$ and
$\mathbf{F}=(F_I, F_Q, F_U)^T$. The RTEs governing this system can
be written as (Eqs. (101) and (104), Chap. X of \citet{Chand1960},
page 249)
\begin{eqnarray}
\label{diffrefl}
\begin{split}
&\displaystyle \mu\frac{d\mathbf{I}(\tau, \mu, \varphi)}{d\tau}= \mathbf{I}(\tau, \mu, \varphi)-\frac{1}{4\pi}
\int_{-1}^{+1}\int_0^{2\pi} \mathbf{P}(\mu, \varphi;\mu', \varphi') \times\\
&\mathbf{I}(\tau, \mu', \varphi')d\mu'd\varphi'-\frac{1}{4}e^{-\tau/\mu_0}\mathbf{P}
(\mu, \varphi;-\mu_0, \varphi_0)\mathbf{F},\\
&\displaystyle \mu\frac{dV(\tau, \mu, \varphi)}{d\tau}= V(\tau, \mu, \varphi)-\frac{3}{8\pi}
\int_{-1}^{+1}\int_0^{2\pi} P_V(\mu, \varphi;\mu', \varphi') \times\\
&V(\tau, \mu', \varphi')d\mu'd\varphi'-\frac{3}{8}e^{-\tau/\mu_0}P_V
(\mu, \varphi;-\mu_0, \varphi_0)F_V,\\
\end{split}
\end{eqnarray}
where $\mathbf{P}$ (given in the Appendix B) is the scattering
matrix (or phase-matrix) for $\mathbf{I}$. Since the scattering
matrix in terms of $I, Q$ is more suitable for sampling, we prefer
to use $I, Q$ rather than $I_l, I_r$, hence the expression of
$\mathbf{P}$ adopted here is different with that of
\citet{Chand1960}. The expression of $P_V$ reads
\begin{eqnarray}
&\displaystyle P_V=\mu\mu'+\sqrt{1-\mu^2}\sqrt{1-\mu'^2}\cos(\varphi'- \varphi).
\end{eqnarray}
From Eq. \eqref{diffrefl} one can see that the emissivity functions
are given by
\begin{eqnarray}
&\displaystyle \mathbf{J}= \frac{1}{4}e^{-\tau/\mu_0}\mathbf{P}
(\mu, \varphi;-\mu_0, \varphi_0)\mathbf{F},\\
&\displaystyle J_V= \frac{3}{8}e^{-\tau/\mu_0}P_V
(\mu, \varphi;-\mu_0, \varphi_0)F_V.
\end{eqnarray}
From which we can get the first sample of the sequence:
$P_1=(\tau_1, \mu_1, \varphi_1)$ and $\mathbf{\Psi}_1$. By sampling
$p(\tau)=e^{-\tau/\mu_0}$, we can directly obtain
$\tau_1=-\mu_0\ln(1-\xi)$. From the expressions of $\mathbf{J}$ and
$J_V$, one can see that $(\mu, \varphi)$ is actually the scattered
direction for an incident direction $(-\mu_0, \varphi_0)$.
Thus $(\mu_1, \varphi_1)$ can be obtained by sampling a common PDF
$f_I$ (given by Eq. \eqref{expression_f_I}) shared by four
emissivity functions. Meanwhile the remaining parts are taken as
values for $\mathbf{\Psi}_1$.

Similarly the next scattering position $\tau_{m+1}$ can be obtained
by using Eq. \eqref{eaq_tau2} as $\tau_m$ is provided. The weight
factor $w$ should be multiplied by the factor $A_1$ for $\mu_m>0$
(and $A_2$ for $\mu_m <0$), where $A_1$ and $A_s$ are given by Eq.
\eqref{eaq_tau1}. Since the scattering sampling procedure of
generating $(\mu_{m+1}, \varphi_{m+1})$ with $(\mu_{m},
\varphi_{m})$ provided is very complicated, we will defer the
relevant discussions to the Appendix \ref{app2}. After $\mu_{m+1},
\varphi_{m+1}$ is obtained, $\mathbf{\Psi}_{m+1}$ should be updated
according to Eq. \eqref{psi_update}. Analogously, once a sequence of
$P_i$ and $\mathbf{\Psi}_{i}$ ($i=1, 2, \cdots$) is generated, the
estimations for $\mathbf{I}_{m}$ and $V_m$ can be obtained as
\begin{eqnarray}
\begin{split}
\left\{
   \begin{array}{ll}
&\displaystyle F^{(X)}_m=
w\cdot f_{X}(\mathbf{\Omega}_{\text{obs}};\mathbf{\Omega}_{m-1})\cdot\Psi^{(I)}_{m-1}\cdot f_{m},\\
& F^{(V)}_m= w\cdot
f_V(\mathbf{\Omega}_{\text{obs}};\mathbf{\Omega}_{m-1})\cdot\Psi^{(V)}_{m-1}\cdot
f_{m},
    \end{array}
  \right.
\end{split}
\end{eqnarray}
where $X$ can be $I, Q, U$, and
\begin{eqnarray}
\begin{split}
 &f_m= \displaystyle\frac{1}{\mu_{\text{obs}}}
 \exp\left(-\frac{\tau_{m}}{\mu_{\text{obs}}}\right),
\end{split}
\end{eqnarray}
is the recording function. Repeating this procedure, the
angular-dependent diffusely reflected spectrum can be eventually
generated.

\begin{figure}[t]
\begin{center}
  \includegraphics[scale=0.48]{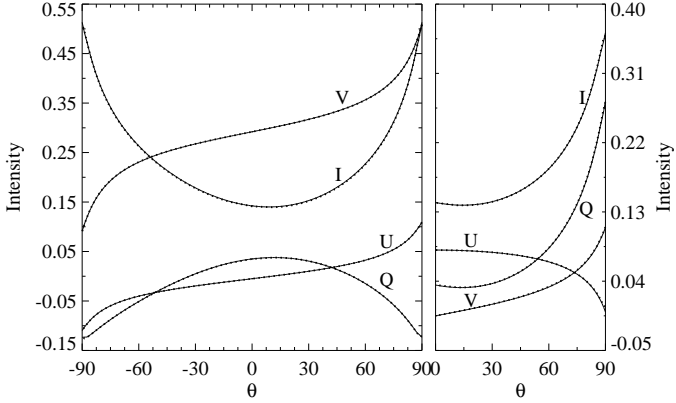}
\caption{SPs of radiations diffusely reflected from a semi-infinite
plane-parallel atmosphere distribute with respect to the polar angle
in the incident plane ($\varphi-\varphi_0=0^\circ, 180^\circ$, left
panel) and the plane perpendicular to it
($\varphi-\varphi_0=90^\circ, 270^\circ$, right panel). The
direction of the incident parallel beam is $\mu_0=0.2, \varphi_0=0$.
And the incident beam is polarized with SPs given by $F_Q=F_I/4$,
$F_U=F_I/4$ and $F_V=F_I/4$. The solid and dotted lines represent
the results calculated by the semi-analytical formula of
\citet{Chand1960} and our MC scheme, respectively. Compare to Figure
24 of \citet{Chand1960}.} \label{diffrefle11}
\end{center}
\end{figure}

\begin{figure}[t]
\begin{center}
  \includegraphics[scale=0.49]{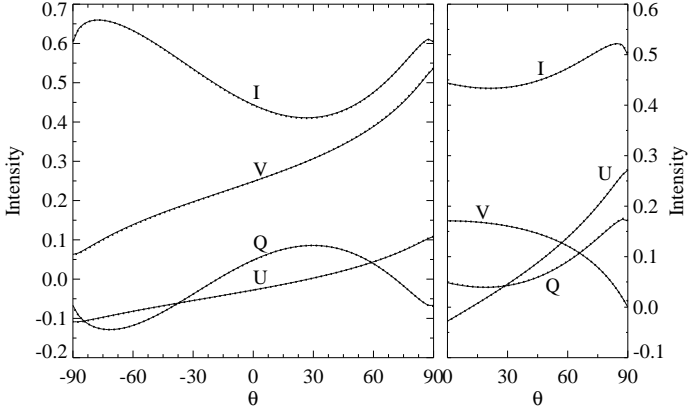}
\caption{Same as in Figure \ref{diffrefle11}, but for $\mu_0=0.5$.
Compare to Figure 25 of \citet{Chand1960}.} \label{diffrefle12}
\end{center}
\end{figure}

\begin{figure}[t]
\begin{center}
  \includegraphics[scale=0.49]{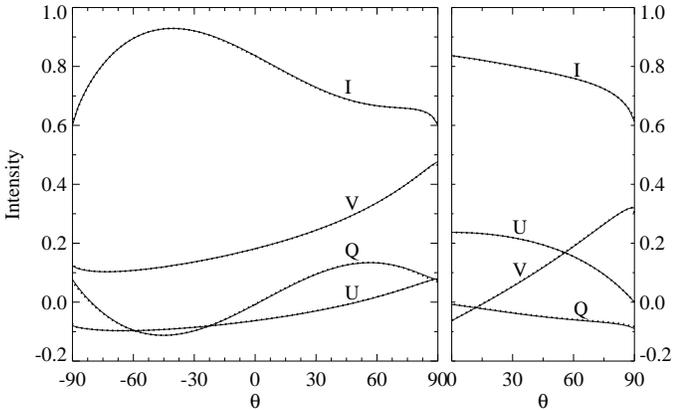}
\caption{Same as in Figure \ref{diffrefle11}, but for $\mu_0=0.8$.
Compare to Figure 26 of \citet{Chand1960}.} \label{diffrefle13}
\end{center}
\end{figure}

We first reproduce the results of Fig. 11 of \citet{Schnittman2013},
which mainly illustrate the distributions of radiations diffusely
reflected forward and backward in the incident plane (where
$\varphi-\varphi_0=0^{\circ}, 180^{\circ}$) and the plane
perpendicular to it (where $\varphi-\varphi_0=90^{\circ},
270^{\circ}$), respectively. Due to the symmetry, the radiations
reflected to the directions at $\varphi-\varphi_0=90^{\circ}$ and
$270^{\circ}$ are exactly the same, we can put them together in the
plotting. The angles of incident beam are $\varphi_0=0$ and
$\mu_0=0.2, 0.5, 0.8$ for three cases and the beam is polarized with
the SPs given by $F_Q=F_I/4$, $F_U=F_I/4$ and $F_V=F_I/4$. The
results are shown in Fig. \ref{diffrefle11}, \ref{diffrefle12} and
\ref{diffrefle13}, in which the semi-analytical results of
\citet{Chand1960} for diffuse reflection are also plotted. They can
be obtained directly from the well know formula of \citet{Chand1960}
given by
\begin{eqnarray}
\label{chandraIlru} &\displaystyle  \mathbf{I}(\tau=0, \mu,
\varphi)=\left(
 \begin{array}{c}
     I_l\\
     I_r\\
     U\\
     V
 \end{array}\right)=\frac{1}{4\mu}\mathbf{Q}\mathbf{S}(\mu, \varphi; -\mu_0, \varphi_0)\left(
 \begin{array}{c}
     F_l\\
     F_r\\
     F_U\\
     F_V
 \end{array}\right),\quad\quad
\end{eqnarray}
where $F_l, F_r, F_U$ are the SPs of the incident radiation. From
these figures, one can see that the two results coincide with each
other very well. Comparing to the cases with unpolarized incident
beams, the polarized incident radiation will give rise to a nonzero
reflected $U$.

\begin{figure}[t]
\begin{center}
  \includegraphics[scale=0.89]{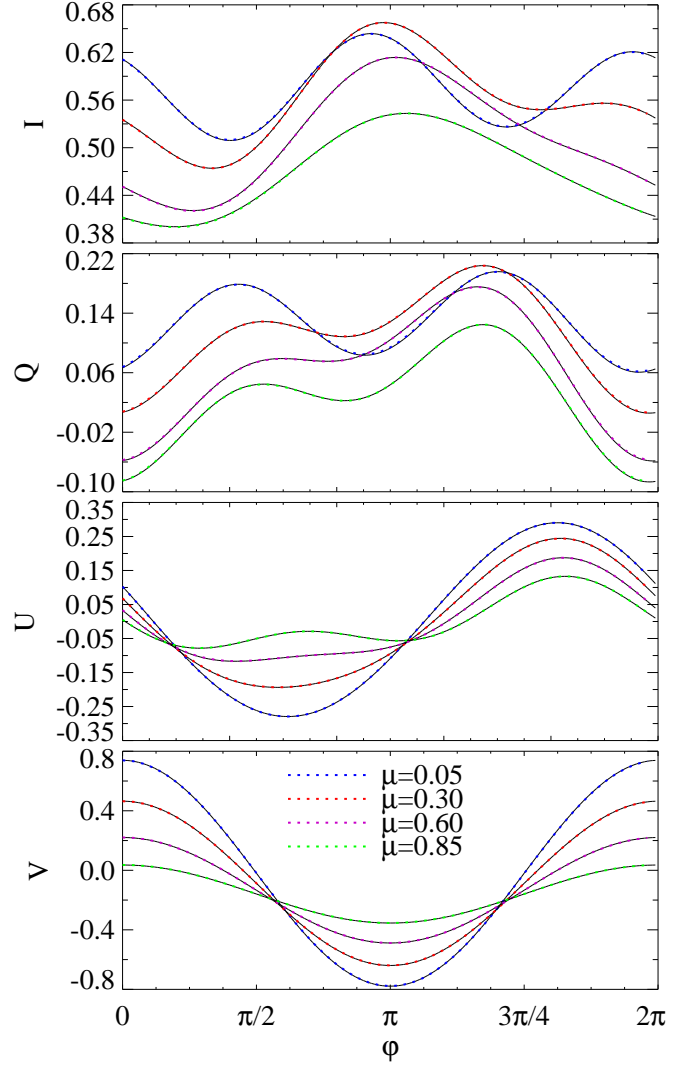}
\caption{The distribution of I, Q, U, V of radiations diffusely
reflected from a plane-parallel atmosphere with respect to the
azimuth angle $\varphi$ for various cosines of viewing angles, i.e.,
$\mu=0.05, 0.3, 0.6, 0.85$, respectively. The incident radiation is
polarized and has a direction of $\mu_0=0.5$ and $\varphi_0=0$. The
solid and dotted lines represent the results calculated by the
semi-analytical formula of Eq. \eqref{chandraIlru} and our MC
scheme, respectively.} \label{diffrefle14}
\end{center}
\end{figure}

\begin{figure}[t]
\begin{center}
  \includegraphics[scale=0.89]{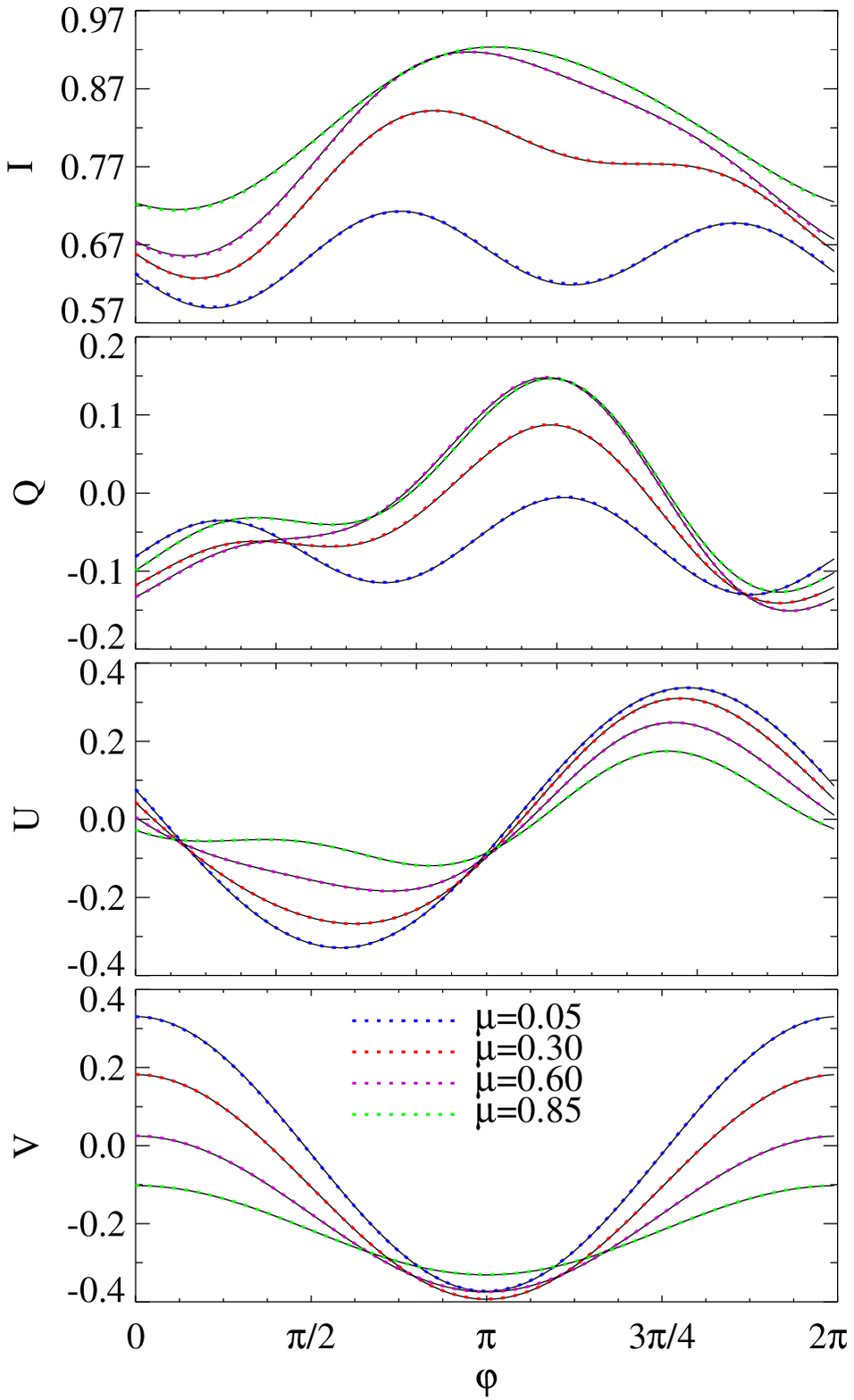}
\caption{Same as in Figure \ref{diffrefle14}, but for $\mu_0=0.8$.}
\label{diffrefle15}
\end{center}
\end{figure}

We proceed to demonstrate the SPs of diffusely reflected radiations
distributed with respect to the azimuth angle $\varphi$ for various
observational angles in Fig. \ref{diffrefle14} and
\ref{diffrefle15}. The incident radiation is also polarized with the
same manner given above. For comparison, we also plot the
semi-analytical results of \citet{Chand1960}. In our calculations,
the maximum optical depth along the vertical direction of the
atmosphere is set to be $\tau=100$. From these Figures one can see
that as $\mu$ decrease (the escaped radiations will travel through a
higher optical depth and be scattered sufficiently by the
atmosphere), the maxima and minima of the SPs show a period of $\pi$
and $2\pi$ approximately, because the functions of $\cos2\varphi$
and $\sin2\varphi$ appear in the scattering matrix. One can find
that these results agree with each other very well.

\subsection{the Transmission through a Plane-Parallel Atmosphere}
\begin{figure*}[t]
\begin{center}
  \includegraphics[scale=0.401]{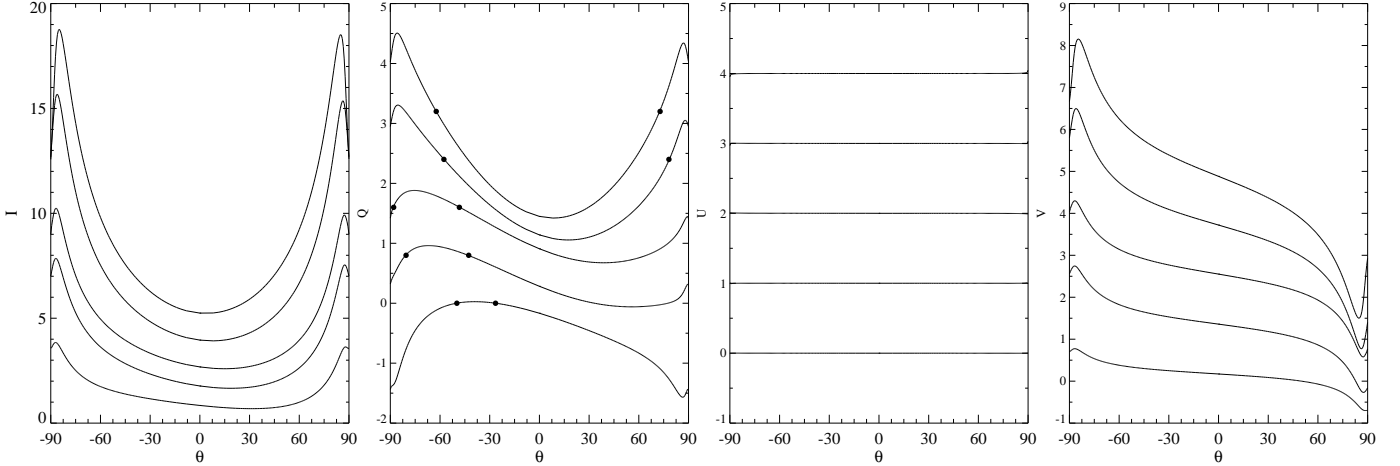}
\caption{The SPs of radiations transmitted through a plane-parallel
atmosphere distribute with respect to the polar angle $\theta$.
Curves from top to bottom in each panel correspond to different
cosines of incident polar angles, $\cos\theta_0=0.1, 0.2, 0.4, 0.5,
0.8$, respectively. In the second panel, the spots indicate the
neutral points where $Q$ changes its signs. Compare to Fig. 27 of
\citet{Chand1960}.} \label{Transmission1}
\end{center}
\end{figure*}

With the preparations given in the last subsection, we can readily
reproduce the SPs distributions of radiations transmitted through a
plane-parallel atmosphere with a finite optical depth $\tau_0$. The
transmission process obeys exactly the same RTEs given by Eqs.
\eqref{diffrefl} and the unpolarized parallel incident radiations
are injected from the bottom boundary ($\tau=\tau_0$). The cosines
of incident polar angles are $\mu_0=0.1, 0.2, 0.4, 0.5, 0.8$
respectively and azimuth angle $\varphi_0=0^\circ$. In order to
compare with the results of Fig. 27 of \citet{Chand1960}, here we
only consider the transmitted radiations lying in the incident
plane, i.e., $\varphi-\varphi_0=0^\circ, 180^\circ$ and
$\theta\in[0^\circ, 90^\circ]$. The result is shown in Fig.
\ref{Transmission1}, where the SPs $U$ and $V$ are also provided.
For the sake of clarity, each curve has been displaced by a distance
along the y-axis. In the second panel, the black spots indicate the
neutral points of $Q$, which correspond to Babinet (Ba), Brewster
(Br) and Arago (A) points respectively. One can see that the U component vanishes, which is
due to the left and right symmetry of the observer that is located in the
incident plane of the beam.
Comparing with Fig. 27 of \citet{Chand1960}, one can find that the
results are consistent.

\subsection{Anisotropic Comptonization in Thermal Plasma}

The inverse Compton scattering is one of most important mechanisms
that can produce high-energy photons, such as X-rays and
$\gamma$-rays, by scattering low-frequency radiations against hot or
relativistic electron gas \citep{younsi2013}. This mechanism has
been extensively studied by analytical method \citep{Haardt1993},
numerical method \citep{Pout1996a} and MC method \citep{Pozd1983,
hua1995, Hua1997, Dolence2009}. In our MC scheme based on Neumann
solution, we have shown that the MCRT amounts to calculate infinite
multiple integrals simultaneously, and the Compton scattering can be
completely regarded as a sampling procedure of a PDF with an
integral form given by Eq. \eqref{eqa5811}. The sampling and
estimation procedures for them have been thoroughly discussed in the
section. \ref{compton_section} and the polarized case are given in
the section. \ref{polarized_Compton_sect}. For scattering sampling,
we always make a Lorentz transformation and get into the rest frame
of the electron, in which the formula of scattering will be
simplified. For estimations, after eliminating the $\delta$-function
in the recording function, we take $C(\nu,
\mathbf{\Omega}\rightarrow\nu',\mathbf{\Omega}_{\text{obs}})$ as a
weight. Usually when $\nu, \mathbf{\Omega}, \mathbf{\Omega}'$ are
specified, the scattered frequency $\nu'$ is uniquely determined,
$C$ is also completely determined. If the scattering of $\nu$ and
$\mathbf{\Omega}$ are independent, we can determine $C$ by
introducing another $\delta$ function
$\delta(\nu'-\nu_{\text{obs}})$ to replace $\nu'$ by
$\nu_{\text{obs}}$. For averaged Compton scattering, the kernel is
given by Eq. \eqref{C_normal}. To determine $C$, we need finish the
integrals in terms of $\gamma, \mu_e$ and $\phi_e$. As
aforementioned, this can be done by the MC method, i.e., we sample
$f_e(P_e)$ directly and take the rest of the integrand as the
weight, which is given by
\begin{eqnarray}
\begin{split}
&\displaystyle w_s=\frac{1}{\sigma_{\text{KN}}(\epsilon)}
\frac{d\sigma_{\text{KN}}}{d\mu_e'\phi_e'},
\end{split}
\end{eqnarray}
Then the recording function becomes
\begin{eqnarray}
   \begin{split}
   \begin{array}{l}
\displaystyle f_{\text{new}} = \displaystyle w_{s}
\exp\left[-\tau(\mathbf{r}, \nu', \mathbf{\Omega}_\text{obs})
\right].
   \end{array}
   \end{split}
\end{eqnarray}

To verify the correctness of this scheme, we first apply it to
calculate the Comptonized spectrum of low-energy radiation by the
thermal plasma in a plane-parallel atmosphere. This Comptonization
process has been calculated by \citet{Haardt1993}. They treated this
transfer process by using a semi-analytical formalism and compared
it with the results obtained from the MC method. This computation is
quite suitable to check our scheme. The low-energy thermal photons
are injected from the bottom surface of the atmosphere. The incident
intensity is angular-dependent and simply given by
\citep{Haardt1993}
\begin{eqnarray}
\begin{split}
&\displaystyle U_0(\mu)=\left\{
   \begin{array}{lrl}
      \displaystyle \frac{6}{3+2b}(\mu+b\mu^2),  \quad &0&<\mu<1; \\
      \displaystyle 0,  \quad  &-1 &<\mu<0, \\
   \end{array}\right.\\
\end{split}
\end{eqnarray}
where $b$ is a constant and set to be $2$ in the practical
calculation. Obviously $U_0(\mu)$ gives rise to a limb-darkening law
for the incident radiations.

\begin{figure*}[t]
\begin{center}
  \includegraphics[scale=0.355]{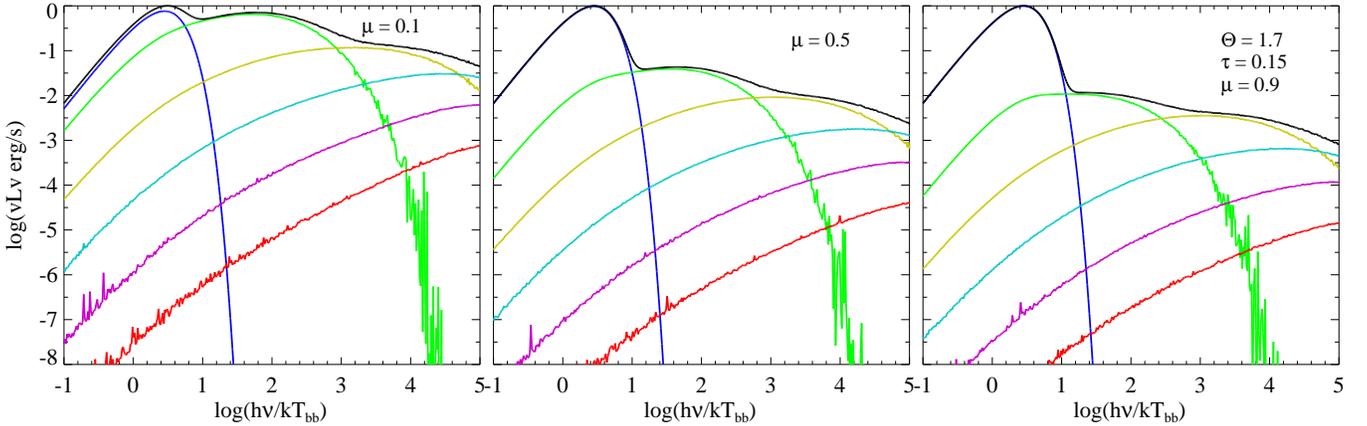}
\caption{The Comptonized spectra of low-energy thermal radiations by
a hot plasma atmosphere. The total spectrum (solid black line) and
each single scattering contributions are shown together. The
dimensionless temperature and optical depth for the hot plasma are
$\Theta=1.7$ and $\tau=0.15$. The cosines of the viewing angles are
$\mu=0.1$, $0.5$ and $0.9$ for panels from left to right,
respectively. The observed photon energies are in unit of
$kT_{\text{bb}}$, which is the temperature of the incident thermal
radiation. Compare to Fig. 2 of \citet{Haardt1993}.} \label{Haardt1}
\end{center}
\end{figure*}

\begin{figure*}[t]
\begin{center}
  \includegraphics[scale=0.355]{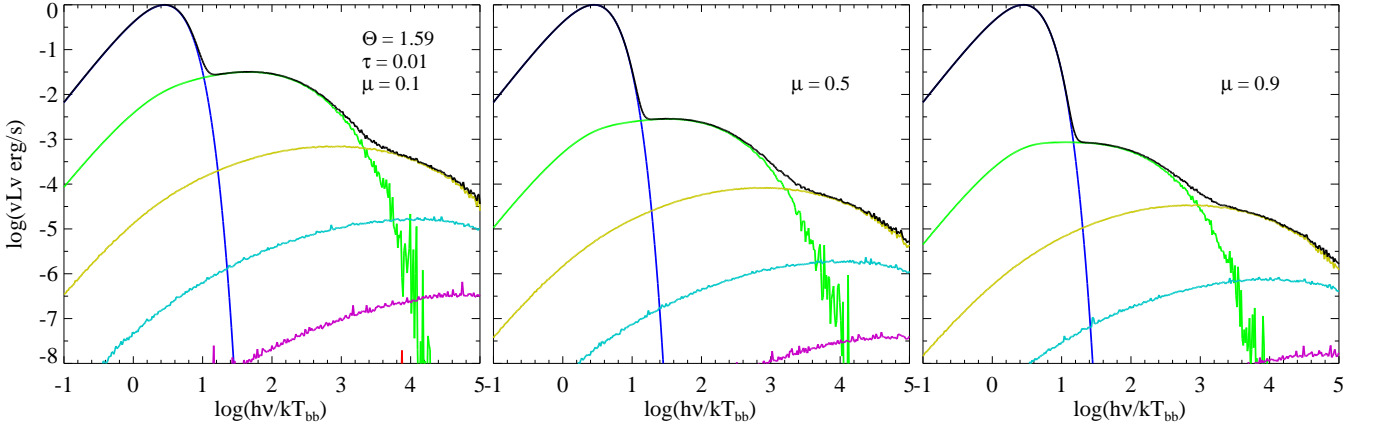}
\caption{Same as in Figure \ref{Haardt1}, but for $\Theta=1.59$ and $\tau = 0.01$.
Compare to Fig. 3 of \citet{Haardt1993}.}
\label{Haardt2}
\end{center}
\end{figure*}

\begin{figure*}[t]
\begin{center}
  \includegraphics[scale=0.355]{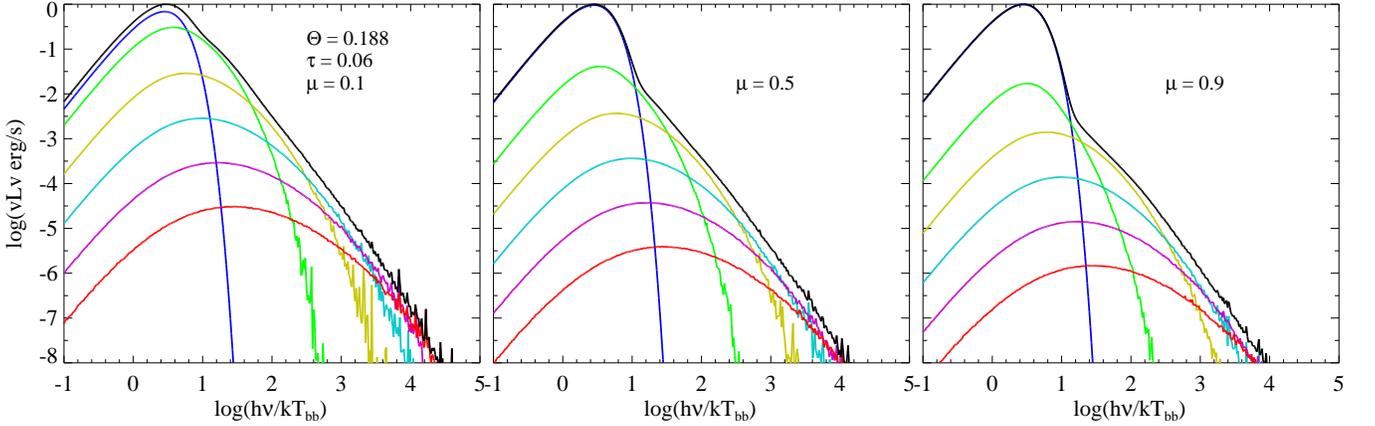}
\caption{Same as in Figure \ref{Haardt1}, but for $\Theta=0.188$ and $\tau = 0.06$.
 Compare to Fig. 4 of \citet{Haardt1993}.}
\label{Haardt3}
\end{center}
\end{figure*}

\begin{figure*}[t]
\begin{center}
  \includegraphics[scale=0.5]{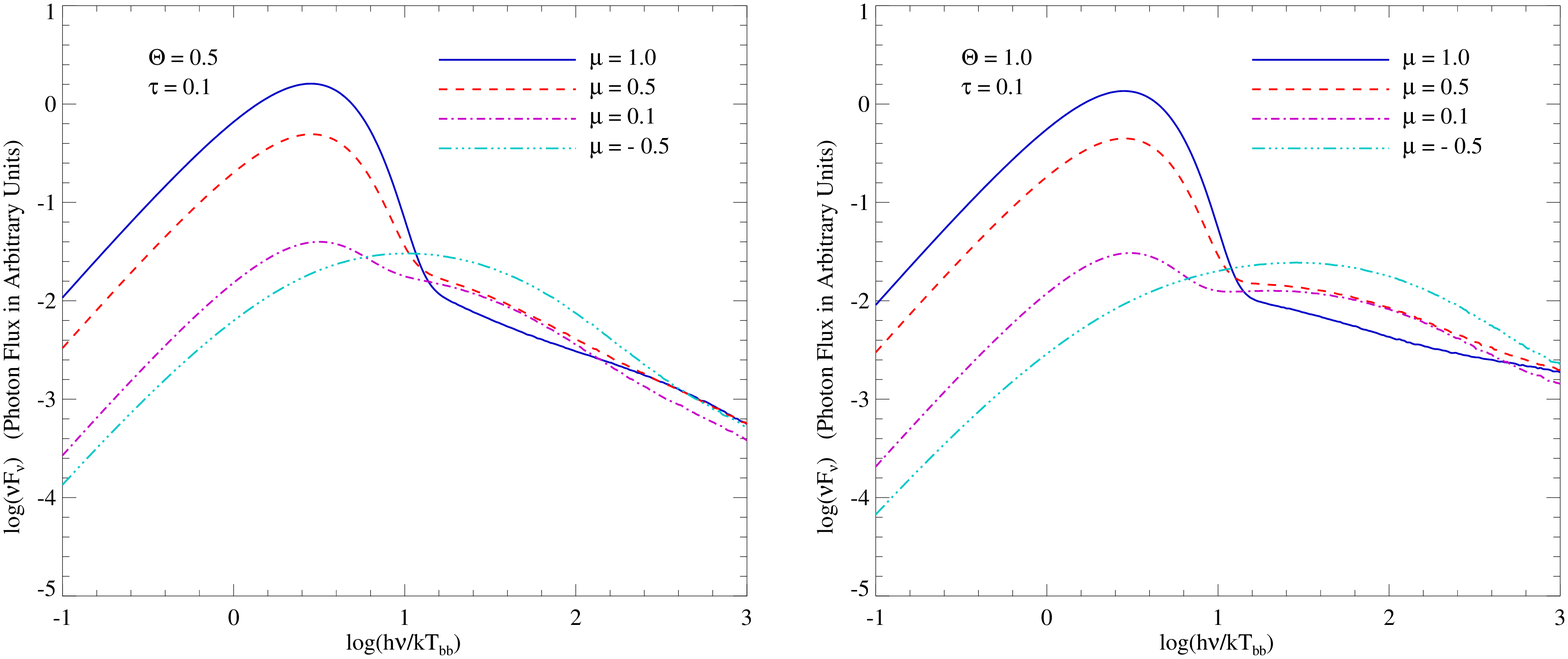}
\caption{The inverse Comptonized spectra of low-energy thermal
radiations by a hot plasma atmosphere for various viewing angles.
The parameters for the hot plasma are $\Theta=0.5$, $\tau = 0.1$ and
$\Theta=1.0$, $\tau = 0.1$ for left and right panels, respectively.
Compare to Figs. 5 and 6 of \citet{Haardt1993}.} \label{Haardt4}
\end{center}
\end{figure*}

Our results is shown in Figs. \ref{Haardt1}-\ref{Haardt4},
corresponding to Figs. 2-6 of \citet{Haardt1993}. Figs.
\ref{Haardt1}-\ref{Haardt3} demonstrate the total Comptonized
spectra together with the spectra of each single scattering. The
spectra vary with different optical depth, plasma temperature and
observational angle. Fig. \ref{Haardt4} shows how the total spectrum
changes with the viewing angles, especially for $\mu<0$,
corresponding to the spectrum reflected back from the atmosphere.
One can see that our results are consistent with that of
\citet{Haardt1993}.

\subsection{The First Order Green's Function For Relativistic Compton Reflection}

\begin{figure}
\begin{center}
   \includegraphics[scale=0.34]{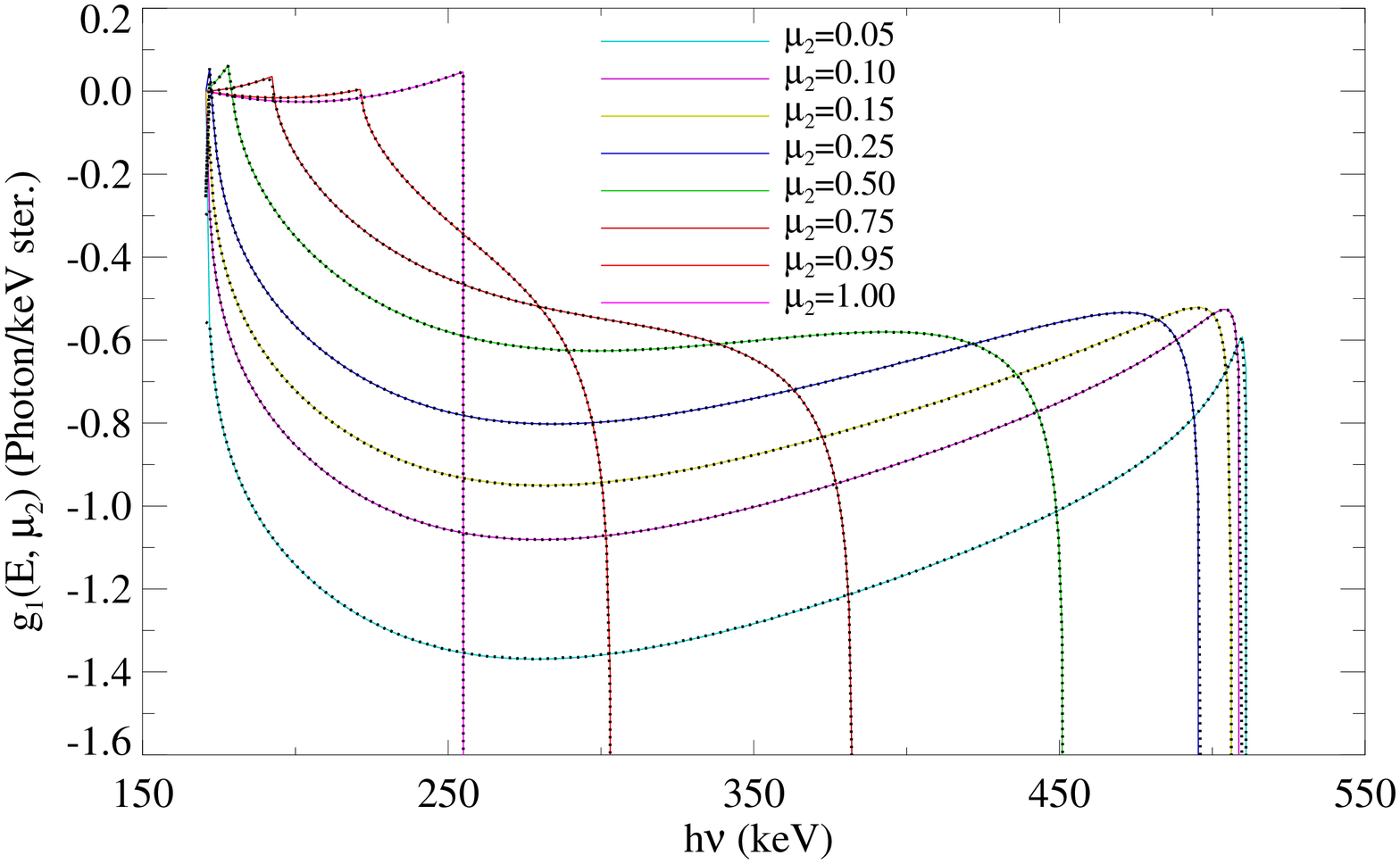}
\caption{The first-order Green's function for monoenergetic incident
photons with 511 keV reflected from a semi-infinite atmosphere of
cold electron. The curves with different colors correspond to
different values of the cosines of viewing angle $\mu_2$. The
solid and dotted lines represent the results calculated by the
analytical formulae of \citet{hua1992} and our MC scheme,
respectively. Compare to Fig. 1 in \citet{hua1992}.} \label{hxm1}
\end{center}
\end{figure}

With the above discussed strategy, we will reproduce the
calculations of \citet{hua1992} for the first order Green's
functions of relativistic Compton reflection by a semi-infinite
plane-parallel atmosphere of cold electrons. \citet{hua1992} have
shown that the first order Green's function for such a system has an
analytical expression and is strongly dependent on the viewing
angles of the atmosphere. Since the cold electrons can be
approximately regarded as static, we do not have to consider any
electron sampling process and Lorentz transformation at all. The
unpolarized and monoenergetic (511 keV) photons are isotropically
injected from all directions to the surface of the atmosphere. The
Green's function can be expressed as an expansion of functions
characterized by the number of scattering times \citep{hua1992}
\begin{eqnarray}
&\displaystyle G(E, \mu_2)=\sum_{n=1}^\infty g_n(E, \mu_2),
\end{eqnarray}
where $E$ is the escaping energy of the photon, $\mu_2$ is the
cosine of the viewing polar angle $\theta_2$. Using the the MC
method, one can easily obtain the numerical result for any function
$g_n$ in this series. In our scheme, we simply have
\begin{eqnarray}
\begin{split}
&g_n(E, \mu_2)=\displaystyle \int\cdots\int S(P_0)K(P_0\rightarrow
P_1)\cdots\\
&\times K(P_{n-1}\rightarrow
P_n)f(P_n)\delta(\mu-\mu_2)dP_0dP_1\cdots dP_n.
\end{split}
\end{eqnarray}
Here we mainly care about the first order $g_1(E, \mu_2)$, i.e.,
\begin{eqnarray}
\begin{split}
&g_1(E, \mu_2)=\displaystyle \iint S(P_0)K(P_0\rightarrow P_1)\\
&\times f(P_1)\delta(\mu-\mu_2)dP_0dP_1,
\end{split}
\end{eqnarray}
which has been analytically integrated by \citet{hua1992}.

For a given viewing angle, there is a cut-off energy \citep{hua1992}
\begin{eqnarray}
&\displaystyle E_{\text{cutoff}}=\frac{511}{2-\sqrt{1-\mu_2^2}}
\text{ keV},
\end{eqnarray}
above which no reflected photon can be seen. The results of $g_1$
for various $\mu_2$ are plotted in Fig. \eqref{hxm1}, where the
analytical results of \citet{hua1992} are also shown for comparison.
One can see that the agreement between these results is quite well.

\subsection{Comptonization of Polarized Radiation from a Reflecting Disk with a Hot Corona}
\label{RT_reflection} In the last two applications, we mainly focus
on the unpolarized Compton scattering dominated RT. In this section
we proceed to discuss the applications with polarization and
calculate the analogous spectra emerging from the hot corona of an
optically thick disk. The effect that radiations can be reflected
from the disk when they irradiate the disk surface is also
considered. The reflected radiation $\mathbf{I}_r$ in the direction
of $\mathbf{\Omega}$ equals the sum of contributions coming from all
other incident directions $\mathbf{\Omega}'$, i.e.,
\citep{pout1996b},
\begin{eqnarray}
\label{Ir_reflection}
\begin{split}
 \mathbf{I}_r(\nu, \mathbf{\Omega})&=\int \mathbf{G}(\nu, \mathbf{\Omega}; \nu',\mathbf{\Omega}')
 \mathbf{I}(\nu', \mathbf{\Omega}')d\nu'd\mathbf{\Omega}'.
\end{split}
\end{eqnarray}
where $\mathbf{G}(\mathbf{\Omega},\nu; \mathbf{\Omega}',\nu')$ is
the Green function obtained by solving the RT process in the disk
\citep{mag1995}. The matrix function $\mathbf{QS}/4\mu$ in Eq.
\eqref{chandraIlru} is actually such a Green function. We will use
it to get the reflected radiations. In Appendix
\ref{app_reflection}, we discussed how to incorporate the
reflection term $\mathbf{I}_r$ into the integral equation and sample
the modified transfer kernel.

As aforementioned the transformation between the incident and
Compton scattered Stokes parameters are given by Eqs.
\eqref{eq:sp_new_old2} and \eqref{eq:sp_new_old3}. Considering the
distribution of electron gas and the unpolarized situation, this
transformation formula should be modified as
\begin{eqnarray}
\label{eqa58}
\begin{split}
&\displaystyle \mathbf{\Psi}'=\frac{1}{4\pi}\int_1^\infty d\gamma N_e(\gamma)\int_{-1}^1d\mu_e(1-\mu_e v)\int_0^{2\pi}d\varphi_e\\
&\times \frac{r^2_e}{\gamma^2(1-\mu_e
v)^2}\left(\frac{\nu'}{\nu}\right)^2
\mathbf{FM}(\varphi)\mathbf{\Psi}.
\end{split}
\end{eqnarray}
Similarly the integrals in the above equation can be evaluated by
sampling a set of $\gamma, \mu_e, \varphi_e$. With them and after
the multiplication of matrices $\mathbf{FM}(\varphi)$ and
$\mathbf{\Psi}$, we have
\begin{eqnarray}
\begin{split}
&\displaystyle  \Psi'^{(X)}=A_0 f_X(\mu', \varphi'; \mu, \varphi)\Psi^{(I)},\\
&\displaystyle  \Psi'^{(V)}=A_0 f_V(\mu', \varphi'; \mu,
\varphi)\Psi^{(V)},
\end{split}
\end{eqnarray}
where $A_0=(r^2_e/2)
(\epsilon'/\epsilon)^2/\sigma_{\text{KN}}(\epsilon)$ and $X=(I, Q,
U)$. Transforming $f_X$ and $f_V$ into the rest frame of the
electron, we have
\begin{eqnarray}
\left\{\begin{array}{ll}
&\displaystyle \widetilde{f_I} =F_0+F_3(q\cos2\Phi'+u\sin2\Phi'),\\
&\displaystyle \widetilde{f_Q} =F_3+F_{33}(q\cos2\Phi'+u\sin2\Phi'),\\
&\displaystyle \widetilde{f_U} =F_{11}(-q\sin2\Phi'+u\cos2\Phi'),\\
&\displaystyle \widetilde{f_V} =F_{22},
\end{array}\right.
\end{eqnarray}
where $q=\Psi^{(Q)} / \Psi^{(I)}, u=\Psi^{(U)}/ \Psi^{(I)}$, and
$(\Psi', \Phi')$ are the scattered polar and azimuth angles in the
rest frame of the electron. We also use $A_0 \widetilde{f_I}$
to construct a PDF for $(\Psi', \Phi')$ and sample it to get the
scattered direction. After transforming it back into the static
frame, we get $(\mu', \varphi')$ and the scattered $\mathbf{\Psi}'$
as
\begin{eqnarray}
\left\{\begin{array}{ll} &\displaystyle \Psi'^{(X)}=\frac{f_X(\mu',
\varphi'; \mu,\varphi)}{f_I(\mu', \varphi'; \mu, \varphi)}\Psi^{(I)},\\
&\displaystyle  \Psi'^{(V)}=\frac{f_V(\mu', \varphi'; \mu,
\varphi)}{f_I(\mu', \varphi'; \mu, \varphi)}\Psi^{(V)}.
\end{array}\right.
\end{eqnarray}
The estimation functions for the Stokes parameters are given by
\begin{eqnarray}
\left\{\begin{array}{ll} &\displaystyle
F_{X}=A_0\widetilde{f_X}(\mathbf{\Omega}_{\text{obs}})\Psi^{(I)}_m\exp\left[-\tau(\nu,
\mathbf{\Omega}_{\text{obs}})\right],\\
&\displaystyle
F_{V}=A_0\widetilde{f_V}(\mathbf{\Omega}_{\text{obs}})\Psi^{(V)}_m\exp\left[-\tau(\nu,
\mathbf{\Omega}_{\text{obs}})\right].
\end{array}\right.
\end{eqnarray}

\begin{figure*}[t]
\begin{center}
\includegraphics[scale=0.65]{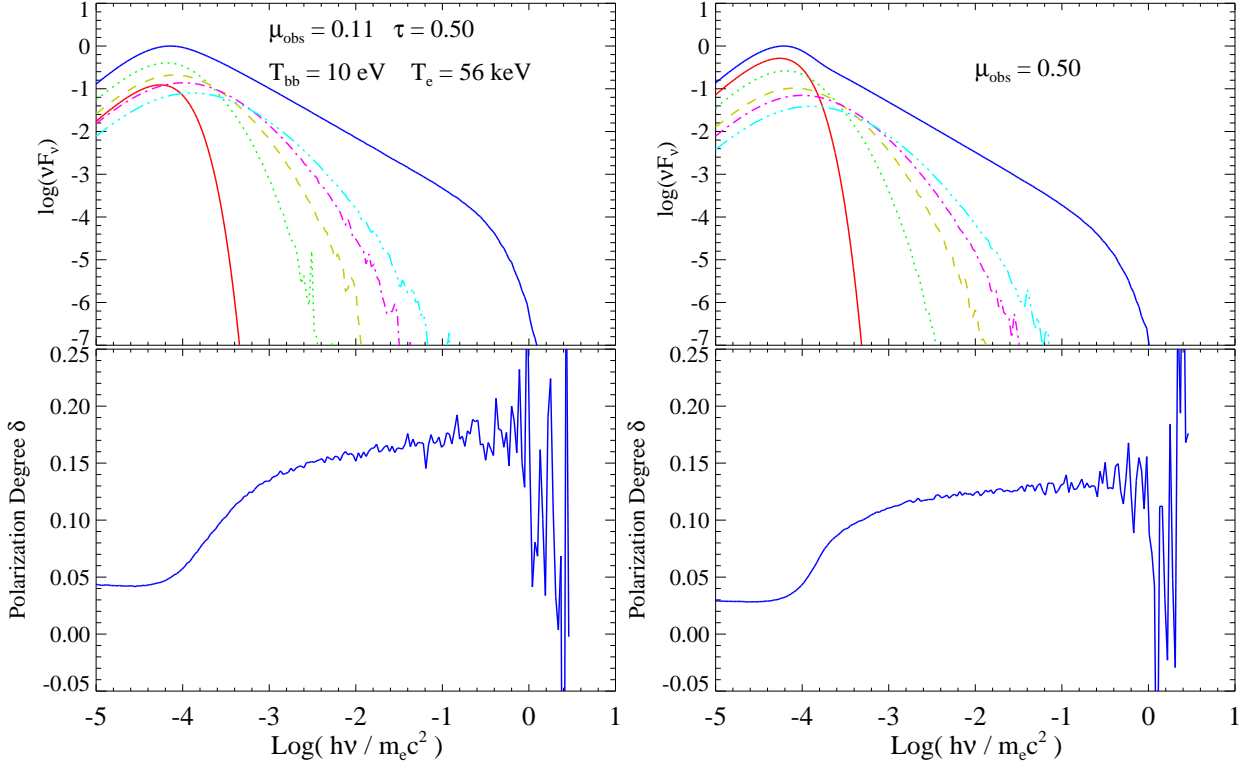}
\caption{The spectra (top panels) and polarization degrees (bottom
panels) of radiations emerging from a plane-parallel corona above an
accretion disk. The solid blue curves correspond to the total
spectra and the solid red, dotted, dashed, dot-dashed,
triple-dot-dashed curves correspond to scattering number of 0, 1, 2,
3, 4, respectively. Compare to Figure 5 of \citet{Pout1996a}.}
\label{reflections1}
\end{center}
\end{figure*}

\begin{figure*}[t]
\begin{center}
\includegraphics[scale=0.65]{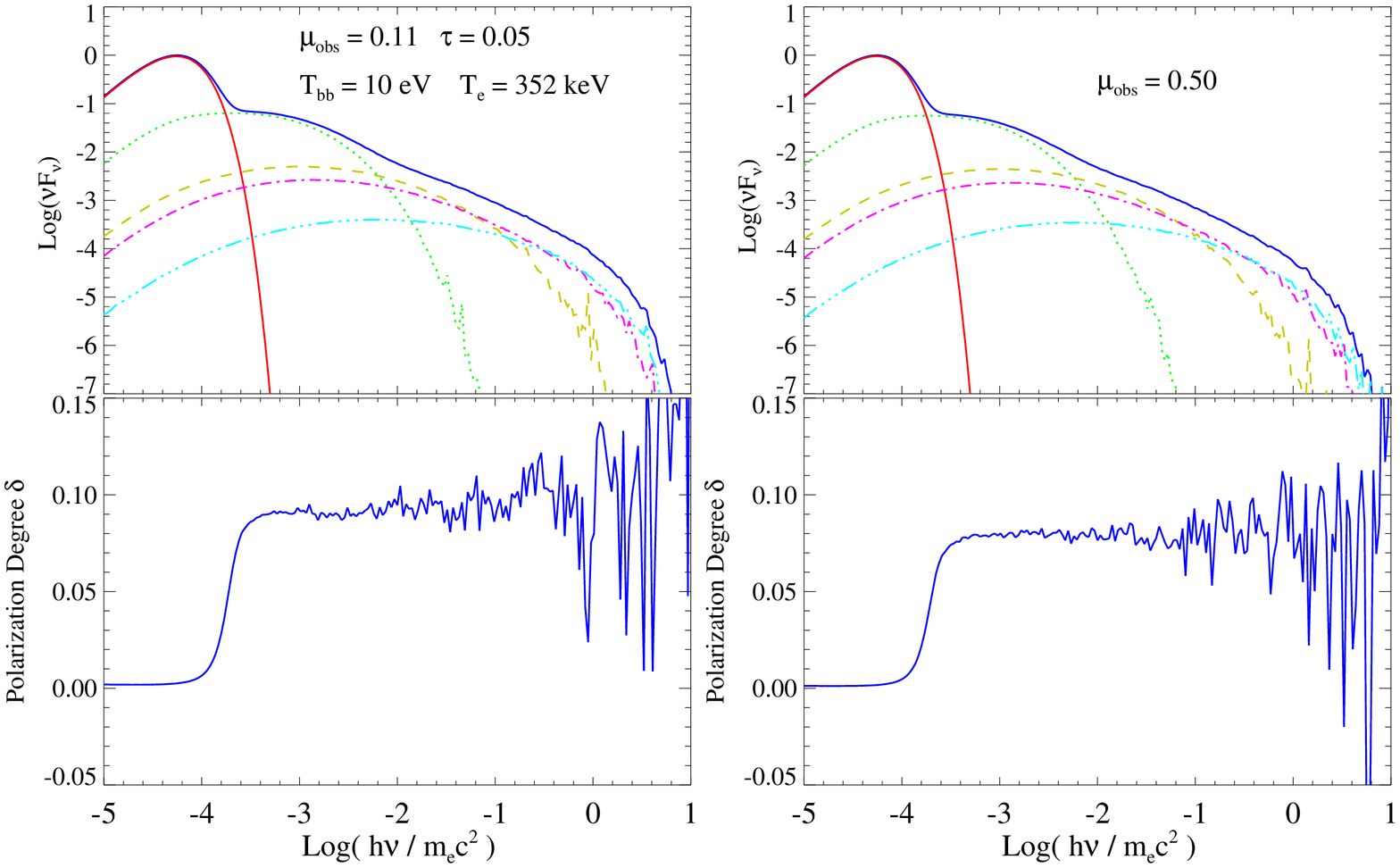}
\caption{The same as in Fig. \ref{reflections1}, but for $T_e=353$
keV and $\tau=0.05$. Compare to Figure 6 of \citet{Pout1996a}.}
\label{reflections2}
\end{center}
\end{figure*}

With these preparations, we can now proceed to complete the
calculations mentioned at the beginning of this section. The source
thermal radiation has a temperature $kT_{bb}=10$ eV and is located
isotropically and uniformly on the disk surface. The radiations
scattered back into the disk will re-emerge with an altered SPs
determined by Eq. \eqref{chandraIlru}. The results are plotted in
Figs. \ref{reflections1} and \ref{reflections2}. The corona
temperature and Thomson optical depth are $T_e$ and
$\tau=h\sigma_Tn_e$, where $h$ is the height of corona. The values
of these parameters correspond to those of Figs. 5 and 6 of
\citet{Pout1996a} (also refer to Figs 12 and 13 of
\citet{Schnittman2013}). One can see that our results are similar
but different with them, especially for the curves corresponding to
each single scattering. In our scheme, since each scattering
(including the reflection) can make contributions to the spectra, we
do not reset the variable of scattering number $N_{\text{scat}}$ to
zero after the photons are reflected from the disk. It is contrast
to the treatment of \citet{Schnittman2013}. The reflection processes
are described by Eq. \eqref{chandraIlru}, where the photoelectric
absorption and energy losses due to Compton recoil are not
considered. Therefore the reflection processes will not change the
energies of photons. We believe that these different treatments give
rise to the discrepancies in the spectra profiles and the
polarization degrees.

One can see that the high energy parts of the spectra, almost
greater than $\sim 100$ keV, are quite noise and oscillate heavily,
because the photons received by the observer at that energy band
have experienced a large averaged scattering number. The weights of
these photons have been attenuated heavily, hence the polarization
signals comprised by these weights are difficult to be resolved
\citep{Schnittman2013}.

\subsection{The Scattering of Polarized Photon Beam by Relativistic Electrons}

In this section we will consider the polarization states of a
polarized photon beam scattered off only once by a relativistic
electron gas that can have an arbitrary distribution function
$N_e(\gamma)$. We denote the energies, polarization and wave vectors
for the incident and scattered photons as $\epsilon,
\pmb{\varepsilon}, \mathbf{k}$ and $\epsilon', \pmb{\varepsilon}',
\mathbf{k}'$ respectively, and the scattering angle between
$\mathbf{k}$ and $\mathbf{k}'$ is $\theta'$. Then the scattered
intensity can be expressed semi-analytically as follows
\citep{bono1970, bono1973a, bono1973b}
\begin{eqnarray}
\label{bonome1}
   \begin{split}
        J_{\pmb{\varepsilon}'}&=\pi \left(\frac{e^2}{4\pi}\right)\frac{\epsilon'}{\epsilon}E_{\text{min}}
       \left\{\left|\pmb{\varepsilon}\cdot\pmb{\varepsilon}'^* + \frac{(\mathbf{k}\cdot\pmb{\varepsilon}'^*)
        ( \mathbf{k}'\cdot\pmb{\varepsilon} ) }{ 1-\cos\theta'}
        \right|^2\Sigma_1\right.  \\
        &\left.+\left|\pmb{\varepsilon}\cdot\pmb{\varepsilon}' + \frac{(\mathbf{k}\cdot\pmb{\varepsilon}') (\mathbf{k}'\cdot\pmb{\varepsilon} ) }
        { 1-\cos\theta'} \right|^2\Sigma_2 + \Sigma_2  \right\},
   \end{split}
\end{eqnarray}
where
\begin{eqnarray}
   \begin{split}
       &\displaystyle \Sigma_1 = \int_0^1 m(\gamma)\left(x^2 - \frac{1}{x^2} + 2 \right)dx,  \\
       &\displaystyle \Sigma_2 = \int_0^1 m(\gamma) \frac{(1 - x^2)^2}{x^2}dx,\\
       &\displaystyle x = \frac{E_{\text{min}}}{\gamma}, \quad \quad E_{\text{min}} = \sqrt{ \frac{1}{2} \frac{\epsilon'}{\epsilon} \frac{1}{(1-\cos\theta')}},
   \end{split}
\end{eqnarray}
and the quantity $m(\gamma)$ is related to $N_e(\gamma)$, and
\begin{eqnarray}
   \begin{split}
      m(\gamma)=\frac{N_e(\gamma)}{\gamma^2}.
   \end{split}
\end{eqnarray}
For simplicity we denote $\displaystyle J_0=\pi
\left(\frac{e^2}{4\pi}\right)\frac{\epsilon'}{\epsilon}E_{\text{min}}$.
Using the base vectors $\Ket{\mathbf{e}_x}, \Ket{\mathbf{e}_y}$, the
polarization vectors $\pmb{\varepsilon}$ and $\pmb{\varepsilon}'$
can be decomposed as
\begin{eqnarray}
   \begin{split}
        \pmb{\varepsilon} = \alpha \Ket{\mathbf{e}_x} + \beta \Ket{\mathbf{e}_y}, \quad \pmb{\varepsilon}' = \alpha' \Ket{\mathbf{e}'_x} +
         \beta' \Ket{\mathbf{e}'_y}.
   \end{split}
\end{eqnarray}
Then with $\alpha, \beta$ and $\alpha', \beta'$, Eq. \eqref{bonome1}
can be recast as \citep{bono1970}
\begin{eqnarray}
\label{bonome2}
   \begin{split}
     J_{\pmb{\varepsilon}'}&=J_0 \left\{c_1\Sigma_1 + c_2\Sigma_2 + \Sigma_2  \right\}.
   \end{split}
\end{eqnarray}
where $c_1=\left|\alpha \alpha'^*-\beta\beta'^* \right|^2$ and $c_2
= \left|  \alpha \alpha'-\beta\beta' \right|^2$. Without loss of
generality, one can choose $\alpha=x, \,\beta=ye^{i\theta}$
(obviously $x^2+y^2=1$), the coefficients $c_1$ and $c_2$
can be written as
\begin{eqnarray}
\label{bonome4}
   \begin{split}
        c_1 &= x^2x'^2-2xx'yy'\cos(\theta-\theta')+y^2y'^2,\\
        c_2 &= x^2x'^2-2xx'yy'\cos(\theta+\theta')+y^2y'^2.
   \end{split}
\end{eqnarray}
Usually we describe the polarization state by SPs: $I, Q, U, V$ or
the normalized SPs: $\xi_1=U/I,\, \xi_2=V/I,\, \xi_3=Q/I$. For
simplicity we define a vector $\mathbf{S}=(1, \xi_3, \xi_1, \xi_2)$.
By using Eq. \eqref{xi2xy}, we can immediately obtain $\xi_i$ from
$x, y$ and $\theta$:
\begin{eqnarray}
 \label{xy2xi123}
   \begin{split}
        & \xi_1 = 2xy\cos\theta, \quad \xi_2 = 2xy\sin\theta,\quad
\xi_3=x^2-y^2,
   \end{split}
\end{eqnarray}
and vice versa
\begin{eqnarray}
   \begin{split}
        &x = \sqrt{ \frac{1+\xi_3}{2}}, \quad y = \sqrt{ \frac{1-\xi_3}{2}},\\
        & \cos\theta = \frac{\xi_1}{\sqrt{ \xi_1^2+\xi_2^2} }, \quad \sin\theta = \frac{\xi_2}{\sqrt{ \xi_1^2+\xi_2^2} }.
   \end{split}
\end{eqnarray}
For any two orthogonal polarization states of
$\pmb{\varepsilon}=(\alpha, \beta)$ and
$\pmb{\varepsilon}_{\perp}=(\alpha_{\perp}, \beta_{\perp})$, since
$\pmb{\varepsilon}\cdot\pmb{\varepsilon}_{\perp}^*=
\alpha\alpha_{\perp}^*+\beta\beta_{\perp}^*=0$, without loss of
generality we can obtain $\pmb{\varepsilon}_{\perp}=(-\beta^*,
\alpha^*)$. Using Eq. \eqref{xi2xy}, one can immediately obtain that
the normalized SPs for $\pmb{\varepsilon}_{\perp}$ is
$\mathbf{S}_{\pmb{\varepsilon}_{\perp}}=(1, -\xi_3, -\xi_1, -\xi_2)$
if the corresponding ones for $\pmb{\varepsilon}$ is
$\mathbf{S}_{\pmb{\varepsilon}}=(1, \xi_3, \xi_1, \xi_2)$. The total
intensity for the scattered radiation is the sum of
$J_{\pmb{\varepsilon}'}$ over any two orthonormal polarization
states, and one can easily obtain \citep{bono1970}
\begin{eqnarray}
   \label{J_i11}
   \begin{split}
       J_{I'} =J_{\pmb{\varepsilon}'}+J_{\pmb{\varepsilon}'_\perp}= J_0 \left\{ \Sigma_1 + 3 \Sigma_2 \right\}.
   \end{split}
\end{eqnarray}
Due to the additive property of SPs \citep{Chand1960}, the
normalized SPs for $J_{I'}$ is
$\mathbf{S}_{I'}=\mathbf{S}_{\pmb{\varepsilon}'}+\mathbf{S}_{\pmb{\varepsilon}'_{\perp}}=(1,
0, 0, 0)$.

Similarly, the scattered intensity for the polarization state of
$Q'$ is given by
\begin{eqnarray}
   \begin{split}
       J_{Q'} & = J_{S'_+} - J_{S'_-}\\
              & = J_0 \left[ (c_1^+ - c_1^-)\Sigma_1 + (c_2^+ - c_2^-) \Sigma_2 \right],
   \end{split}
\end{eqnarray}
where $\mathbf{S}'_\pm=(1, \pm1, 0, 0)$. The corresponding
coefficients $c_1^{\pm}$ and $c_2^\pm$ are obtained from Eq.
\eqref{bonome4}. Similarly, the scattered intensities for
polarization states $U'$ and $V'$ are given by
\begin{eqnarray}
   \begin{split}
       J_{U'} & = J_{S'_+} - J_{S'_-}\\
        &=J_0 \left[ (c_1^+ - c_1^-)\Sigma_1 + (c_2^+ - c_2^-) \Sigma_2 \right],\\
       J_{V'}  & = J_{S'_+} - J_{S'_-}\\
        &=J_0 \left[ (c_1^+ - c_1^-)\Sigma_1 + (c_2^+ - c_2^-) \Sigma_2
        \right],
   \end{split}
\end{eqnarray}
where the coefficients $c_1^{\pm}$ and $c_2^\pm$ are obtained from
the normalized SPs $\mathbf{S}'_\pm=(1, 0, \pm1, 0)$ for $U'$ and
$\mathbf{S}'_\pm=(1, 0, 0, \pm1)$ for $V'$, respectively. Since any
polarization state is completely determined by SPs, therefore once
$J_{I'}, J_{Q'}, J_{U'}, J_{V'}$ are obtained, the scattered
polarization state is also determined.

\begin{figure*}
\begin{center}
 \includegraphics[scale=0.37]{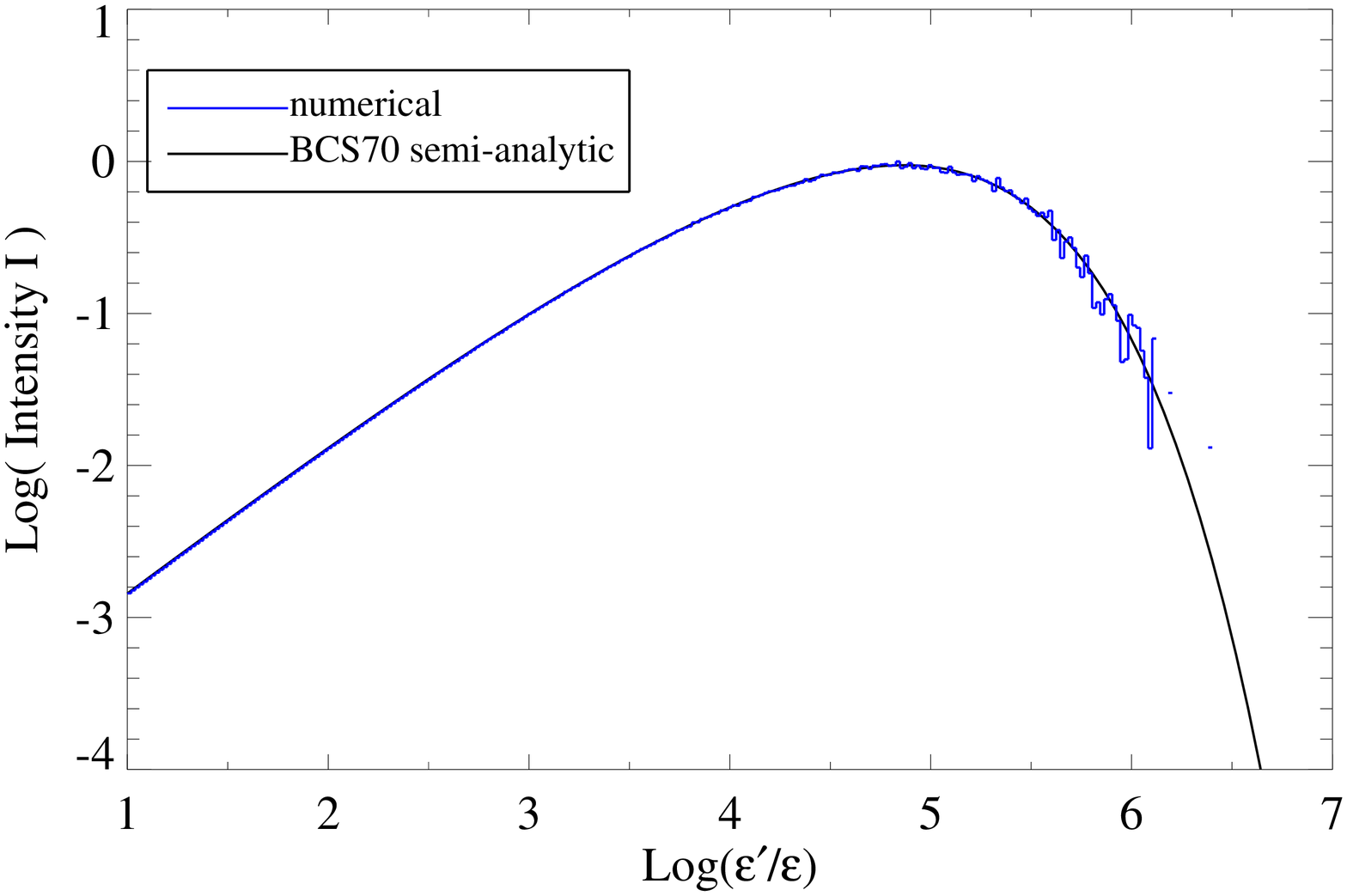}
 \includegraphics[scale=0.37]{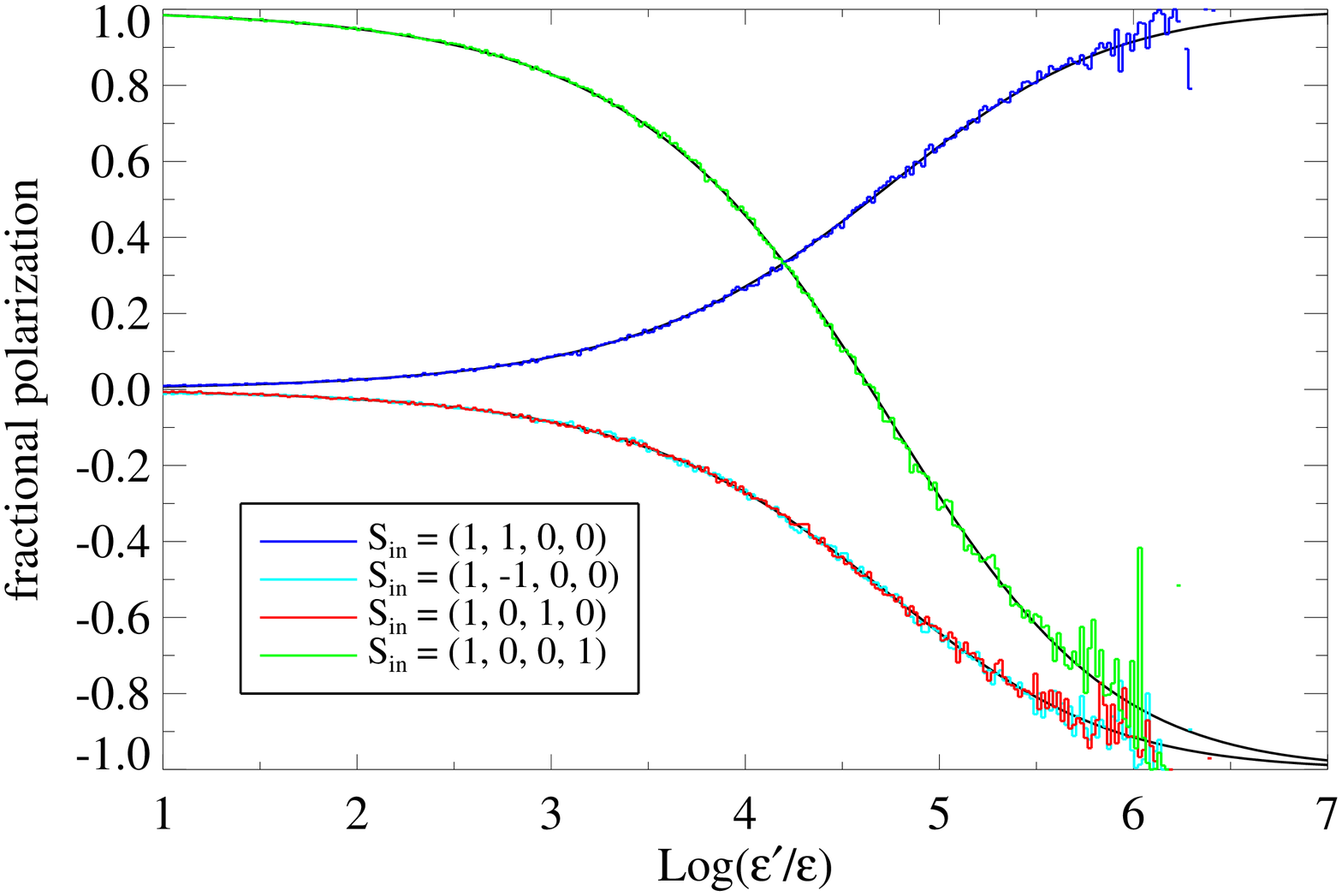}
\caption{The intensity (left) and normalized SPs (right) of photons
scattered off a group of thermal relativistic electron gas with
various incident polarizations $\mathbf{S}_\text{in}$. The
dimensionless temperature of the electron gas is $\Theta_e=100$. The
incident energy is monoenergetic and
$\epsilon=h\omega/m_ec^2=2.5\times10^{-11}$. The solid lines and
histograms represent the results obtained by the semi-analytical
formulae of \citet{bono1970} and our MC scheme, respectively.
Compare to Fig. 2 of \citet{mosc2020}.} \label{mosc1}
\end{center}
\end{figure*}

With these preparations, we can readily obtain the scattered
polarization states semi-analytically, which can be used to verify
our MC scheme. We first reproduce the results in Fig. 2 of
\citet{mosc2020}, where four polarized incident beams are scattered
off by a group of hot electron gas distributed according to the
relativistic Maxwell function (see Eq. \eqref{maxwelldis}). The
dimensionless temperature of the gas is $\Theta_e=kT_e/m_ec^2=100$.
The incident beams with fixed energy $\epsilon=2.5\times10^{-11}$
propagate along the positive $z$-axis and the viewing direction is
$(\theta', \varphi')=(85^\circ, 0)$. The normalized SPs of the four
incident beams are $\mathbf{S}_{\text{in},1}=(1, 1, 0, 0)$,
$\mathbf{S}_{\text{in},2}=(1, -1, 0, 0)$,
$\mathbf{S}_{\text{in},3}=(1, 0, 1, 0)$, and
$\mathbf{S}_{\text{in},4}=(1, 0, 0, 1)$, respectively. The scattered
intensities for all beams are the same and independent on the
polarizations of the incident beams (see Eq. \eqref{J_i11}). The
intensity (left panel) and fractional polarizations (right panel) of
the scattered photons distributed with respect to the photon energy
are shown in Fig. \ref{mosc1}, where only the nonvanishing
normalized SPs are plotted. One can see that the numerical results
produced by our MC scheme agree with the semi-analytical results of
\citet{bono1970} very well, especially in the low energy band. The
noise of the high energy part is relatively large due to the sparse
productions of energetic photons comparing to that of the low energy
ones.

\begin{figure*}
\begin{center}
 \includegraphics[scale=0.37]{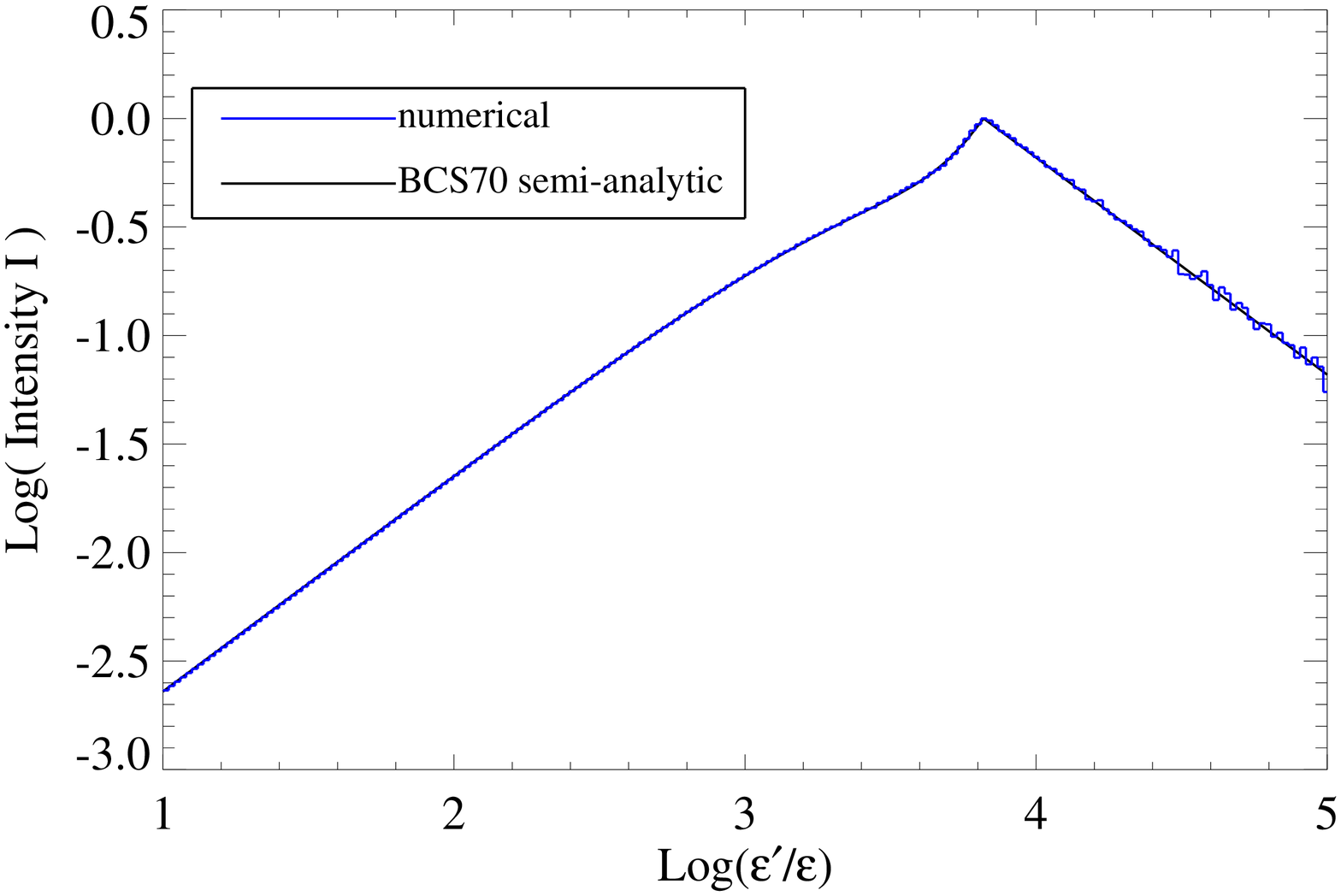}
 \includegraphics[scale=0.37]{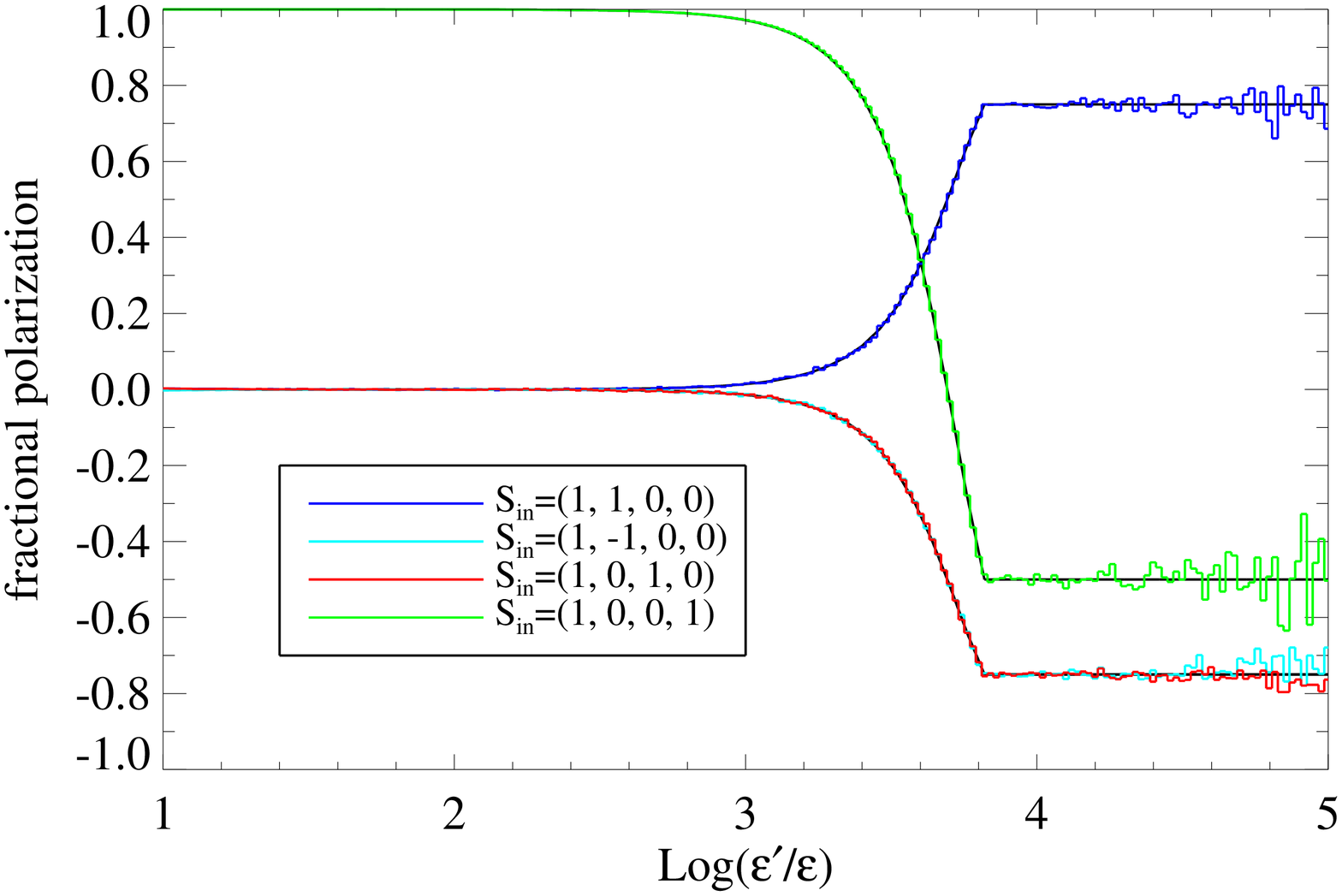}
\caption{Same as in Fig. \ref{mosc1}, but the electron gas is
distributed in a power law profile given be Eq. \eqref{Ne_power},
where $\gamma_1=60.0$, $\alpha=3$ and
$\epsilon=2.5\times10^{-11}$.} \label{mosc2}
\end{center}
\end{figure*}

\begin{figure*}[t]
\begin{center}
 \includegraphics[scale=0.37]{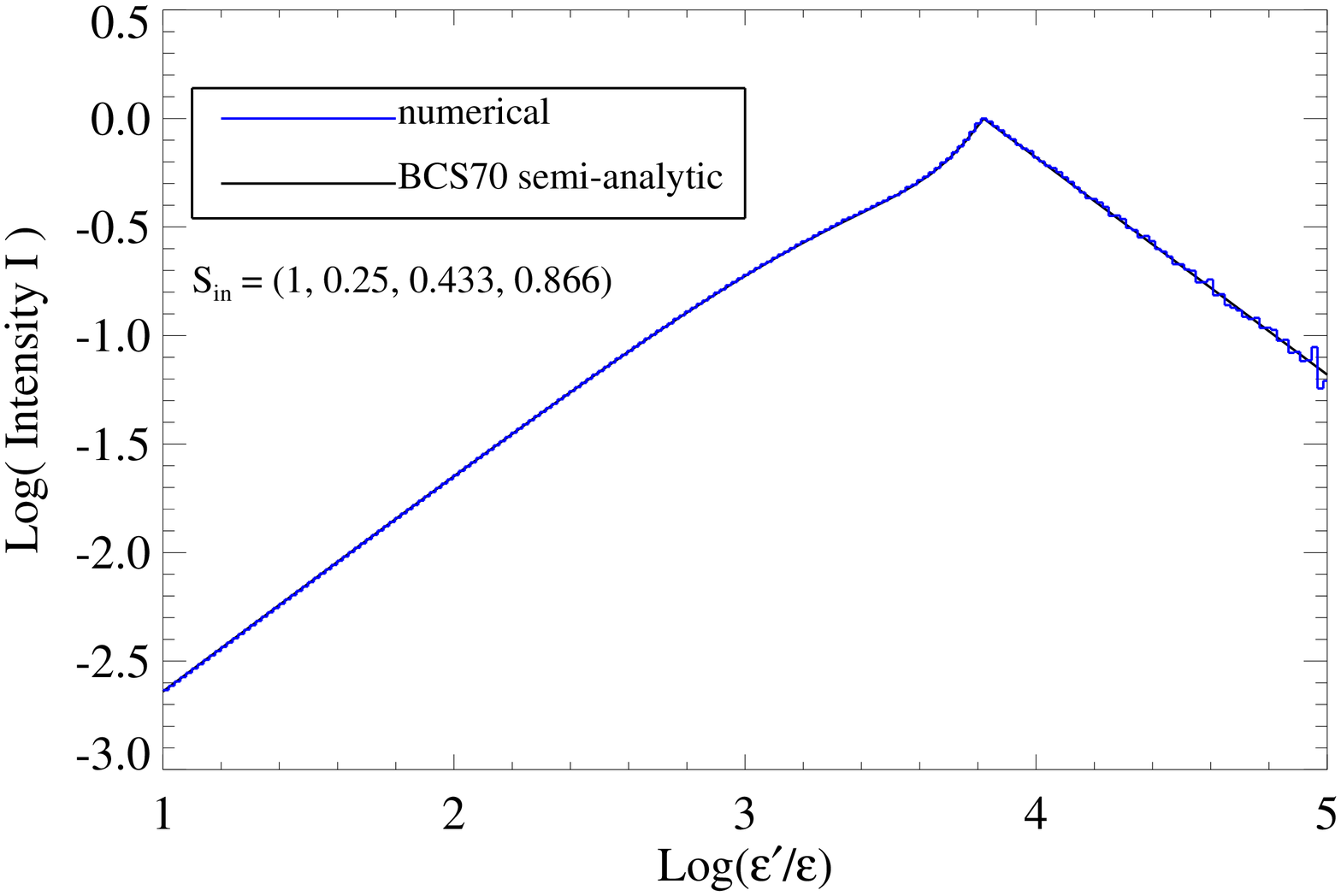}
 \includegraphics[scale=0.37]{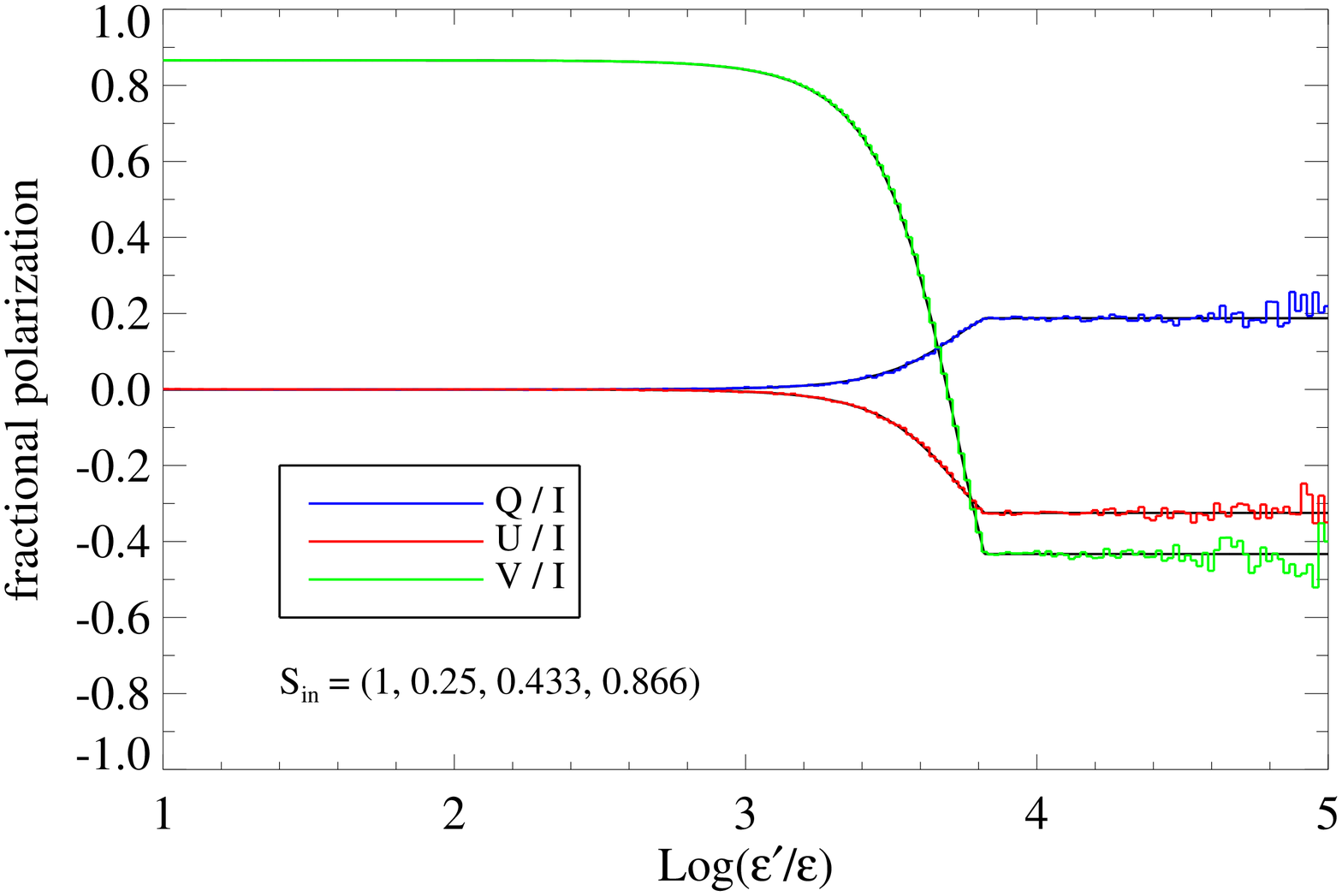}
\caption{Same as in Fig. \ref{mosc2}, but for
$S_{\text{in}}=(1,0.25, 0.433, 0.866)$.} \label{mosc3}
\end{center}
\end{figure*}

Next we consider the case where the electron gas is distributed in a
power law profile, i.e.,
\begin{eqnarray}
\label{Ne_power}
   N_e(\gamma) = K\gamma^{-\alpha}, \quad \gamma \ge \gamma_1,
\end{eqnarray}
where $K= (\alpha-1)/\gamma_1^{1-\alpha}$ is the normalization
factor and $\alpha$ is the power law index. Then
$m(\gamma)=K\gamma^{-(\alpha+2)}$, one can immediately
obtain the expressions for quantities $\Sigma_1$ and $\Sigma_2$ as
(\citet{bono1970})
\begin{eqnarray}
   \label{bcseq1}
   \begin{split}
   \Sigma_1=\frac{K}{E^{\alpha+1}_{\text{min}}}x_1^{\alpha+1}\left[\frac{x_1^4}{\alpha+5}-\frac{1}{\alpha+1}+\frac{2x_1^2}{\alpha+3}\right],\\
   \Sigma_2=\frac{K}{E^{\alpha+1}_{\text{min}}}x_1^{\alpha+1}\left[\frac{x_1^4}{\alpha+5}+\frac{1}{\alpha+1}-\frac{2x_1^2}{\alpha+3}\right],
   \end{split}
\end{eqnarray}
where (notice that $E_{\text{min}}$ is dependent on the energies of
scattered photons)
\begin{eqnarray}
    x_1=\left\{
    \begin{array}{lr}
        \displaystyle 1,   &\quad \quad \text{    if   }  E_{\text{min}}\ge\gamma_1,\\
        \displaystyle \frac{E_{\text{min}}}{\gamma_1},   &\quad\quad\text{    if   } E_{\text{min}}<\gamma_1.
    \end{array}\right.
\end{eqnarray}
Our results in comparison with the analytic ones of \citet{bono1970}
are shown in Figs. \ref{mosc2} and \ref{mosc3}. In these
calculations we set $\gamma_1=60.0$, the power law index $\alpha=3$
and the incident monoenergy $\epsilon=2.5\times10^{-11}$. Using the
inverse CDF method, we can easily sample $\gamma$ from
$N_e(\gamma)$, i.e., $\gamma_\xi=\gamma_1(1-\xi)^{1/(1-\alpha)}$.
From Eq. \eqref{bcseq1} one can see that there exist a turning point
$\epsilon'_{\text{tr}}$ for the scattered photon energy $\epsilon'$,
above which all of the fractional polarizations: $\xi_3, \xi_2,
\xi_1$ are constants (but notice that the SPs $I, Q, U, V$ are not).
From the condition $E_{\text{min}}=\gamma_1$ one can easily obtain
that $\epsilon'_{\text{tr}}/\epsilon=2(1-\cos\theta')\gamma_1^2$. In
Figs. \ref{mosc2} and \ref{mosc3}, where $\theta'=85^\circ$ and
$\gamma_1=60.0$, we have
$\log_{10}(\epsilon'_{\text{tr}}/\epsilon)\approx3.818$. The
normalized SPs of the four incident beams in Fig. \ref{mosc2} are
the same with those in Fig. \ref{mosc1}. While in Fig. \ref{mosc3},
the incident SPs are $\mathbf{S}_{\text{in}}=(1, 0.25, 0.433,
0.866)$ and all of the scattered SPs are nonvanishing.

\subsection{Polarized RT with Faraday Rotation and Conversion}
\label{testofRT_FRC} 
\begin{figure*}
\begin{center}
 \includegraphics[scale=0.45]{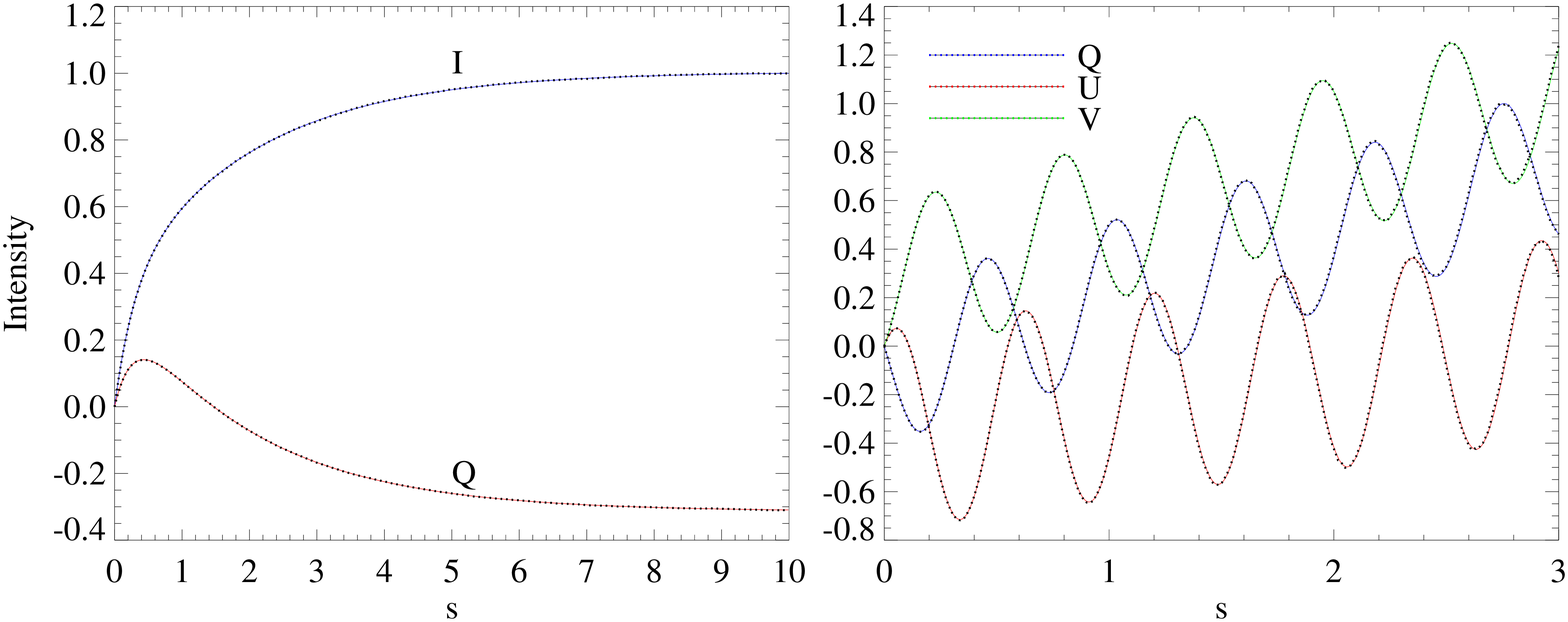}
\caption{The tests of MC scheme for RT with Faraday rotation and
conversion incorporated against the analytical solutions. Left
panel: only the SPs $I$ and $Q$ are involved, and the parameters
are: $\alpha_I=2.0, \alpha_Q=1.5, j_I=2.1, j_Q=1.2$ and
$\alpha=3.0$. Right panel: the Figure illustrates the evolution of
SPs $Q$, $U$ and $V$. The parameters are: $\alpha_{IQUV}=0.0,
\rho_Q=7.5, \rho_U=3.4, \rho_V=7.2, j_Q=-10.717, j_U=9.033,
\rho_V=9.0587$ and $\alpha=200.0$. In the two panels, the solid and
dotted lines represent the analytical and MC numerical results,
respectively.} \label{FRC1}
\end{center}
\end{figure*}

In section \ref{RTwithFRC} we discussed a
scheme to solve the polarized RT with Faraday rotation and
conversion incorporated. Now we will test its validation by applying
it to solve two RT problems that have analytic solutions presented
in Appendix \ref{RT_FRC_solutions}. Similar tests have been carried
out by \citet{dexter2016} and \citet{mosc2018} using alternative
numerical methods. We emphasize that if there has no scattering in
these RT problems, the MC method just provides another choice to
solve them and has no any advantages comparing to these numerical
methods. The advantage of our method lies in solving the RT with
scattering. The results of the two tests are demonstrated in Fig.
\ref{FRC1}, where the solid and dotted lines represent the
analytical and our numerical results, respectively. One can see that
they are consistent with each other very well.

\section{DISCUSSION}
\label{sec5_discussion}

In this paper we have proposed a new MC scheme based on Neumann
series solutions of Fredholm integral equations of second kind to
solve the RT problems with scattering incorporated. From this
perspective, the problem of RT is equivalent to evaluate infinite
terms of multiple integrals simultaneously and the MC method is
naturally introduced since it has a remarkable efficiency in
evaluating multiple integrals. Due to the special structure of
Neumann solution, the procedure of integral evaluation amounts to
generate random sequences in the phase space. The combination of
these sequences with their corresponding weight functions can
immediately give the estimations for the observational quantities.

We have reviewed and redescribed the procedure to solve an RTE by
the MC method completely and systematically from the perspective of
integral equation and its Neumann solution. We particularly
emphasize that all complicated and abstract MC sampling algorithms
for the transport and scattering of a photon can be understood and
explained mathematically. Meanwhile we try to ignore the
corresponding physical explanations. Such a treatment not only makes
the MC method to be more understandable but also can avert mistakes
caused by physical intuitions. Most importantly it enables one to
have the flexibility of choosing various sampling PDFs and
corresponding weight functions conveniently. These choices are
equivalent but with different computational efficiency and accuracy.
Particularly the widely adopted conventional photon tracing scheme
can be regarded as a special choice, where the PDF for position
transport has definitions in the whole space.

We emphasize the importance of recording function which connects the
Neumann solution with the observational quantities directly.
Especially, any delta function (which is exactly the reason why the
photon tracing scheme has a low computational efficiency, since
tremendous calculations are inevitably abandoned) contained in the
recording function can be eliminated in advance. After that we can
choose anther recording function, which enables that any sample in a
random sequence can make contribution to the quantity that is under
calculation. This estimation strategy can significantly improve the
calculation accuracy, especially for systems with axial or spherical
symmetry. Usually under the same condition our results have a
relatively higher precision comparing to that of the photon tracing
scheme. In some ideal situations, the precision is even comparable
with the semi-analytical results. But for each estimation, our
scheme requires to calculate the recording function and total
optical depth from the scattering site to the boundary of the
region. It will inevitably occupy additional computational
resources.

This new scheme can be directly extended to deal with polarized
radiative transfer processes, especially for that the Faraday
rotation and conversion effects are incorporated. The polarizations
are described by the SPs and the RTEs become a set of
differentio-integral equations. The key point is that one should
choose common transport PDFs (including position transport and
scattering) that is shared by all components of the SPs, meanwhile
their updating ways from the incident direction to the scattered
direction and estimation functions are different. One of the
advantages of our scheme in dealing with the polarized Compton
scattering is that we do not have to carry out the Lorentz
transformations in terms of SPs between different references. This
procedure that involves complicated SPs transformations and
rotations \citep{kraw2012} now can be avoided.

According to the scheme, a public available MCRT code, named Lemon,
has been developed with Fortran. In the development, we find that
the Object-Oriented programming is very suitable and convenient to
implement our scheme. We first define a basic particle class, which
contains the primary members and methods, such as coordinates,
momentum, energy, weights, etc.. Then we extend this class gradually
by class inheritance and adding new members and methods to the
subclass that can implement various special functions. Each
extension corresponds one function and the relevant variables and
subroutines of which are saved in a single module. These functions
include initiating all parameters, starting a sequence, position
transport, photon scattering and quantity estimation and so on. As
one of the functions is modified, any other will not be affected.
Finally we can instantiate these subclasses and obtain different
objects. Calling the object's methods we can generate a scattering
sequences and estimate the observational quantity as the generation
goes on.

In order to verify the correctness of the scheme and the code, we
have reproduced a lot of classic results and made comparisons with
them, which have demonstrated the validation and advantages of our
scheme. Since the algorithms for sampling various PDFs are crucial
for MC method, they have been discussed in detail and the proofs for
some of them are also provided in the appendix.

MCRT has an intrinsic parallelization property due to the
calculations of random sequences are independent from each other. In
order to speed up the calculations, the MPI parallel scheme has been
incorporated into the code. In the implementation of
parallelization, all of the MPI processes are treated equally, i.e.,
they are initiated with the same parameters, load the same data
files (that contain the tabulated total scattering cross section,
the temperature and the electron number density distributions, etc.)
and share the same emissivity, scattering and absorption
coefficients. Which means that each MPI process requires the access
to the full radiating domain. And this requirement can be easily
fulfilled for all of our test examples, since they are relatively
simple (or the radiating domain can be described semi-analytically)
and do not involve any applications to complicated numerical
simulations. The sizes of these data files are small and can be
easily loaded by each process.

Without any applications to numerical simulations is really the
shortage of our code testing. Considering the extreme significance
and widely applications of magnetohydrodynamic simulations in
nowadays, we really hope that we can apply our code to these
simulations in the near future. Numerical simulations with large
simulation domains can no longer be load by each process. To
overcome these difficulties, new parallelization algorithms and
techniques are in desperate needed (e.g., down sampling of the
simulation grid and domain decomposition). And the structure of MPI
parallelization implemented in the present code should be redesigned
completely.

Anther limit of our code testing is that all of the test problems
are confined to static configurations. However the fluid flows in
numerical simulations usually have large nonuniform velocities that
even can be comparable to the speed of light. So can our code handle
the RT processes in fluid flow with irregular velocity field
consistently? The answer is yes or in principle it can. The
procedure can be stated as follows.

First we must choose a global and static reference, since the
Neumann solution consists of integrals defined over the entire
radiative domain. Such reference obviously exists, in which the
velocity field of the fluid flow is also appropriately defined on a
grid. Next our scheme requires the mathematical expressions for all
relevant quantities of the fluid flow (such as the emissivity,
absorption and scattering coefficients, the mass and particle number
density, temperature distribution) to be explicitly given in this
frame. These quantities are well defined in the comoving (or rest)
frame of the fluid flow. So simply through a Lorentz transformation,
we can obtain these expressions in the global static frame (one
should also notice that in order to get the values of these
quantities at an arbitrary position from the simulation data that
defined on a grid, a suitable interpolation scheme should be
introduced). The Lorentz transformations for these quantities have
been extensively studied and can be found in many textbooks (e.g.,
in the appendix of \citet{Pom1973}). With these quantities are
specified, we can write down the RTE immediately and obtain the
Neumann solution as well. Then the following steps are just routines
that have been discussed in the previous sections.

Another problem needs be carefully treated when dealing with RT in
optically thick media with our scheme. In this special case, the
aforementioned criterion (as the weight of these photons falling
below a threshold) to truncate a scattering sequence will become
extremely inefficient and therefore should be rejected. The reason
can be stated as follow. Since the optically thick region tends to
trap any photon that travels into it, the samples of a scattering
sequence could be disproportionately distributed over this region.
Due to the factor $\exp(-\tau)$ in the recording function, the large
optical depth will heavily attenuate the contributions made by these
samples to the observed flux. Thus the whole contribution of the
sequence is quite low. Meanwhile the evaluation of $\tau$ for each
sample is time consuming in optical thick region, especially for
post-processing of finite volume simulations with large grids. The
situation can be even worse if the weight decreases slowly, such as
in cases where the true absorption can be neglected or no absorption
exists at all. If we still require the sequence to be truncated as
the weight is smaller than the threshold, the scattering times will
be a quite large number, which makes the computational cost for
evaluating $\tau$ becomes unacceptable. The cost almost offsets the
limited gain in the lower variance.

We think the best way to alleviate this difficulty is to truncate
the scattering sequence as soon as possible. Once the photon is
scattered into high optical depth region, the frequent scattering
will trap it and prevent it to escape from the region. Considering
the high computational costs and low contributions made by the
subsequent samples, we can safely truncate the sequence once the
optical depth of a sample is larger than a threshold.

Another useful truncation standard is related to the scattering
numbers. For systems with low scattering optical depth, we
approximately have $\exp(-\tau_s)\approx1+\tau_s$ and
$A=1-\exp(-\tau_s)\approx\tau_s$, then $I_m\propto A_1A_2\cdots
A_m\exp(-\tau_s)\approx\tau_s^m$, which will decrease rapidly as $m$
increases. So we can truncate the sequence as $m$ is greater than a
particular number. For example, from Fig. 18, one can see that the
contributions made by photons that have been scattered for more than
4 times can be safely rejected.

In present paper, the discussions are mainly restricted to the
classic RT problems in flat spacetime. Our scheme is dependent
heavily on the specific forms of RTEs, which should be given
explicitly to obtain the Neumann solution and the transport kernel.
In curved spacetime, the RTEs are usually given in a covariant form
\citep{younsi2012, gammie2012, bron2020}, but the scattering
incorporated RTEs are rarely provided completely and explicitly,
especially for the polarized situations. One exception is in
cosmology regime, where in order to study the polarized radiative
transfer and perturbations of CMB in the cosmic metrics, the RTEs
with polarized Compton scattering have been extensively discussed
\citep{koso1996, Ports2004, weinberg2008}. For the MCRTs with
scattering surround Black holes, instead of writing down the RTEs
and starting from the Neumann solutions, one prefers to generate
photons and trace them until to the endings \citep{Dolence2009,
Schnittman2013, zhangwd2019}, and the polarizations are usually
described by the so-called polarization vectors, which are parallel
transported along geodesics. In the future, we hope to extend our
scheme from flat to curved spacetime and incorporate it with the
results of GRMHD simulations to mimic more real RT processes, which
are crucial to interpret the polarized observations generated by
accreting black holes.

Finally we want to point out a difference between flat and curved
spacetime which may cause difficulties for our scheme. In flat
spacetime, once the line of sight of the observer,
$\mathbf{\Omega}_{\text{obs}}$, is provided, the scattered direction
$\mathbf{\Omega}'$ in the scattering kernel $C(\nu,
\mathbf{\Omega}\rightarrow\nu', \mathbf{\Omega}')$ can be replaced
by $\mathbf{\Omega}_{\text{obs}}$ directly and $C(\nu,
\mathbf{\Omega}\rightarrow\nu', \mathbf{\Omega}_{\text{obs}})$ will
be taken as a weight for estimations. While in curved spacetime, due
to the light bending, we can not carry out such a replacement
anymore, unless we determine the geodesic connecting the observer
and the scattering point and then $\mathbf{\Omega}'$ should be
replaced by the unit tangent vector of the geodesic at the
scattering point rather than $\mathbf{\Omega}_{\text{obs}}$. It
implies that for each estimation of our scheme in curved spacetime,
a geodesic with two fixed points (i.e., observer at infinity and the
scattering point) needs be determined additionally. Fortunately,
some excellent codes already exist for computing the photon
geodesics in curved spacetime. Especially the public available code
geokerr \citep{dexter2009} can compute the geodesics between two
points in Kerr metric, even without full integrations.

\acknowledgments
\section*{Acknowledgments}
We thank a lot for the anonymous referee's valuable comments and
suggestions on our manuscript, which have improved the quality of
the manuscript and the results greatly. YXL thanks Yuan Zun-li,
Zhang Guo-bao, Hou Xian and Zhang Wen-da for helpful discussions.
Especially, we thank Zhang Guo-bao and Hou Xian for their
generosities of allowing us to use their personal workstations
freely, without which the calculations in this manuscript are
impossible to be finished. We also thank the Yunnan Observatories
Supercomputing Platform, on which our code is partly tested. We
acknowledge the financial supports from the National Natural Science
Foundation of China (grant Nos. 11573060, 11661161010, U1838116, 11673060,
U2031111, U1931204, 12073069) and Yunnan Natural Science Foundation (Nos. 2019FB008).


\begin{appendix}

\section{the proof of the algorithm for position transport}
\label{app1} Here we present a proof for Algorithm \ref{algo1} that
randomly generates a scattering distance $s$. What we need to prove
is that the probability $p(s)$ for a sample $s$ is selected by this
algorithm is proportional to $\sigma_s(s)\exp\left(-\int_0^{s}
\sigma_s(l)dl\right)$, where
$\sigma_s(s)=\sigma_s(\mathbf{r}+s\mathbf{\Omega}, \nu,
\mathbf{\Omega})$ is the scattering coefficient. According to the
algorithm, we know that an $s$ can be expressed as the sum of $s_i$:
$s=s_1+s_2+\cdots+s_n$, where $n=1, 2, \cdots$, and $s_i$ are
samples of a random variable $l$ with the PDF given by
$\displaystyle\widetilde{p}_1(l)=
\sigma_{\text{max}}\exp(-\sigma_{\text{max}}l)/N,\, 0\le l\le
s_{\text{max}}$, and $N=1-\exp(-\sigma_{\text{max}}s_{\text{max}})$
is the normalization factor. Suppose $p_n(s)$ is the probability
that the $s_i$ ($i=1,\cdots,n$) are generated and their sum is
accepted as a sample of $s$. Then the total probability
corresponding an $s$ is selected can be expressed as
\begin{eqnarray}
 p(s) = \sum^\infty_{n=1}p_n(s).
\end{eqnarray}
Obviously for $n=1$, we have
\begin{eqnarray}
\displaystyle p_1(s)=\widetilde{p}_1(s)\frac{\sigma_s(s)}{\sigma_{\text{max}}}=\frac{\sigma_s(s)}{N}\exp(-\sigma_{\text{max}}s).
\end{eqnarray}
As $n=2$, we have $s=s_1+s_2$, and the probability of generating $s_1,
s_2$ by sampling $\widetilde{p}_1(l)$ and accepting them can be written as
\begin{eqnarray}
   \begin{split}
  &p(s_1) \displaystyle= \widetilde{p}_1(s_1)\left(1-\frac{\sigma_s(s_1)}{\sigma_{\text{max}}}\right)\widetilde{p}_1(s_2)
  \frac{\sigma_s(s_1+s_2)}{\sigma_{\text{max}}}N \displaystyle=\frac{\sigma_s(s)}{N}\exp(-\sigma_{\text{max}}s) [\sigma_{\text{max}}-\sigma_s(s_1)].
   \end{split}
\end{eqnarray}
Notice that $s_1$ can actually take any value on interval $(0, s)$
and each value corresponds a probability $p(s_1)$, thus the total
probability should be
\begin{eqnarray}
   \displaystyle p_2(s) =\int_0^sp(s_1)ds_1=\frac{\sigma_s(s)}{N}\exp(-\sigma_{\text{max}}s) \int_0^s[\sigma_{\text{max}}-\sigma_s(s_1)]ds_1.
\end{eqnarray}
Similarly, for $n=3$, we have
\begin{eqnarray}
   \begin{split}
  p(s_1, s_2)& \displaystyle= \widetilde{p}_1(s_1)\left(1-\frac{\sigma_s(s_1)}{\sigma_{\text{max}}}\right)\widetilde{p}_1(s_2)
  \left(1-\frac{\sigma_s(s_1+s_2)}{\sigma_{\text{max}}}\right)N\widetilde{p}_1(s_3)\frac{\sigma_s(s)}{\sigma_{\text{max}}}N\\
&\displaystyle=\frac{\sigma_s(s)}{N}\exp(-\sigma_{\text{max}}s) [\sigma_{\text{max}}-\sigma_s(s_1)] [\sigma_{\text{max}}-\sigma_s(s_1+s_2)].
   \end{split}
\end{eqnarray}
Notice that $s_1$ can also take any value on interval $(0, s)$ and
$s_2$ on $(0, s-s_1)$, thus we have
\begin{eqnarray}
   \begin{split}
  &p_3(s)\displaystyle=\int_0^sds_1\int_{0}^{s-s_1}ds_2 p(s_1, s_2) = \frac{\sigma_s(s)}{N}\exp(-\sigma_{\text{max}}s)  \int_0^sds_1\int_{0}^{s-s_1}ds_2
  [\sigma_{\text{max}}-\sigma_s(s_1)] [\sigma_{\text{max}}-\sigma_s(s_1+s_2)].
   \end{split}
\end{eqnarray}
Then the expression for $p_n(s)$ can be obtained directly as
\begin{eqnarray}
   \begin{split}
  &p_n(s)\displaystyle=\frac{\sigma_s(s)}{N}\exp(-\sigma_{\text{max}}s)  \int_0^sds_1\int_{0}^{s-s_1}ds_2
  \cdots\int_{0}^{s-(s_1+\cdots+s_{n-2})}ds_{n-1}f(s_1)f(s_1+s_2) \cdots f(s_1+\cdots+s_{n-1}),
   \end{split}
\end{eqnarray}
where $f(s) = \sigma_{\text{max}}-\sigma(s)$. We introduce a set of
new variables $y_1, \cdots, y_{n-1}$ defined by
\begin{eqnarray}
y_1=s_1, \,\,\,\,y_2=s_1+s_2, \,\,\,\,\cdots, \,\,\,\,y_{n-1}=s_1+\cdots+s_{n-1},
\end{eqnarray}
we have $s_1=y_1, \,\,s_2=y_2-y_1, \,\,\cdots,
\,\,s_{n-1}=y_{n-1}-y_{n-2}$. One can easily show that the Jacobian
of this transformation is $J=\left|
\partial(s_1,\cdots,s_{n-1})/\partial(y_1,\cdots,y_{n-1})\right|=1$.
So the volume element is unchanged, i.e., $dy_1dy_2\cdots
dy_{n-1}=ds_1ds_2\cdots ds_{n-1}$. Then $p_n(s)$ can be rewritten as
\begin{eqnarray}
  &p_n(s)\displaystyle=\frac{\sigma_s(s)}{N}\exp(-\sigma_{\text{max}}s)
  \int_0^sdy_1\int_{y_1}^{s}dy_2\cdots \int_{y_{n-2}}^{s}dy_{n-1}f(y_1)f(y_2) \cdots f(y_{n-1}).
\end{eqnarray}
Notice that
\begin{eqnarray}
\label{identity}
\int_0^{s}dy_1\int_{y_1}^sdy_2\cdots\int_{y_{n-2}}^{s}dy_{n-1}
f(y_1) \cdots f(y_{n-1})=\frac{1}{(n-1)!}\left[\int_0^sf(y)dy\right]^{n-1},
\end{eqnarray}
we have
\begin{eqnarray}
p_n(s) \displaystyle=\frac{\sigma_s(s)}{N}\exp(-\sigma_{\text{max}}s)\frac{1}{(n-1)!}
\left[\int_0^s[\sigma_{\text{max}}-\sigma_s(y)]dy\right]^{n-1}.
\end{eqnarray}
Eventually we arrive at
\begin{eqnarray}
   \begin{split}
p(s) &= \sum_{n=1}^{\infty} p_{n}(s) = \frac{\sigma_s(s)}{N}\exp(-\sigma_{\text{max}}s)\sum_{n=1}^{\infty}
\frac{1}{(n-1)!}\left[ \int_0^s[\sigma_{\text{max}}-\sigma_s(y)]dy\right]^{n-1}\\
&=\frac{\sigma_s(s)}{N}\exp(-\sigma_{\text{max}}s) \exp\left[\sigma_{\text{max}}s- \int_0^s\sigma_s(y)dy\right]
=\frac{\sigma_s(s)}{N} \exp\left[ - \int_0^s\sigma_s(y)dy\right],
   \end{split}
\end{eqnarray}
i.e., $p(s)\propto \sigma_s(s)\exp\left(-\int_0^{s} \sigma_s(l)dl\right)$.

\section{The Sampling Algorithm For PDFs with Triangular functions}
\label{app1_1}

\begin{figure}[t]
\begin{center}
  \includegraphics[scale=0.72]{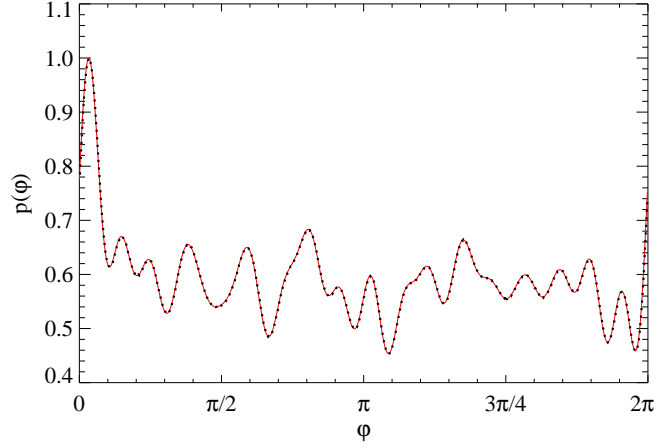}
\caption{A testing example of Algorithm. \ref{algo2} for sampling a
PDF given by Eq. \eqref{mu_tri_func1}. The coefficients $a_i$ and
$b_i$ are all random numbers and $M=40,\, N=30$. The function
$p(\varphi)$ is plotted in red solid line, while the normalized
sample numbers generated by the algorithm is plotted in black dotted
line.} \label{appendix1s}
\end{center}
\end{figure}

When dealing with scattering processes, one often encounters PDFs
that consist of triangular functions, for example
$p(\varphi)=a_0+a_1\cos\varphi+a_2\cos2\varphi+b_1\sin\varphi+b_2\sin2\varphi$.
If one uses the inverse CDF method to sample such an PDF, a
transcendental equation which is analogous to Kepler equation but
more complicated needs to be solved. However, we can construct a new
algorithm that utilizes the properties of piecewise central symmetry
and the odd and even of triangular functions to sample a more
general triangular function PDF $p(\varphi)$ given by
\begin{eqnarray}
  \label{mu_tri_func1}
  p(\varphi)=a_0+\sum_{m=1}^{M} a_m\cos m\varphi+\sum_{n=1}^N b_n\sin n\varphi,
\end{eqnarray}
where $M$ and $N$ can be any positive integers, the coefficients
$a_i$ and $b_i$ are given in a way that $p(\varphi)$ is always
positive. We introduce some new constants defined as
\begin{eqnarray}
 p_m=\frac{|a_m|}{a_0}, \quad p_0= \displaystyle 1-\sum_{m=1}^{M} p_m, \quad  \displaystyle P_m=\sum_{i=0}^mp_i,
\end{eqnarray}
which satisfy the conditions: $0\le p_i < 1$ and
$p_0+p_1+\cdots+p_M=1$. Then $p(\varphi)$ can be recast as
\begin{eqnarray}
  p(\varphi)=F(\varphi) + H(\varphi),
\end{eqnarray}
where
\begin{eqnarray}
  &\displaystyle F(\varphi)=a_0\left[p_0+\sum_{m=1}^{M} p_m(1+s_m\cos m\varphi)\right], \quad\displaystyle H(\varphi)=\sum_{n=1}^N b_n\sin n\varphi,
\end{eqnarray}
and $s_m=\text{sign}(a_m)$. The sampling procedure of $p(\varphi)$
is provided in Algorithm. \ref{algo2}, where $\xi_i$ are random
numbers, $n=[\varphi_t/(\pi/m)]$ and $[x]$ represents the function
that gets the maximum integer that is smaller than $x$. A test
example of this algorithm is given in Fig. \ref{appendix1s}.

\begin{algorithm}
   \label{algo2}
    \caption{Triangular PDF Sampling Algorithm}
    \textcircled{1}  Let $\varphi_t = 2\pi\cdot\xi_1$ \\
    \If{$\left(\displaystyle 0 \le \xi_2 < P_{0} \right)$}{
        $\varphi_\xi=\varphi_t$
    }
    \ElseIf{$\left(\displaystyle P_{m-1}\le \xi_2 < P_{m} \right)$}{
        \If{$2\xi_3 \le 1+s_m\cos m\varphi$}{
            $\varphi_\xi=\varphi_t$
        }
        \Else{$\varphi_\xi=(2n+1)\pi/m-\varphi_t$}
    }
    \If{$\left(\displaystyle\xi_4-1 \le \frac{H(\varphi_\xi) }{F(\varphi_\xi)
    }\right)$}{$\quad\varphi =\varphi_\xi$}
    \Else{$\varphi =2\pi-\varphi_\xi$}
\end{algorithm}

\section{The Sampling Procedures For Rayleigh's Scattering matrix in a static frame}
\label{app2}

In section \ref{diffuseReflec}, we discussed the angular dependent
SPs of radiations that are diffusely reflected from a plane-parallel
semi-infinite atmosphere. The equations describing the RT in such a
system are given by Eqs. \eqref{diffrefl}, in which the Rayleigh's
scattering matrix $\mathbf{P}$ is given by (Eqs. (220)-(224) of
\citep{Chand1960} Chap. I, page 42)
\begin{eqnarray}
  \displaystyle \mathbf{P}(\mu, \varphi;\mu',\varphi')=\mathbf{Q}[\mathbf{P}^{(0)}(\mu, \mu')+
\sqrt{1-\mu^2}\sqrt{1-\mu'^2}\mathbf{P}^{(1)}(\mu, \varphi;\mu',\varphi')+
\mathbf{P}^{(2)}(\mu, \varphi;\mu',\varphi')],
\end{eqnarray}
where
\begin{eqnarray}
  \displaystyle \mathbf{Q}= \left(
  \begin{array}{cccc}
  \displaystyle  1 &  \quad0 &  \quad0 &  \quad0 \\
  \displaystyle  0 &  \quad1 &  \quad0 &  \quad0 \\
  \displaystyle  0 &  \quad0 &  \quad2 &  \quad0 \\
  \displaystyle  0 &  \quad0 &  \quad0 &  \quad2 \\
   \end{array} \right), \quad\quad
  \displaystyle \mathbf{P}^{(0)}(\mu, \mu')=\frac{3}{4}\left(
  \begin{array}{cccc}
  \displaystyle 2(1-\mu^2)(1-\mu'^2)+\mu^2\mu'^2 & \quad\mu^2 &  \quad 0 &  \quad 0 \\
  \displaystyle  \mu'^2 &  \quad1 &  \quad0 &  \quad0 \\
  \displaystyle  0 &   \quad0 &  \quad0 &  \quad0 \\
  \displaystyle  0 &  \quad0 &  \quad0 &  \quad\mu\mu' \\
   \end{array} \right),
\end{eqnarray}

\begin{eqnarray}
  \displaystyle \mathbf{P}^{(1)}(\mu, \varphi;\mu',\varphi')=\frac{3}{4}\left(
  \begin{array}{cccc}
  \displaystyle 4\mu\mu'\cos(\varphi'-\varphi) & \quad0 &  \quad 2\mu\sin(\varphi'-\varphi) &  \quad 0 \\
  \displaystyle  0 &  \quad0 &  \quad0 &  \quad0 \\
  \displaystyle  -2\mu'\sin(\varphi'-\varphi) &   \quad0 &  \quad\cos(\varphi'-\varphi)&  \quad0 \\
  \displaystyle  0 &  \quad0 &  \quad0 &  \quad\cos(\varphi'-\varphi) \\
   \end{array} \right),
\end{eqnarray}

\begin{eqnarray}
  \displaystyle \mathbf{P}^{(2)}(\mu, \varphi;\mu',\varphi')=\frac{3}{4}\left(
  \begin{array}{cccc}
  \displaystyle \mu^2\mu'^2\cos2(\varphi'-\varphi) & \quad -\mu^2\cos2(\varphi'-\varphi) &
        \quad \mu^2\mu'\sin2(\varphi'-\varphi) &  \quad 0 \\
  \displaystyle  -\mu'^2\cos2(\varphi'-\varphi) &  \quad\cos2(\varphi'-\varphi) &
        \quad-\mu'\sin2(\varphi'-\varphi)  &  \quad0 \\
  \displaystyle  -\mu\mu'^2\sin2(\varphi'-\varphi) &  \quad\mu\sin2(\varphi'-\varphi) &
        \quad\mu\mu'\cos2(\varphi'-\varphi)  &  \quad0 \\
  \displaystyle  0 &  \quad0 &  \quad0 &  \quad 0 \\
   \end{array} \right).
\end{eqnarray}
Notice that the above matrices are defined with respect to SPs $I_l,
I_r, U, V$ rather than $I, Q, U, V$. In MCRT, we prefer to use $I,
Q, U, V$ to describe the polarizations instead, so we need to make a
transformation to get a new matrix $\widetilde{\mathbf{P}}$ that is
defined in terms of $I, Q, U, V$. One can show that the elements of
the new matrix are given by
\begin{eqnarray}
 \left\{
  \begin{array}{llll}
\displaystyle \widetilde{P}_{11} = ( P_{11} + P_{12} + P_{21} + P_{22})/2, &\quad
           \widetilde{P}_{12}= ( P_{11} - P_{12} + P_{21} - P_{22})/2, &\quad
           \widetilde{P}_{13}=  P_{13} + P_{23} , &\quad \widetilde{P}_{14}=0, \\
\displaystyle \widetilde{P}_{21}= ( P_{11} + P_{12} - P_{21} - P_{22})/2, &\quad
           \widetilde{P}_{22}=  ( P_{11} - P_{12} - P_{21} + P_{22})/2, &\quad
           \widetilde{P}_{23}=  P_{13} - P_{23} , &\quad\widetilde{P}_{24}=0,\\
\displaystyle \widetilde{P}_{31}=  ( P_{31} + P_{32} )/2, &\quad
           \widetilde{P}_{32}=  ( P_{31} - P_{32} )/2,&\quad
           \widetilde{P}_{33}=  P_{33},&\quad\widetilde{P}_{34}=0, \\
  \widetilde{P}_{41}=0, &\quad\widetilde{P}_{42}=0, &\quad\widetilde{P}_{43}=0,
        &\quad\widetilde{P}_{44}=P_{44}.
   \end{array} \right.
\end{eqnarray}
The updating relationship for the quantity
$\mathbf{\Psi}=(\Psi^{(I)}, \Psi^{(Q)}, \Psi^{(U)}, \Psi^{(V)})^T$
from $m$ to $m+1$ can be obtained directly by multiplying
$\mathbf{\Psi}$ with $\widetilde{\mathbf{P}}$ from the left side,
i.e.,
\begin{eqnarray}
 \left.
  \begin{array}{ll}
\displaystyle   \Psi^{(I)}_{m+1}=\widetilde{P}_{11}\Psi^{(I)}_{m} + \widetilde{P}_{12}\Psi^{(Q)}_{m}
      + \widetilde{P}_{13}\Psi^{(U)}_{m} ,\quad
&\displaystyle  \Psi^{(Q)}_{m+1} =\widetilde{P}_{21}\Psi^{(I)}_{m} +
\widetilde{P}_{22}\Psi^{(Q)}_{m}
      + \widetilde{P}_{23}\Psi^{(U)}_{m} ,\\
\displaystyle   \Psi^{(U)}_{m+1} =\widetilde{P}_{31}\Psi^{(I)}_{m} +
\widetilde{P}_{32}\Psi^{(Q)}_{m}
      + \widetilde{P}_{33}\Psi^{(U)}_{m} ,\quad
&\displaystyle  \Psi^{(V)}_{m+1} =  \widetilde{P}_{44}\Psi^{(V)}_{m}.
   \end{array} \right.
\end{eqnarray}
Or in a more compact form
\begin{eqnarray}
  \begin{array}{ll}
  \displaystyle  \Psi^{(I)}_{m+1}= f_I(\mu, \varphi;\mu', \varphi')\Psi^{(I)}_{m} ,\quad
 &\displaystyle  \Psi^{(Q)}_{m+1} =f_Q(\mu, \varphi;\mu', \varphi')\Psi^{(I)}_{m} ,\\
  \displaystyle  \Psi^{(U)}_{m+1} =f_U(\mu, \varphi;\mu', \varphi')\Psi^{(I)}_{m} ,\quad
 &\displaystyle  \Psi^{(V)}_{m+1} =f_V(\mu, \varphi;\mu', \varphi')\Psi^{(V)}_{m},
   \end{array}
\end{eqnarray}
where
\begin{eqnarray}
 \left\{
  \begin{array}{ll}
   f_I(\mu, \varphi;\mu', \varphi') = &\displaystyle\frac{3}{32\pi}\left[ 2(1-\mu^2)(1-\mu'^2)(1+Q_{m}) + (1+\mu^2)(1+\mu'^2-(1-\mu'^2)Q_{m}) \right. \\
      & \displaystyle + 4\mu\mu'\sqrt{1-\mu^2}\sqrt{1-\mu'^2}(1+Q_{m})\cos(\varphi'-\varphi)+ 4\mu\sqrt{1-\mu^2}\sqrt{1-\mu'^2} U_{m} \sin(\varphi'-\varphi) \\
      & \left. + (1-\mu^2)(1-\mu'^2)-(1-\mu^2)(1+\mu'^2)Q_{m}\cos2(\varphi'-\varphi)  -2(1-\mu^2) \mu' U_{m}\sin2(\varphi'-\varphi) \right],\\
   f_Q(\mu, \varphi;\mu', \varphi') = &\displaystyle\frac{3}{32\pi} [ (1-\mu^2)(1-\mu'^2)(2+3Q_{m})
          -(1-\mu^2)(1+\mu'^2) + 4\mu\sqrt{1-\mu^2}\sqrt{1-\mu'^2} U_{m} \sin(\varphi'-\varphi)  \\
      & + 4\mu\mu'\sqrt{1-\mu^2}\sqrt{1-\mu'^2}(1+Q_{m}) \cos(\varphi'-\varphi)  \\
      & + (1+\mu^2)(\mu'^2-1+(1+\mu'^2)Q_{m})\cos2(\varphi'-\varphi)  + 2(1+\mu^2)\mu'\sin2(\varphi'-\varphi)],\\
    f_U(\mu, \varphi;\mu', \varphi') = &\displaystyle \frac{3}{32\pi} [ 4\sqrt{1-\mu^2}\sqrt{1-\mu'^2}U_m
        \cos(\varphi'-\varphi) -4\mu'(1+Q_m\mu)\sin(\varphi'-\varphi)\\
      & + 2\mu\mu'U_m \cos2(\varphi'-\varphi) - 2\mu(1+\mu'^2(1+Q_m))\sin2(\varphi'-\varphi)], \\
    f_V(\mu, \varphi;\mu', \varphi') = &\displaystyle\frac{3}{8\pi} [  \mu\mu'  +\sqrt{1-\mu^2}\sqrt{1-\mu'^2}\cos2(\varphi'-\varphi) ],
   \end{array}\right.
\end{eqnarray}
and $Q_m=\Psi^{(Q)}_{m}/\Psi^{(I)}_{m}$ and
$U_m=\Psi^{(U)}_{m}/\Psi^{(I)}_{m}$. As aforementioned in the Section.
\ref{section4.3}, we will take $f_I(\mu, \varphi;\mu',
\varphi')$ as the PDF to sample the scattered direction $(\mu,
\varphi)$ when the incident direction $(\mu', \varphi')$ are
provided. We can recast it into a more compact form given as follows
\begin{eqnarray}
\label{expression_f_I} \displaystyle f_I(\mu, \varphi;\mu',
\varphi') = \frac{3}{32\pi}(f_1 +g_2\cos\varphi+
g_3\sin\varphi+g_4\cos2\varphi+g_5\sin2\varphi ),
\end{eqnarray}
where
\begin{eqnarray}
 \left\{
  \begin{array}{l }
  \displaystyle f_1(\mu, \mu') = 2(1-\mu^2)(1-\mu'^2)(1+Q_{m}) + (1+\mu^2)[1+\mu'^2-(1-\mu'^2)Q_{m}],\\
  \displaystyle  f_2(\mu, \mu') = 4\mu\mu'\sqrt{1-\mu^2}\sqrt{1-\mu'^2}(1+Q_{m}),\\
  \displaystyle  f_3(\mu, \mu') = 4\mu\sqrt{1-\mu^2}\sqrt{1-\mu'^2} U_{m}, \\
  \displaystyle  f_4(\mu, \mu') = (1-\mu^2)(1-\mu'^2)-(1-\mu^2)(1+\mu'^2)Q_{m}, \\
  \displaystyle  f_5(\mu, \mu') = -2(1-\mu^2) \mu' U_{m}, \\
  \displaystyle  g_2 = f_2\cos\varphi' + f_3\sin\varphi',
\quad\quad\quad g_3 = f_2\sin\varphi' - f_3\cos\varphi',\\
  \displaystyle g_4 = f_4\cos2\varphi' + f_5\sin2\varphi',
 \quad\quad g_5 = f_4\sin2\varphi' - f_5\cos2\varphi'.
   \end{array} \right.
\end{eqnarray}
Then one can readily show that PDF $p(\mu, \varphi)=f_I(\mu,
\varphi;\mu', \varphi')$ is automatically normalized. Integrating out
$p(\mu, \varphi)$ over $\varphi$, we get the marginal PDF of $\mu$
as
\begin{eqnarray}
  \begin{array}{l}
\displaystyle p(\mu) = \int_0^{2\pi} p(\mu, \varphi)d\varphi =\frac{3}{16}f_1(\mu, \mu')=
     \frac{3}{16}(c_1\mu^2+c_2),
   \end{array}
\end{eqnarray}
where $c_1= 1+\mu'^2 - (1-\mu'^2)Q_m - 2(1-\mu'^2)(1+Q_m)$ and $c_2=
1+\mu'^2 - (1-\mu'^2)Q_m + 2(1-\mu'^2)(1+Q_m)$. Hence we can use the
inverse CDF method to get the scattered $\mu$, which is the real
root of the equation $c_1\mu^3+3c_2\mu+c_1+3c_2-16\xi=0$ on interval
$[-1, 1]$, where $\xi$ is a random number. When $\mu$ is obtained,
we can immediately get the PDF for $\varphi$ by using the Bayes
formula $p(\varphi|\mu)=p(\mu, \varphi)/p(\mu)$ as
\begin{eqnarray}
\displaystyle p(\varphi|\mu) = \frac{1}{2\pi f_1}(f_1 + g_2\cos\varphi+ g_3\sin\varphi+g_4\cos2\varphi+g_5\sin2\varphi).
\end{eqnarray}
Now the inverse CDF method seems unfeasible to sample
$p(\varphi|\mu)$, since it will give rise to a transcendental
equation in terms of $\varphi$ and more complicated than the Kepler
equation. But utilizing the symmetrical and periodic properties of
triangular functions, we have constructed an Algorithm. \ref{algo2}
that can circumvent this difficulty. In order to sample
$p(\varphi|\mu)$ through Algorithm. \ref{algo2}, we introduce some
relevant quantities given by
\begin{eqnarray}
  \begin{array}{l}
\displaystyle  F(\varphi)=\frac{1}{2\pi f_1}(f_1 + g_2\cos\varphi+ g_4\cos2\varphi), \quad H(\varphi)=\frac{1}{2\pi f_1}(g_3\sin\varphi + g_5\sin2\varphi), \\
\displaystyle p_0=1-p_1-p_2,\quad p_1= \frac{|g_2|}{f_1}, \quad p_2=\frac{|g_4|}{f_1}, \quad  s_1=\text{sign}(g_2),\quad s_2=\text{sign}(g_4).
   \end{array}
\end{eqnarray}
With $\mu$ and $\varphi$ are obtained, we can simply update the
components of scattered $\mathbf{\Psi}$ as
\begin{eqnarray}
 \label{psi_update}
  \begin{array}{ll}
  \displaystyle  \Psi^{(I)}_{m+1} = \Psi^{(I)}_{m} ,\quad &\displaystyle  \Psi^{(Q)}_{m+1} =
  \frac{f_Q(\mu, \varphi;\mu', \varphi')}{f_I(\mu, \varphi;\mu', \varphi')}\Psi^{(I)}_{m} ,\\
  \displaystyle  \Psi^{(U)}_{m+1} =\frac{f_U(\mu, \varphi;\mu', \varphi')}
     {f_I(\mu, \varphi;\mu', \varphi')}\Psi^{(I)}_{m} ,\quad
   &\displaystyle  \Psi^{(V)}_{m+1} = \frac{f_V(\mu, \varphi;\mu', \varphi')}
    {f_I(\mu, \varphi;\mu', \varphi')}\Psi^{(V)}_{m}.
   \end{array}
\end{eqnarray}

\section{The Analytic Solutions for RT equations with Faraday Rotation and Conversion}
\label{RT_FRC_solutions} Now we give the analytical solutions for
two RT problems including Faraday effects. In the first case, only
the $j_{IQ}$ and $\alpha_{IQ}$ are presented and the RTE reads as
\begin{eqnarray}
\label{eq:symchrotroneq1}
\begin{split}
       \frac{d}{ds}
    \begin{pmatrix}
        I\\ Q
    \end{pmatrix}=
    \begin{pmatrix}
       j_I\\j_Q
    \end{pmatrix}-
    \begin{pmatrix}
       \alpha_I & \alpha_Q\\
       \alpha_Q & \alpha_I
    \end{pmatrix}
    \begin{pmatrix}
       I\\ Q
    \end{pmatrix},
\end{split}
\end{eqnarray}
the analytic solution of which can be written as
\begin{eqnarray}
\begin{split}
    I(s) = \frac{j_+}{2\alpha_+}(1-\text{e}^{-\alpha_+ s}) + \frac{j_-}{2\alpha_-}(1-\text{e}^{-\alpha_- s})+
     \frac{1}{2} \text{e}^{-\alpha_+ s}(I_0+Q_0) + \frac{1}{2} \text{e}^{-\alpha_- s}(I_0 - Q_0 ),\\
    Q(s) = \frac{j_+}{2\alpha_+}(1-\text{e}^{-\alpha_+ s}) - \frac{j_-}{2\alpha_-}(1-\text{e}^{-\alpha_- s})+
    \frac{1}{2} \text{e}^{-\alpha_+ s}(I_0+Q_0) - \frac{1}{2} \text{e}^{-\alpha_- s}(I_0 - Q_0 ),
\end{split}
\end{eqnarray}
where $j_\pm=j_I\pm j_Q$, $\alpha_\pm=\alpha_I\pm \alpha_Q$, and
$I_0, Q_0$ are the initial values of $I$ and $Q$ at $s=0$. In the
second case, all of the absorption coefficients $\alpha_{IQUV}$ and
emissivity $j_I$ vanish, the RTE can be written as
\begin{eqnarray}
\label{eq:symchrotroneq2}
    \frac{d}{ds}
    \begin{pmatrix}
        Q \\ U \\ V
    \end{pmatrix}=
    \begin{pmatrix}
       j_Q\\j_U \\j_V
    \end{pmatrix}-
    \begin{pmatrix}
       0       &  \rho_V & -\rho_U  \\
       -\rho_V & 0       &  \rho_Q  \\
        \rho_U & -\rho_Q &  0  \\
    \end{pmatrix}
    \begin{pmatrix}
       Q \\ U \\ V
    \end{pmatrix}.
\end{eqnarray}
Integrating the above equations directly, we have \citep{mosc2018}
\begin{eqnarray}
 \left\{
\begin{split}
     Q(s) =& \frac{\rho_Q}{\rho^2}\rho\cdot\pmb{J}s+(j_Q\rho^2-\rho_Q\rho\cdot\pmb{J})\frac{\sin(\rho s)}{\rho^3}+
           (\rho_U j_V - \rho_V  j_U )\frac{1-\cos(\rho s)}{\rho^2}\\
           &+ Q_0\cos(\rho s)+2\frac{\rho_Q(\rho\cdot\pmb{S})}{\rho^2}\sin^2(\rho s/2)+\frac{\rho_UV_0-\rho_VU_0}{\rho}\sin(\rho s),\\
     U(s) =&\frac{\rho_U}{\rho^2}\rho\cdot\pmb{J}s+(j_U\rho^2-\rho_U\rho\cdot\pmb{J})\frac{\sin(\rho s)}{\rho^3}+
           (\rho_V j_Q - \rho_Q  j_V )\frac{1-\cos(\rho s)}{\rho^2}\\
           &+ U_0\cos(\rho s)+2\frac{\rho_U(\rho\cdot\pmb{S})}{\rho^2}\sin^2(\rho s/2)+\frac{\rho_VQ_0-\rho_QV_0}{\rho}\sin(\rho s),\\
     V(s) =& \frac{\rho_V}{\rho^2}\rho\cdot\pmb{J}s+(j_V\rho^2-\rho_V\rho\cdot\pmb{J})\frac{\sin(\rho s)}{\rho^3}+
           (\rho_Q j_U - \rho_U  j_Q )\frac{1-\cos(\rho s)}{\rho^2}\\
           &+ V_0\cos(\rho s)+2\frac{\rho_V(\rho\cdot\pmb{S})}{\rho^2}\sin^2(\rho s/2)+\frac{\rho_QU_0-\rho_UQ_0}{\rho}\sin(\rho s),
\end{split}
\right.
\end{eqnarray}
where $\rho^2=\rho_Q^2+\rho_U^2+\rho_V^2$,
$\rho\cdot\pmb{S}=\rho_QQ_0+\rho_UU_0+\rho_VV_0$,
$\mathbf{\rho}\cdot\pmb{J}=\rho_Qj_Q+\rho_Uj_U+\rho_Vj_V$, and $Q_0,
U_0, V_0$ are the initial values of $Q, U, V$ at $s=0$.

\section{The Neumann solution as the disk reflection effect
considered} \label{app_reflection}

Now we discuss what should the sampling procedure be modified when
the reflection term $\mathbf{I}_r$ given by Eq.
\eqref{Ir_reflection} is incorporated. Although the discussion is
presented in a plane-parallel geometrical system, it is possible to
extend it to any other systems. Since the reflected radiations can
reenter the corona, they can be regarded as the seed photons that
will initiate another round of RT process. It means that the RT
problem now becomes an iterative one. Assume that
$\mathbf{J}(\tau_0, \mathbf{\Omega}, \nu)$ is the primary radiations
emitted by the disk, the corresponding radiative flux generated by
$\mathbf{J}$ is $\mathbf{\Psi}^{(0)}$, which obviously satisfies the
integral equation
\begin{eqnarray}
\begin{split}
& \mathbf{\Psi}^{(0)}(P)=\mathbf{J}(P)+\int \mathbf{K}(P'\rightarrow
P)\mathbf{\Psi}^{(0)}(P')dP'.
\end{split}
\end{eqnarray}
Their Neumann solutions are
\begin{eqnarray}
\begin{split}
& \mathbf{\Psi}^{(0)}(P)=\sum_{m=0}^{\infty}
\mathbf{\Psi}^{(0)}_m(P), \quad
\mathbf{\Psi}^{(0)}_0(P)=\mathbf{J}(P),
\quad\mathbf{\Psi}^{(0)}_1(P)=\int \mathbf{K}(P'\rightarrow P)\mathbf{J}(P')dP'=\mathbf{J}\mathbf{K} , \cdots,\\
& \mathbf{\Psi}^{(0)}_{m}(P)=\int\mathbf{K}(P_{m-1}\rightarrow
P)\mathbf{K}(P_{m-2}\rightarrow P_{m-1})\cdots
\mathbf{K}(P_{0}\rightarrow P_1)\mathbf{J}(P_1)dP_0\cdots
dP_{m-1}=\mathbf{J}\mathbf{K}\cdots\mathbf{K}=\mathbf{J}\mathbf{K}^m.
\end{split}
\end{eqnarray}
For simplicity, we denote the multiple integrals in terms of
any $P$ in a compact form. Using $\mathbf{\Psi}^{(0)}$, we can
immediately obtain the radiations escaped from the up and down
surfaces of the corona respectively as
\begin{eqnarray}
\begin{split}
\begin{array}{lr}
  \displaystyle\mathbf{I}^{(0)}_{\text{up}}(0,\mathbf{\Omega}',\nu')=\sum_{m=0}^\infty \mathbf{I}^{(0)}_{\text{up},m}
  = \sum_{m=0}^\infty\int_0^{\tau_0}\mathbf{J}\mathbf{K}^m
 \exp\left(-\frac{\tau'}{\mu'}\right)\frac{d\tau'}{\mu'}, &\quad \mu'>0,\\
  \displaystyle\mathbf{I}^{(0)}_{\text{dw}}(\tau_0,\mathbf{\Omega}',\nu')=\sum_{m=0}^\infty \mathbf{I}^{(0)}_{\text{dw},m}
  = \sum_{m=0}^\infty\int_0^{\tau_0}\mathbf{J}\mathbf{K}^m
 \exp\left(-\frac{\tau_0-\tau'}{|\mu'|}\right)\eta(-\mu')\frac{d\tau'}{|\mu'|},&\quad
 \mu'<0,
 \end{array}
\end{split}
\end{eqnarray}
where $\tau_0$ is the optical depth of the corona. Substituting
$\mathbf{I}^{(0)}_{\text{dw}}$ into Eq. \eqref{Ir_reflection} gives
the radiation intensity reflected from the disk as
\begin{eqnarray}
\label{temp_eq1}
\begin{split}
 \mathbf{I}^{(1)}_r(\tau_0,\mathbf{\Omega}, \nu)=\sum_{m=0}^\infty \mathbf{I}^{(1)}_{r,m}(\tau_0,\mathbf{\Omega}, \nu),
\end{split}
\end{eqnarray}
where
\begin{eqnarray}
\label{temp_eq1}
\begin{split}
  \mathbf{I}^{(1)}_{r,m}(\tau_0,\mathbf{\Omega}, \nu) = \int\int_0^{\tau_0} \mathbf{G}(\nu'\rightarrow\nu, \mathbf{\Omega}'\rightarrow \mathbf{\Omega})
\mathbf{J}\mathbf{K}^m\displaystyle p(\tau_0,
\tau')\eta(-\mu')\frac{d\tau'}{|\mu'|}d\mathbf{\Omega}'d\nu'=
\mathbf{J}\mathbf{K}^mp\eta\mathbf{G},
\end{split}
\end{eqnarray}
and $\displaystyle p(\tau_0,
\tau')=\exp\left(-\frac{\tau_0-\tau'}{|\mu'|}\right)$. Now each
$\mathbf{I}^{(1)}_{r,m}(\tau_0,\mathbf{\Omega}, \nu)$ can be taken
as the incident radiation from the bottom surface of the corona, and
the corresponding $\mathbf{\Psi}^{(1)}_m$ satisfies:
\begin{eqnarray}
\begin{split}
& \mathbf{\Psi}^{(1)}_m(P)=\mathbf{I}^{(1)}_{r,m}\mathbf{K}(P)+\int
\mathbf{K}(P'\rightarrow P)\mathbf{\Psi}^{(1)}_m(P')dP'.
\end{split}
\end{eqnarray}
The Neuman solution of $\mathbf{\Psi}^{(1)}_m$ is
$\mathbf{\Psi}^{(1)}_{m,n}=\mathbf{I}^{(1)}_{r,m}\mathbf{K}\cdot\mathbf{K}^n=\mathbf{J}\mathbf{K}^mp\eta\mathbf{G}\mathbf{K}\cdot\mathbf{K}^n$.
Similarly we can get that the reflected radiations generated by
$\mathbf{\Psi}^{(1)}_{m,n}$ is
$\mathbf{I}^{(2)}_r(\tau_0,\mathbf{\Omega},
\nu)=\mathbf{J}\mathbf{K}^mp\eta\mathbf{G}\mathbf{K}\cdot\mathbf{K}^np\eta\mathbf{G}$
and its Neuman solution is given by
$\mathbf{\Psi}^{(2)}_{m,n,l}=\mathbf{J}\mathbf{K}^mp\eta\mathbf{G}\mathbf{K}\cdot\mathbf{K}^np\eta\mathbf{G}\mathbf{K}\cdot\mathbf{K}^l$.
We can repeat this procedure to obtain the Neumann solution
$\mathbf{\Psi}^{(i)}_{m,n,l,\cdots}$ for radiations reflected for
arbitrary times. Then the total Neumann solution for this reflection
RT process can be written as
\begin{eqnarray}
 \left\{
\begin{split}
 \mathbf{\Psi}^{(0)}&=\sum_{m=0}^\infty\mathbf{JK}\cdot\mathbf{K}^m =\mathbf{JK} +\mathbf{JK}\cdot\mathbf{K}+\cdots\mathbf{JK}\cdot\mathbf{K}^m+\cdots\\
\mathbf{\Psi}^{(1)}&=\sum_{m=0,n=0}^\infty\mathbf{JK}\cdot\mathbf{K}^m\cdot(p\eta\mathbf{GK})\cdot\mathbf{K}^n,\\
&=\mathbf{JK}(p\eta\mathbf{GK})+\mathbf{JK}(p\eta\mathbf{GK})\mathbf{K}+\mathbf{JK}(p\eta\mathbf{GK})\mathbf{K}^2+
\cdots\mathbf{JK}(p\eta\mathbf{GK})\mathbf{K}^m+\cdots\\
&+\mathbf{JK}^2(p\eta\mathbf{GK})+\mathbf{JK}^2(p\eta\mathbf{GK})\mathbf{K}+\mathbf{JK}^2(p\eta\mathbf{GK})\mathbf{K}^2+
\cdots\mathbf{JK}^2(p\eta\mathbf{GK})\mathbf{K}^m+\cdots\\
& \cdots \\
&+\mathbf{JK}^m(p\eta\mathbf{GK})+\mathbf{JK}^m(p\eta\mathbf{GK})\mathbf{K}+\mathbf{JK}^m(p\eta\mathbf{GK})\mathbf{K}^2+\cdots+
\mathbf{JK}^m(p\eta\mathbf{GK})\mathbf{K}^m+\cdots\\
& \cdots, \\
\mathbf{\Psi}^{(2)}&=\sum_{m=0,n=0,l=0}^\infty\mathbf{JK}\cdot\mathbf{K}^m\cdot(p\eta\mathbf{GK})\cdot\mathbf{K}^n\cdot(p\eta\mathbf{GK})\cdot\mathbf{K}^l,\\
\mathbf{\Psi}^{(3)}&=\sum_{m=0,n=0,l=0,k=0}^\infty\mathbf{JK}\cdot\mathbf{K}^m\cdot(p\eta\mathbf{GK})\cdot\mathbf{K}^n
\cdot(p\eta\mathbf{GK})\cdot\mathbf{K}^l\cdot(p\eta\mathbf{GK})\cdot\mathbf{K}^k,\cdots,
\end{split}
\right.
\end{eqnarray}
and the total Neumann solution is $\mathbf{\Psi} =
\mathbf{\Psi}^{(0)} + \mathbf{\Psi}^{(1)} + \mathbf{\Psi}^{(3)}+
\cdots+\mathbf{\Psi}^{(m)}+\cdots$. Combining those terms with equal
power index of $\mathbf{K}$, $\mathbf{\Psi}(P)$ can be
reexpressed as:
\begin{eqnarray}
\begin{split}
  \mathbf{\Psi}(P)&=\sum_{k=0}^\infty \mathbf{J}\mathbf{K}\cdot(\mathbf{K}+p\eta\mathbf{GK})^k,
\end{split}
\end{eqnarray}
where
\begin{eqnarray}
\begin{split}
   (\mathbf{K}+p\eta\mathbf{GK})^2 =&\mathbf{K}^2+\mathbf{K}(p\eta\mathbf{GK})+(p\eta\mathbf{GK})\mathbf{K}+(p\eta\mathbf{GK})^2,\\
   (\mathbf{K}+p\eta\mathbf{GK})^3 =&\mathbf{K}^3+\mathbf{K}^2(p\eta\mathbf{GK})+\mathbf{K}(p\eta\mathbf{GK})\mathbf{K}+(p\eta\mathbf{GK})\mathbf{K}^2+
     \mathbf{K}(p\eta\mathbf{GK})^2\\
     &+(p\eta\mathbf{GK})\mathbf{K}(p\eta\mathbf{GK})+(p\eta\mathbf{GK})^2\mathbf{K}+(p\eta\mathbf{GK})^3,\cdots.\\
\end{split}
\end{eqnarray}
Of course we have
\begin{eqnarray}
\begin{split}
  &\mathbf{J}\mathbf{K}(P)= \int \mathbf{K}(P_0\rightarrow P) \mathbf{J}(P_0)dP_0,\\
 & \mathbf{J}\mathbf{K}\cdot(\mathbf{K}+p\eta\mathbf{GK})(P)=
\mathbf{J}\mathbf{K}\mathbf{K}(P)+\mathbf{J}\mathbf{K}p\eta\mathbf{GK}(P)
   =\int \mathbf{K}(P_1\rightarrow P)\int \mathbf{K}(P_0\rightarrow
P_1)\mathbf{J}(P_0)dP_0dP_1  \\
 &\quad\quad\quad\quad\quad\quad\quad\quad\quad\quad+\int\mathbf{K}(P'_1\rightarrow P)
\mathbf{G}(P_1\rightarrow P'_1)
  p(\tau_0,\tau_1)\eta(-\mu_1)\int \mathbf{K}(P_0\rightarrow P_1)\mathbf{J}(P_0)dP_0 dP'_1
dP_1, \cdots.
  &  
\end{split}
\end{eqnarray}
Comparing to the case without reflection, the
Neumann solution is almost the same except that the transport kernel now
becomes $\mathbf{K}+p\eta\mathbf{GK}$. For upward propagating
radiations, we always have $\eta(-\mu)\equiv0$, then the transport
kernel is $\mathbf{K}$. Thus only the cases with $\mu<0$ need to be
considered and the transport kernel becomes
$\mathbf{K}+p\mathbf{GK}$. Suppose that the initial position, direction
and frequency of the radiation are $\tau', \,
\mathbf{\Omega}'=(\mu', \varphi'), \,\nu'$ (notice $\mu'<0$),
the transport kernel is given by $\mathbf{K}=\mathbf{C}(\nu'\rightarrow \nu,
\mathbf{\Omega}'\rightarrow \mathbf{\Omega})\cdot
T_s(\tau'\rightarrow\tau)\cdot w_s$, where $\mathbf{C}$ and
$T_s=\exp[-(\tau-\tau')/|\mu'|]/A$ are the scattering and position
transport kernels respectively, and $w_s=A=1-p(\tau_0, \tau')$ is
the weight. Thus we have
\begin{eqnarray}
\begin{split}
 \mathbf{K}+p\mathbf{GK}=&T_s(\tau'\rightarrow\tau)\mathbf{C}(\nu'\rightarrow \nu, \mathbf{\Omega}'\rightarrow \mathbf{\Omega}|\tau)\cdot (1-p)\\
+&\mathbf{G}(\nu'\rightarrow \nu'', \mathbf{\Omega}'\rightarrow
\mathbf{\Omega}''|\tau_0)T_s(\tau_0\rightarrow\tau)w''_s
\mathbf{C}(\nu''\rightarrow \nu, \mathbf{\Omega}''\rightarrow
\mathbf{\Omega}|\tau)\cdot p,
\end{split}
\end{eqnarray}
where $w''_s=1-\exp(-\tau_0/\mu'')$. From them, we can obtain the
sampling procedure for the transport with reflection. We first
generate a random variable $\xi$, if $\xi\le p$, then the radiation
will be reflected and we sample $\mathbf{G},
T_s(\tau_0\rightarrow\tau)$ and $\mathbf{C}$ in turn to get the new
point: $(\tau, \mathbf{\Omega}, \nu)$; otherwise if $\xi > p$ we
sample $T_s(\tau'\rightarrow\tau)$ and $\mathbf{C}$ instead. This is
exactly the revised sampling procedure as reflection effects are
considered in our MC scheme.

From the above discussions one can see that the sampling procedure
can be obtained correctly from mathematical deductions rather than
physical intuitions. It can help us to avoid some unnecessary
mistakes.
\\


\end{appendix}



\end{document}